\documentclass[11pt]{article}

\usepackage{authblk}  
\usepackage{graphicx} 
\usepackage{epsfig}
\usepackage{dcolumn}  
\usepackage{bm}       
\usepackage{amsmath}
\usepackage{amssymb}

\usepackage{multirow}
\usepackage{booktabs}

\newcommand{\beq}{\begin{equation}} \newcommand{\eeq}{\end{equation}}
\newcommand{\bea}{\begin{eqnarray}} \newcommand{\eea}{\end{eqnarray}}

  \newcommand
{\Romannumeral}[1]{\uppercase\expandafter{\romannumeral#1}}

\newcommand{\be}{\begin{enumerate}} \newcommand{\ee}{\end{enumerate}}
\newcommand{\bi}{\begin{itemize}} \newcommand{\ei}{\end{itemize}}
\newcommand{\ba}{\begin{array}} \newcommand{\ea}{\end{array}}
\newcommand{\bc}{\begin{center}} \newcommand{\ec}{\end{center}}
\newcommand{\bt}{\begin{tabular}} \newcommand{\et}{\end{tabular}}

\def\lsim{\mathrel{\rlap{\lower4pt\hbox{\hskip1pt$\sim$}}
    \raise1pt\hbox{$<$}}}           
\def\gsim{\mathrel{\rlap{\lower4pt\hbox{\hskip1pt$\sim$}}
    \raise1pt\hbox{$>$}}}           

\newcommand{\tr}{\mathop{\rm tr}}           
\newcommand{\half}{\textstyle {1\over2} \displaystyle}    
\newcommand{\third}{\textstyle {1\over3} \displaystyle}   
\newcommand{\Dslash}{{\hbox{D}\kern-0.6em\raise0.15ex\hbox{/}}} 

\renewcommand{\et}{\eta}

\begin{document}

\setlength{\oddsidemargin}{0cm} \setlength{\baselineskip}{7mm}

\begin{normalsize}\begin{flushright}

July 2017 \\

\end{flushright}\end{normalsize}

\begin{center}
  
\vspace{5pt}

{\Large \bf Vacuum Condensate Picture of Quantum Gravity}

\vspace{30pt}

{\sl Herbert W. Hamber}
$^{}$\footnote{HHamber@uci.edu. Based on invited lectures given at the
2015 {\it Coral Gables} (Miami) International Conference on Particle Physics and at the
2015 {\it Strongly Interacting Field Theories} Workshop in Jena, Germany.} 
\\
Department of Physics and Astronomy \\
University of California \\
Irvine, California 92697-4575, USA \\

\vspace{10pt}

\end{center}

\begin{center} {\bf ABSTRACT } \end{center}

\noindent

In quantum gravity perturbation theory in Newton's constant $G$ is known to be badly 
divergent, and as a result not very useful.
Nevertheless, some of the most interesting phenomena in physics are often associated with non-analytic behavior in the coupling constant and the existence of nontrivial quantum condensates.
It is therefore possible that pathologies encountered in the case of gravity are more likely the result of inadequate analytical treatment, and not necessarily a reflection of some intrinsic insurmountable problem. 
The nonperturbative treatment of quantum gravity via the Regge-Wheeler 
lattice path integral formulation reveals the existence of a new phase involving a 
nontrivial gravitational vacuum condensate, and a new set of scaling exponents characterizing 
both the running of $G$ and the long-distance behavior of invariant correlation functions. 
The appearance of such a gravitational condensate is viewed as analogous to the (equally nonperturbative) gluon and chiral condensates known to describe the physical vacuum of QCD.
The resulting quantum theory of gravity is highly constrained, and its physical predictions are
found to depend only on one adjustable parameter, a genuinely nonperturbative scale 
$\xi$ in many ways analogous to the scaling violation parameter $\Lambda_{\bar MS} $ of QCD.
Recent results point to significant deviations from classical gravity on distance scales
approaching the effective infrared cutoff set by the observed cosmological constant.
Such subtle quantum effects are expected to be initially small on current cosmological scales, 
but could become detectable in future high precision satellite experiments.




\vfill

\pagestyle{empty}

\newpage


\pagestyle{plain}

\section{Introduction}

\label{sec:intro}

\vskip 20pt

Like QED (Quantum Electrodynamics) and QCD (Quantum Chromodynamics) quantum gravity is, 
in principle, a unique theory. 
In the Feynman path integral approach, only two key ingredients are needed to formulate the quantum theory, the gravitational action and the functional measure over metrics. 
For gravity, the action is given by the Einstein-Hilbert term augmented by a cosmological constant.
Additional higher derivative terms are consistent with general covariance, but nevertheless only 
affect the physics at very short distances, and will not be considered further here. 
The other key ingredient is the functional measure for the metric field, which in the case of gravity describes an integration over all four metrics with weighting given by the DeWitt form.
As in most other cases where the Feynman path integral can be written down (including non-relativistic
quantum mechanics), the proper definition of integrals requires the introduction of a lattice,
so as to properly account for the known fact that quantum paths are nowhere differentiable.
It is therefore a remarkable aspect that, at least in principle, the resulting quantum theory of 
gravity does not seem to require any additional extraneous ingredients, besides the ones 
mentioned above.
Indeed some time ago, Feynman was able to show that Einstein's theory is unique, invariably
arising from the consistent quantization of a massless spin two particle.

At the same time, gravity has been known to present some rather difficult inherent problems.
The first one is related to the fact that the theory is intrinsically nonlinear, since gravity gravitates. 
In addition, perturbation theory in Newton's constant $G$ is useless, since the resulting
series is badly divergent (much more so than in QED and QCD), which makes the theory
not perturbatively renormalizable. 
It is also a known fact that the gravitational action is affected by a conformal instability, 
which makes at least the Euclidean path integral potentially divergent. 
Last but not least, additional, genuinely gravitational, technical complications arise due 
to the fact that physical distances between spacetime points are dependent  on the metric, 
which is a fluctuating dynamical quantum entity.

Serious divergences that appear in perturbation theory originate from the fact that the gravitational
action leads to vertices which are proportional to a momentum squared. 
When these vertices are inserted into diagrams, they give rise to ultraviolet divergences which get 
increasingly worse as the order of perturbation theory is increased. 
The lack of perturbative renormalizability therefore leads to two main, alternative and
clearly mutually exclusive, conclusions. 
One states that the quantum theory of gravity does not exist due to these cascading 
perturbative divergences, and consequently an enlarged, improved theory should be 
investigated instead.
Enlarged theories that attempt to make quantum gravity perturbatively renormalizable
include $N=8$ supergravity, and supersymmetric strings in ten spacetime dimensions.
The other alternative path, followed here, is that the usual diagrammatic methods of 
QED and QCD fail for gravity because perturbation theory is incomplete or invalid, 
presumably due to a more complex analytic structure in the coupling constant, 
thereby leading to gravity not being renormalizable in the usual perturbative sense.
The possibility exists therefore (and is further supported by several well known examples in physics)
that perturbation theory in $G$ fails because physically relevant quantities ($n$-point functions, quantum averages, functions describing the running of $G$ with scale etc.) are 
non-analytic at $G$ equal zero. 
\footnote{The validity of the perturbative approach to gravity is sometimes supported by the 
fallacious argument that in some sense ``gravity is weak''.  That is certainly true when gravity is compared to the other fundamental forces on laboratory scales. 
Nonetheless, unlike QED and QCD, the gravitational coupling is dimensionful which makes such weak coupling arguments invalid, or at least naive, when referred to gravity as its own self-sufficient theory.
Ultimately, the real question is whether large quantum gravitational field fluctuations, which 
cannot be excluded a priori from the path integral, are physically important or not.}

Indeed there are many physically very interesting and deep phenomena 
which cannot be explained, or even studied, using perturbation theory alone. 
One example is QCD, were gluons and quarks are confined with a chromoelectric string tension
known to be non-analytic (in the form of an essential singularity) in the gauge coupling.
Again, in a superconductor, the correct ground state is described by Cooper pairs bound together 
by a weak electron-phonon interaction. 
The latter leads to a gap in the energy spectrum close to the Fermi surface, which is known
to be non-analytic in the fundamental electron-phonon coupling constant. 
In a superfluid, the quantum condensate density is non-analytic in the coupling as well, and so is 
the screeing mechanism in a degenerate Coulomb gas, where the Thomas-Fermi screening length
is known to be non-analytic in the charge.
In this last model, the correct charge screening mechanism is not reproduced to any finite order in perturbation theory.
Regardless, it is easily obtained by resumming infinitely many so-called ring diagrams.
Additional, physically relevant examples include homogeneous turbulence (which is described by nontrivial Kolmogoroff scaling exponents) and order-disorder transitions in ferromagnets and related systems.
The latter exhibit spontaneous symmetry breaking, dimensional transmutation, nontrivial scaling dimensions, and the appearance of a non-vanishing field condensate in the ordered phase.
Related to the last example is the case of a self-interacting scalar field above four dimensions,
where, on the one hand, the theory is known to be perturbatively non-renormalizable with 
the kind of escalating ultraviolet divergencences described earlier.
Yet one can prove rigorously, by using the lattice path integral formulation, that the model
reduces to a non-interacting (Gaussian) theory at large distances, with low energy scattering
amplitudes vanishing as an inverse power of the ultraviolet cutoff.

In many cases, the common thread among these widely different theories and physical phenomena
is the existence of some sort of vacuum condensate, which generally turns out to be a non-analytic 
function of the relevant fundamental coupling constant.
The origin of these non-analiticities can often be traced back to the fact that the physical ground state
is, in the end, fundamentally different from the original unperturbed or free field ground state.
So, the true ground state is qualitatively different from the unperturbed ground state 
which, initially, forms the starting point for perturbation theory.
Physically, a significant rearrangement of the vacuum will often not just involve small 
perturbations, and generally cannot be obtained by perturbative methods, which implicitly assume smooth changes, and thus the existence of a Taylor series in the relevant coupling.
In this framework, the failure of perturbation theory is seen more as a reflection on the
fundamental inadequacy of the mathematical methods used, and not necessarily as a 
shortcoming of the underlying fundamental theory per se.

If gravity is not perturbatively renormalizable, then what are the alternatives? 
In fact perturbatively non-normalizable theories have been theoretically rather well understood since
the early seventies, when the modern renormalization group approach (based on momentum 
slicing, scaling dimensions and multidimensional coupling constant flow) was invented to account for more subtle and complex behavior in quantum field theory \cite{wil72,wil75,par73,par76,sym69,sym70}.
Moreover, a number of significant examples exist of theories which are not perturbatively 
renormalizable, and nevertheless give rise to physically acceptable and interesting theories, 
and for which very detailed and accurate physical predictions can be produced.
Most often these involve models formulated in less than four dimensions,
which are thus, generally, more relevant to statistical field theory than to particle physics or gravitation. 
Indeed, in a statistical field theory context, it is often possible to bypass the limitations of perturbation theory, by resorting to additional, but complementary, approximation and expansion methods.
These methods include Wilson's $2+\epsilon$ and $4-\epsilon$ expansions, 
the large $N$ expansion, weak and strong coupling expansions (sometimes referred to as the low and high temperature expansion), 
partial resummation methods, and finally a combination of all of the above methods paired with 
high accuracy direct numerical evaluations of the original path integral or partition function
(for an overview, see for example \cite{par81,zin02,itz91,car96,bre10}).

One of the reason why these more powerful methods are eventually capable of providing useful (and ultimately correct) physical information about the systems studied, lies in the fact that they are able to 
access new nontrivial strong coupling fixed points of the renormalization group, which are often
not at all visible nor accessible in weak coupling perturbation theory. 
In other words, the common thread among many of the models that are perturbatively non-renormalizable - but which in the end turn out to be physically acceptable and relevant - is the 
existence of a nontrivial strong coupling renormalization group fixed point. 
Furthermore, in support of the legitimacy of such a more sophisticated approach one should mention, 
as an example, the fact that exquisitely detailed predictions for a class of perturbatively 
non-renormalizable theories, namely the $O(N)$ non-linear $\sigma$-model \cite{bre76,bre76a}
in three space dimensions, now provide the second most accurate test of quantum field theory \cite{gui98,lip03}, after the QED predictions for the $g-2$ anomalous magnetic moment of the electron.

Therefore, it seems reasonable to apply the very same (and by now well established and very successful)
methods to one more perturbatively non-renormalizable theory, namely the quantum theory 
of gravity in four dimensions \cite{tho74,vel75,tho78}.
One first notes that a controlled non-perturbative approach clearly requires a useful and explicit 
ultraviolet regulator, and the only known reliable way to  evaluate non-perturbatively 
the Feynman path integral in four dimensional quantum field theories is via the lattice formulation. 
Indeed, as shown in detail already by Feynman for non-relativistic quantum mechanics, 
the very definition of the path integral (which sums over all paths, known to
be generally nowhere differentiable) requires the introduction of a lattice
discretization, due to the Wiener path nature of quantum trajectories \cite{fey65}.
One shining example of the success and reliability of the lattice approach is the 
elucidation of the subtle mechanism of confinement and chiral symmetry breaking in QCD.

The explicit introduction of a lattice achieves two purposes: one is to provide an explicit discretization 
(which is required in order to define in an explicit, as opposed to formal, way what is meant 
by the sum over all paths); and second to give a necessary regularization (in the sense of taming the ubiquitous field theoretic short distance divergences) of the quantum path integral. 
Additional advantages of the path integral formulation, present both in the case of gauge theories and gravity, are the existence of a manifestly covariant formulation, and the known fact that no gauge fixing is in principle required (as first shown by Wilson in the gauge theory case \cite{wil74}) outside the 
traditional framework of perturbation theory. 
Sometimes it is possible to rely on some sort of saddle point expansion around a smooth 
solution to the classical field equations \cite{haw79,gib77}, however it is also generally recognized that dominant paths which contribute to the path integral are nowhere differentiable, 
and ultimately can only be accounted for properly in a controlled discretized formulation. 
While it is certainly possible to evaluate the gravitational path integral using perturbation theory, the latter is not always the only avenue open, and is seen in fact as rather restrictive for the
reasons outlined herein. 
The case of the non-linear sigma model shows rather clearly that results derived
using perturbation theory alone can be entirely misleading, and do not capture correctly the 
underlying physics of the ground state and key aspects related to the existence
of an order-disorder transition.
Also, the lattice formulation for gauge theories and gravity presented somewhat of a novelty thirty
years ago, but is now extensively tested for QCD, scalar field theories, and a variety of spin systems.
In addition, today one can rely on over thirty years experience in an array of both analytical and numerical calculations, and in their fruitful mutual interplay.
 
Unless one desires to reinvent entirely new and ad hoc methods, the natural prototype for dealing with 
genuine non-perturbative aspects of gravity is Wilson's lattice formulation of QCD. 
Indeed, while QCD is perturbatively renormalizable, it is well known that in this case perturbation 
theory is largely useless at low energies, where confinement effects take over and fundamentally
modify the physical picture of the vacuum state.
One key aspect of the lattice gauge theory is that, in order to preserve a form of exact local invariance
(and related quantum Ward identities), the formulation requires an integration over gauge fields
with a nontrivial (but uniquely determined by local gauge invariance) Haar measure. 
Then, in the lattice framework, confinement is an almost immediate and easily visualized consequence
of large field fluctuations at strong coupling.

QCD is a hard theory to solve, and many deep insights have come from the lattice formulation. 
It cannot be stressed enough that one important outcome of the lattice calculations is that the physical vacuum bears little resemblance to the perturbative vacuum, due to significant nonlinearities and nontrivial field condensation effects.
The former exhibits a rich spectrum of hardons and glueballs, chromoelectric and quark field vacuum condensates, all of which are ultimately non-analytic in the gauge coupling $g$, and
cannot be reproduced by perturbative methods.
Indeed, to this day, Wilson's lattice theory provides the only convincing evidence for confinement and chiral symmetry breaking in QCD and, more generally, in non-Abelian gauge theories.
In addition, the lattice theory allows credible calculations of the running of alpha strong versus energy, which compare rather well with current experimental data.

For a quantum theory of gravity, the Feynman path integral again represents a 
natural starting point \cite{fey,dew67,haw79,gib77}. 
It is therefore rather fortunate that an elegant lattice formulation for gravity was written down by Regge and Wheeler in the early sixties, and is, not unexpectedly, based on the key concept of a dynamical lattice
\cite{reg61,whe64} (for a recent overview see \cite{book} and references therein).
The main features of this theory can be summarized as follows.
It incorporates a continuous local invariance, completely analogous to the diffeomorphism invariance
of the continuum theory.
As already pointed out originally by Regge, the local invariance of the lattice theory then leads to a 
lattice analog of the Bianchi identities, and thus to corresponding Ward identities in the 
quantum version.
It also puts within reach of computation problems in classical general relativity which 
are in practical terms beyond the power of analytical methods; 
this last aspect was perhaps one of the main motivations initially (in the early sixties) for a discrete 
formulation of General Relativity.
Furthermore, like most lattice field theories, it affords in principle any desired level of 
accuracy by a sufficiently fine subdivision of spacetime, allowing eventually a reconstruction
of the original continuum theory.
The resulting Regge-Wheeler lattice theory of gravity is generally known as simpicial quantum gravity,
for the simple reason that it is based on a construction of space-time out of geometric simplices, four dimensional analogues of triangles and tetrahedra.
In this formulation, curvature is described by angles,
metric components are replaced by edge lengths, and the relevant geometric quantities can be calculated from the values of the edge lengths to give local lattice volumes, angles and local curvatures. 
In other words, local curvature is completely determined by an assignment of edge lengths and by
how each edge is locally connected to neighboring edges (the incidence matrix). 
It is then possible to write down the lattice analog of the local volume element, 
of the local Riemann tensor, of the scalar curvature, and therefore ultimately, of 
invariant terms such as the Einstein-Hilbert action. 

As a consequence, the first key ingredient of a discretized form for the Feynman path integral 
for gravity, namely the action, is provided by the Regge-Wheeler theory. 
Furthermore, since the path integral involves an integration over all four metrics, and since the
metric is locally related to the lattice edge lengths squared, 
the implication is that the analog of the DeWitt functional measure over continuum metrics
turns into an integration over all lattice edge lengths squared (with some suitable
volume inequality constraints, so as to guarantee a sensible geometric interpretation).
For ordinary field theories, the rigorous construction of the Feynman path integral often 
involves a Wick rotation to complex spacetime, and the same procedure can be achieved 
in the context of gravity as well, both in the continuum and on the lattice.

While it is possible in some cases to proceed with a Lorentzian signature \cite{fey,dew67,tho74,vel75,tho02} (in the continuum, and on the lattice for example by the 
use of a discretized Wheeler-DeWitt equation), it is generally accepted that the 
Euclidean formulation provides a mathematically more sound description of the Feynman 
path integral \cite{haw79,gib77}. 
In addition, such a formulation generally relies on weights involving positive real probabilities, 
which then allows the use of established numerical probabilistic methods.
It is certainly possible that, in the context of gravity, the Lorentzian and Euclidean theories
belong to two different universality classes, and give rise to two entirely different sets of
renormalization group beta functions and scaling exponents. 
This would be rather unique, since no other instance of such an occurrence is known. 
Nevertheless the evidence so far suggests that basic results in the Lorentzian and Euclidean
lattice theories agree quite well.
An explicit test of this statement lies in the ongoing comparison of results for universal scaling
dimensions obtained in the two formulations.
A recent example of a Lorentzian formulation of lattice quantum gravity, also based on the 
Regge-Wheeler discretization, is an exact solution of the lattice Wheeler-DeWitt 
equations in $2+1$ dimensions discussed in \cite{hw11,htw12,ht16}.
In addition, it is possible to force quantum gravity to become perturbatively renormalizable
by formally expanding about two dimensions (essentially Wilson's $2+ \epsilon$ 
expansion applied to the case of gravity). 
In this approach, it seems quite clear that universal results such as scaling dimensions
are expected to be identical for the Lorentzian and Euclidean signatures to all 
orders in perturbation theory \cite{epsilon,wei79,eps,aid97}.



\vskip 40pt

\section{Regularized Path Integral for Quantum Gravity}

\label{sec:ave}

\vskip 20pt

One usually considers as the starting point for a nonperturbative formulation of quantum gravity
a suitably discretized form of the Feynman path integral, initially for pure gravity without matter
fields, which can then be added at a later stage.
In the continuum the path integral is given formally by \cite{fey,haw79}
\beq
Z_C \; = \; \int d \mu [ g_{\mu\nu} ] \;
\exp \left \{ - I [ g_{\mu\nu} ] \right \}  \;\; ,
\label{eq:z_cont}
\eeq
with the Einstein-Hilbert gravitational action
\beq
I [ g_{\mu\nu} ] \; = \;  \int d^4 x \, \sqrt g \, \Bigl ( \lambda_0 - { k \over 2 } \, R \, + \, \cdots \Bigr ) \; .
\label{eq:ac_cont}
\eeq
The dots here indicate possible matter and higher derivative terms, with the latter
getting generated, for example, by radiative corrections as they arise already in the 
framework of perturbation theory.
The functional integration over metrics is done using the DeWitt diffeomorphism invariant 
measure \cite{dew67}
\beq
\int d \mu [ g_{\mu\nu} ] \; = \; \int \prod_x \;  \prod_{ \mu \ge \nu } \, d g_{ \mu \nu } (x) \; .
\label{eq:meas_cont}
\eeq
In the above expression $k^{-1} \equiv 8 \pi G $, with $G$ the bare
Newton's constant and $\lambda_0$ a bare cosmological constant.
In the following we will consider almost exclusively the case of no
higher derivative $R^2$-type terms, and no dynamical matter (quenched approximation).

The continuum Feynman path integral given above is generally ill-defined (the integration is 
dominated by non-differentiable Wiener paths), and so it needs to be formulated more precisely by 
introducing a suitable discretization, as is done in both non-relativistic quantum mechanics
and quantum field theory \cite{fey65}.
This last step is particularly crucial for nonperturbative gravity calculations,
where the nontrivial invariant measure over the $g_{\mu\nu}$'s has been
shown to play an important role.
In the 60's Regge and Wheeler proposed an elegant discretization of
the classical gravitational action \cite{reg61,whe64}, which forms the basis for the 
lattice formulation of quantum gravity used here; 
early references include \cite{cms82,rowi81,hw84,har85,lesh84}.
Once the measure and the path integral have been transcribed on the lattice,
the ultimate goal then becomes to recover the original 
continuum theory of Eq.~(\ref{eq:z_cont}) in the limit of a suitably small 
lattice spacing. 
It is known that taking this limit is a rather subtle affair, and, in order for it to be taken correctly, 
it will require the full machinery of the modern (Wilson) renormalization group.

A suitable starting point is therefore the following discrete
form for the Euclidean Feynman path integral for pure gravity
\beq
Z_L \; = \; \int d \mu [ l^2 ] 
\exp \left \{  - I [l^2 ] \right \}  \;\; ,
\label{eq:z_latt} 
\eeq
with a compactly written lattice gravitational action
\beq
I [ l^2 ] \; = \; \sum_h \, \Bigl ( \lambda_0 \, V_h - k \, \delta_h A_h \, + \, \cdots \Bigr )
\label{eq:ac_latt} 
\eeq
and lattice integration measure
\beq
\int d \mu [ l^2 ] \; = \;
\prod_{ ij } \, dl_{ij}^2 \; \Theta [l_{ij}^2]  \; .
\label{eq:meas_latt} 
\eeq
In these last expressions the sum over hinges $h$ in four dimensions corresponds to a sum 
over all lattice triangles with area $A_h$, with the deficit angle
$\delta_h$ describing the curvature around them \cite{reg61,whe64}.
The $\Theta$-function constraint appearing in the discrete measure ensures that
the triangle inequalities and their higher dimensional analogs are satisfied by all simplices.
The discrete gravitational measure in $Z_L$ of Eq.~(\ref{eq:z_latt}) can then be
viewed as a regularized version of the original DeWitt continuum functional measure 
of Eq.~(\ref{eq:meas_cont}).
A bare cosmological constant term with $\lambda_0 >0 $ is essential for the convergence of the path
integral, since for bare $\lambda_0  \leq 0$ the Euclidean path integral is clearly divergent \cite{hw84,lesh84}.

It is a rather useful fact that the lattice edge lengths are locally related in a simple way to 
the continuum metric.
In terms of the edge lengths $l_{ij} $ attached to a four-dimensional simplex $s$ one has
for the induced metric within that simplex
\beq
g_{ij} (s) \; = \; \half \left ( l_{0i}^2 + l_{0j}^2 - l_{ij}^2 \right ) \;\; ,
\label{eq:latmet}
\eeq
where the four-simplex here is based at the point $0$.
This last result then provides the needed connection between the continuum metric
$g_{\mu\nu} (x) $ and the lattice squared edge lengths degrees of freedom $l^2_i$;
the latter is essential in establishing a clear and unambiguous relationship between 
lattice and continuum operators, just as in the case of Yang-Mills theories
on the lattice.
Appropriate lattice analogs of various curvature invariants can then be written
down, making use of the well-understood correspondences
\bea
\sqrt{g} \, (x) \; & \to & \; 
\sum_{{\rm hinges} \, h \, \supset \, x } \; V_h 
\nonumber \\
\sqrt{g} \, R (x) \; & \to & \; 
2 \sum_{{\rm hinges} \, h \, \supset \, x } \; \delta_h A_h
\nonumber \\
\sqrt{g} \, R_{\mu\nu\lambda\sigma} \, R^{\mu\nu\lambda\sigma} (x) \; & \to & \;
4 \sum_{{\rm hinges} \, h \, \supset \, x } \; ( \delta_h A_h )^2 / V_h  \;\; .
\label{eq:ops}
\eea
In the above expressions the hinges $h$ correspond to triangles in four dimensions.
A detailed discussion of such operators, as well as additional four-dimensional curvature
invariants, can be found for ex. in \cite{book} and references therein.
In evaluating the lattice path integral, by whatever means, the continuum
functional integration over metric
is thus replaced by a finite-dimensional integration over squared edge lengths,
which become the fundamental variables in the discrete theory.
The general aim of the calculation is then to evaluate
the lattice path integral either approximately or exactly by numerical means,
by performing a properly weighted sum over all lattice field configurations.

In lattice field theories it is customary to deal with dimensionless
quantities \cite{par81,zin02,itz91,car96,bre10}, and here this rather well-established procedure is 
followed again, for obvious reasons.
The bare coupling constants $\lambda_0$ and $G$ appearing in
the continuum theory are expressed from the start
in units of a fundamental lattice  cutoff $\Lambda = 1/a$;
without such a cutoff the continuum theory is generally ill defined \cite{fey65}.
The latter is then set equal to one, so that all observable
quantities, correlators and couplings are expressed in units of this
fundamental  cutoff.
In the end the actual value for the  cutoff (say in $cm^{-1}$) is determined
by comparing suitable physical quantities.
Furthermore, the functional integral depends on several bare coupling constants, but it is important 
to note that in the absence of matter the theory only depends
on {\it one} bare parameter, the dimensionless coupling $ k / \sqrt{\lambda_0} $.
This is easily seen, for example, from the fact that in $d$ dimensions a constant rescaling of the metric
\beq
g_{\mu\nu} = \omega \, g_{\mu\nu}'
\label{eq:metric_scale}
\eeq
turns the cosmological constant term $ \lambda_0 \sqrt{g} $
into $ \lambda_0 \, \omega^{d/2} \, \sqrt{g'} $, 
so that a subsequent rescaling of the bare coupling constants
\beq
G \rightarrow \omega^{-d/2+1} \, G \; , \;\;\;\; 
\lambda_0 \rightarrow \lambda_0 \, \omega^{d/2}
\eeq
leaves the dimensionless combination $G^d \lambda_0^{d-2}$ unchanged.
One concludes that only the latter combination has a physical meaning in pure gravity;
in particular one can always suitably chose the scale $\omega = \lambda_0^{-2/d}$
so as to adjust the volume term to acquire a unit coefficient.
This ability to rescale the field variables (the metric) so as to reabsorb certain
renormalizations of the couplings is an absolutely crucial, and physically quite
consequential, aspect of quantum gravity and which can be easily lost by an 
overly crude regularization procedure.
Without any loss of generality it is therefore entirely legitimate to set the bare cosmological
constant $\lambda_0 = 1$ in units of the  cutoff \cite{lesh84}.
The latter contribution then controls the scale for the edge lengths, and thus
the overall scale in the problem.
\footnote{In the continuum diagrammatic treatment a similar key result can be derived.
There one can show that the renormalization of $\lambda_0$ is gauge-  and
scheme-dependent;  only the renormalization of Newton's constant $G$ is
unaffected by the choice of gauge conditions \cite{aid97,lambda}.
Physically, these results simply express the fact that $\lambda_0$ controls the
overall spacetime volume, and that, in a renormalization group context,
a "running volume" is meaningless, or at least somewhat contradictory.}


It is clear by now that to accurately study the physical consequences of the theory 
requires the full machinery of quantum field theory and the renormalization group.
Nevertheless, some key information about the behavior of physical correlations
can already be obtained indirectly from averages of local diffeomorphism invariant operators.
Also, it will often be convenient to continue to use the continuum language (as opposed
to the lattice one) to discuss such quantities; 
in most cases the two languages are interchangeable, with the lattice one providing
a more precise and thus less ambiguous (short-distance regulated) expression.
Consider for example the average local curvature
\beq
{\cal R} (G) \; \sim \;
{ < \int d^4 x \, \sqrt{ g } \, R(x) >
\over < \int d^4 x \, \sqrt{ g } > } \;\;\; .
\label{eq:avr_cont}
\eeq
The above quantity describes the parallel transport of vectors around 
infinitesimal loops and is, by construction, manifestly diffeomorphism invariant.
An appropriate lattice transcription reads \cite{hw84,lesh84}
\beq
{\cal R} (G) \; \equiv \; 
< \! l^2 \! > { < 2 \; \sum_h \delta_h A_h > \over < \sum_h V_h > } \;\;\; .
\label{eq:avr_latt} 
\eeq
A second quantity of physical interest is the fluctuation in the local curvature
\beq
\chi_{\cal R}  (G) \; \sim \;
{ < ( \int d^4 x \, \sqrt{g} \, R )^2 > - < \int d^4 x \, \sqrt{g} \, R >^2
\over < \int d^4 x \, \sqrt{g} > } \;\; .
\label{eq:chi_cont}
\eeq
The latter is directly related to the invariant curvature correlation 
function at zero momentum \cite{lesh84,cor94}, see later below.
On the lattice the previous quantity takes on the form
\beq
\chi_{\cal R}  (G) \; \equiv \; 
{ < (\sum_h 2 \, \delta_h A_h)^2 > -  < \sum_h 2 \, \delta_h A_h >^2 \over < \sum_h V_h > } \;\; .
\label{eq:chi_latt}
\eeq
Moreover, in the functional integral formulation of Eqs.~(\ref{eq:z_cont}) and (\ref{eq:z_latt})
the average curvature ${\cal R} (G) $ and its fluctuation $\chi_{\cal R} (G) $ can also
be obtained by taking derivatives with respect to $k$ of the lattice partition function $Z_L $ in Eq.~(\ref{eq:z_latt}).
On the lattice one has from the definition of the path integral
\beq
{\cal R} (G) \, \sim \, \frac{1}{<\!V\!>} \, \frac{\partial}{\partial k} \log Z_L \;
\label{eq:avr_z} 
\eeq
as well as
\beq
\chi_{\cal R}  (G) \, \sim \, \frac{1}{<\!V\!>} \, \frac{\partial^2}{\partial k^2} \log Z_L \;  .
\label{eq:chir_z} 
\eeq
Exact scaling relationships then arise between various quantities, such as the
ones in Eqs.~(\ref{eq:avr_z}) and (\ref{eq:chir_z}), and these can later be used to derive 
scaling relations and check for mathematical consistency.



\vskip 40pt

\section{Diffeomorphism Invariant Gravitational Correlation Functions}

\label{sec:corr}

\vskip 20pt

Generally in a quantum theory of gravity physical distances between any two
points $x$ and $y$ in a fixed background geometry are determined from the metric
\beq
d(x,y \, \vert \, g) \; = \; \min_{\xi} \; \int_{\tau(x)}^{\tau(y)} d \tau 
\sqrt{ \textstyle g_{\mu\nu} ( \xi )
{d \xi^{\mu} \over d \tau} {d \xi^{\nu} \over d \tau} \displaystyle } \;\; .
\eeq
Because of quantum fluctuations, the latter depends,
in the lattice case, on the specific edge length configuration considered.
Correlation functions of local operators need to account for this
fluctuating distance, and as a result these correlations are 
computed at some fixed geodesic distance between 
a given set of spacetime points \cite{lesh84,cor94}.
On a given lattice this process involves constructing a 
complete table of distances between any two lattice points,
and then computing from it the required two point functions.
In addition, in gravity one generally requires that the local
operators entering the correlation function should be coordinate scalars.
In principle one could also {\it smear} such operators over a small region
of spacetime with an assigned linear size \cite{npb400,ham15}.
It is then possible to also consider {\it nonlocal} gravitational observables, in
analogy to what is done in Yang-Mills theories, by defining the 
gravitational analog of the Wilson loop.
The latter carries information about
the parallel transport of vectors around large loops, and therefore
about large scale curvature \cite{modacorr,modaloop,npb400,loops,modapot,lines}, 
and will be discussed later.

In a quantum theory of gravity a fundamental two-point correlation function 
is the one associated with the scalar curvature, 
\beq
G_R (d) \; \sim \; < \sqrt{g} \; R(x) \; \sqrt{g} \; R(y) \;
\delta ( | x - y | -d ) >_c \; .
\label{eq:corr_cont}
\eeq
with physical points $x$ and $y$ separated by a given fixed geodesic distance $d$.
On the lattice it has the corresponding form 
\beq
G_R (d) \; \equiv \; < \sum_{ h \supset x } 2 \, \delta_h A_h \;
\sum_{ h' \supset y } 2 \, \delta_{h'} A_{h'} \;
\delta ( | x - y | -d ) >_c \; .
\label{eq:corr_latt}
\eeq
For the curvature correlation at fixed geodesic distance 
one expects at short distances (i.e. distances much shorter
than the gravitational correlation length $\xi$ to be introduced below) 
a power law decay
\beq
< \sqrt{g} \; R(x) \; \sqrt{g} \; R(y) \; \delta ( | x - y | -d ) >_c
\;\; \mathrel{\mathop\sim_{d \; \ll \; \xi }} \;\; 
d^{- 2 n}  \;\;\;\; ,
\label{eq:corr_pow}
\eeq
with the power law here characterized by a universal exponent $n$.
\footnote{
It is preferable here not to use the notation $\Delta$ for the conformal
dimension $n$, as this would generate confusion later on with the
Laplacian operator.}
How $n$ is related by scaling to other calculable universal critical
exponents [in particular to the exponent $\nu$ of Eq.~(\ref{eq:xi_g})]
is discussed further below [see for ex. Eq.~ (\ref{eq:n_pow})].
Alternatively, the short distance correlation function expression of Eq.~(\ref{eq:corr_pow}) can
be expressed in momentum space, using the formal Fourier transform
result valid in $d$ dimensions
\beq
\int d^d x \; e^{- i p \cdot x}  { 1 \over x^{2 n} }
\; = \; 
{ \pi^{d/2} \, 2^{d - 2 n} \, \Gamma ( { d - 2 n \over 2 } ) \over
\Gamma (n) } \; { 1 \over p^{ d - 2 n} } \;\; .
\label{eq:corr_fourier}
\eeq
On the other hand, for sufficiently strong coupling (large $G$, or small
$k$) fluctuations in different spacetime regions largely decouple: 
the kinetic or derivative term in Eqs.~(\ref{eq:z_cont}) or
(\ref{eq:z_latt}) is responsible for coupling fluctuations in different
spacetime regions, and in the action it comes with a coefficient $1/G$.
In this regime one then expects a faster, exponential decay, 
controlled by a nonperturbative correlation length $\xi$
\beq
< \sqrt{g} \; R(x) \; \sqrt{g} \; R(y) \; \delta ( | x - y | -d ) >_c
\;\; \mathrel{\mathop\sim_{d \; \gg \; \xi }} \;\;
e^{ - d / \xi } \;\;\;\; .
\label{eq:corr_exp}
\eeq
So the fundamental gravitational correlation length $\xi$ can be defined unambiguously 
by the long-distance decay of the connected invariant curvature 
correlations at fixed geodesic distance $d$.
Then the behavior in Eq.~(\ref{eq:corr_pow}) is expected to hold at
short distances $ d \ll \xi $, whereas
the behavior in Eq.~(\ref{eq:corr_exp}) is expected to hold
at much larger distances, $ d \gg \xi $.
In either case, in order to reach a sensible lattice continuum limit
the physical distances involved need to be much larger than the
fundamental average lattice spacing $l_0$,  $ d, \xi \gg l_0 $ (the
so-called scaling limit).

Consistency between the two expressions in Eqs.~(\ref{eq:corr_exp}) and (\ref{eq:corr_pow})
is eventually regained from the fact that in the vicinity of the critical point
a superposition of many exponentials are expected to add up to a power.
This is seen, for example, from the spectral (Lehmann) representation of the
two point function,  with spectral function 
$ \rho (\mu ) \, = \, A \, \mu^{\alpha -1} / \Gamma (\alpha)$.
Then 
\beq
G_R (d) \; = \; \int_m^\infty \, d \mu \; \rho (\mu) \; e^{- \mu \, d} \; = \;
A \cdot {\Gamma (\alpha, m \, d) \over \Gamma (\alpha )} \cdot { 1 \over d^\alpha } \;\; .
\label{eq:spectral}
\eeq
In the limit of a small infrared cutoff $m \equiv 1/\xi$ the above result simplifies to a
power law plus small corrections,
\beq
G_R (d) \; \simeq \; {A \over d^\alpha } \; 
\left [ \, 1 \, +\, 
{ (m \, d)^\alpha  \left [ ( m \, d -1) \alpha -1 \right ] \over \Gamma ( 2+ \alpha ) }
\, + \, \cdots \right ]
\;\; \mathrel{\mathop\sim_{\alpha \, \rightarrow \, 2 }} \;\;
{A \over d^2 } \; 
\left [ \, 1 \, - \,  { m^2 d^2 \over 2 } + O(m^4)  \; \right ]  \;\; .
\label{eq:spectral_m}
\eeq
In the last expression the known value for the gravitational
curvature correlation function, $\alpha=2\, (4-1/\nu)=2$ \cite{ham15}, 
has been inserted.\footnote{
In \cite{ham15} the most recent numerical results $\nu = 0.334(4) $ 
and thus $ 2 n = 2.01(7)$ are given.}
Note that, due to the dimensions of the curvature correlation function,
$A$ has to have dimensions of one over length squared, $A \sim A_0 / a^2 $ with
$a$ the lattice spacing and $A_0$ some dimensionless constant.
Another key result is the fact, used here later on, that the local curvature fluctuation 
of Eq.~(\ref{eq:chi_cont})  is directly related to the connected curvature correlation of
Eq.~(\ref{eq:corr_cont}) at zero momentum
\beq
\chi_{\cal R} \; 
\sim \; { \int d^4 x \int d^4 y < \sqrt{g(x)} \, R(x) \; \sqrt{g(y)} \, R(y) >_c
\over < \int d^4 x \sqrt{g(x)} > } \;\;\; .
\label{eq:chi_corr}
\eeq
This simple observation allows one to compute the exponents $\nu$ and $n$ more easily (and 
much more accurately) from the above expression than say from the distance-dependence 
of the correlation function itself.
A second useful consequence of such relations, and specifically of the result of  
Eq.~(\ref{eq:chi_corr}), is that the power $n$ in Eq.~(\ref{eq:corr_pow}) is related to the
correlation length exponent $\nu$ in four dimensions by $n=4-1/\nu$
[see Eq.~(\ref{eq:n_pow}) later on].
Numerical evaluations of the path integral so far are consistent with
$\nu=1/3$, which then leads simply to $n=1$ in Eq.~(\ref{eq:corr_pow}).
\footnote{
One can contrast this power with what one obtains in weak field perturbation theory
$ < \! \sqrt{g} R (x) \sqrt{g} R (y) \! >_c 
\; \sim \; < \! \partial^2 h (x) \, \partial^2 h (y) \! > 
\; \sim \; 1/ \vert x-y \vert^{d+2}$, 
which is quite different from the result in Eq.~(\ref{eq:corr_pow}) with
$n=4-1/\nu = 1$,
unless $\nu = 2/(d-2) $, which is nevertheless correct for $d$ close to two,
where Einstein gravity becomes perturbatively renormalizable, and
corrections to free field behavior become small.
}

An important and central feature of the lattice nonperturbative treatment is the
existence of a critical point in $G$, located at $G_c$.
The latter is interpreted as corresponding to a non-trivial fixed point in
renormalization group language, see for ex. \cite{ham15} and references
therein.
Furthermore it is known that the weak coupling phase $G < G_c$ is
{\it nonpertubatively unstable} on the lattice: it corresponds to a branched
polymer phase with no sensible continuum limit \cite{hw84,lesh84}
(it is generally understood that such instabilities are usually quite difficult, if not impossible,
to detect in a perturbative, or weak field, treatment).
In accordance with this important result, in the following only the physical 
strong gravity phase for $G>G_c$ will be considered further.

In general, in the vicinity of such a nontrivial fixed point, one expects for the 
fundamental correlation length $ \xi =1/m $ a power law divergence
\beq
\xi (G) \;\; \mathrel{\mathop\sim_{G \rightarrow G_c }} \;\;
A_\xi \, \Lambda^{-1} \, | \, G ( \Lambda ) - G_c |^{ -\nu } \;\; ,
\label{eq:xi_g}
\eeq
with $\Lambda = 1 /a$ the inverse lattice spacing,
$ A_\xi $ the correlation length amplitude, $G_c$ the
critical point in the bare coupling $G$, and $\nu$ a universal
exponent characterizing the divergence of $\xi$ at the critical point.
At the fixed point $G_c$ the theory regains scale invariance (due to the divergence
of $\xi$), and in its vicinity one can then reconstruct the original, regularized continuum theory.
In some ways $\xi^{-1} =m $ can be viewed as a nonperturbative renormalized mass,
analogous to the dynamically generated (but nevertheless gauge invariant) scale in 
Yang-Mills theories.
For extensive reviews on the general subject of renormalization group scaling see, for example, \cite{par81,itz91,zin02,car96,bre10}.
There is by now a rather well established body of knowledge in quantum field
theory and statistical field theory on this subject, and thus no obvious or
apparent reason why its basic tenets should not apply to gravity as
well, with quantum gravity describing essentially the unique theory of a massless 
spin two particle coupled to a covariantly conserved energy-momentum tensor.
\footnote{
It is a well-established fact that for theories with a nontrivial 
fixed point \cite{wil72,wil75}, 
the long distance (and thus infrared) universal scaling 
properties are uniquely determined, up to subleading corrections
to exponents and scaling amplitudes, by the (generally nontrivial) scaling 
dimensions obtained via renormalization group methods in the vicinity
of the  fixed point \cite{par81,zin02,itz91,car96,bre10}. 
These sets of results form the basis of universal
predictions for, as an example, the perturbatively nonrenormalizable
nonlinear sigma model \cite{bre76,bre76a}.
The latter gives one of the
most accurate tests of quantum field theory \cite{gui98,lip03}, after the $g-2$ prediction
for $QED$
(for a comprehensive set of references, see \cite{zin02,book}, and
references therein). 
It is also a well-established fact of modern renormalization group
theory that in lattice $QCD$
the scaling behavior of the theory in the vicinity of 
the asymptotic freedom  fixed point unambiguously determines 
the universal nonperturbative scaling properties of
the theory, as quantified by physical
observables such as hadron masses, vacuum gluon and chiral condensates,
decay amplitudes, the QCD string tension etc. \cite{hag10,fod12}.}

One consequence of the renormalization group 
scaling relations in the vicinity of the fixed point, such as the scaling
behavior \cite{scaling} for the singular part of the free energy
\beq
F_{sing} \; = \; - { 1 \over V } \, \log Z_{L\, sing} \sim \xi^{-d}  \;\; ,
\label{eq:f_sing}
\eeq
is to allow a precise determination of the 
correlation length exponent $\nu$ in Eq.~(\ref{eq:xi_g})
and associated quantities, such as amplitudes and corrections
to scaling.
From the lattice one finds for the critical value of $G$
\beq
G_c \; \equiv \; { 1 \over 8 \pi k_c } \; = \; 0.623042(25) \;\; .
\label{eq:Gc}
\eeq
and $  \nu = 0.334(4) $ which is consistent with the conjectured exact value $\nu = 1/3$ 
for pure quantum gravity in four dimensions \cite{ham15}.
After restoring dimensions, this in turn fixes the lattice spacing $a$, and thus the value for the 
 cutoff (in four dimensions $G$ has dimensions of a length squared),
\footnote{That the physical $G$ is actually very close to $G_c$ will be discussed later below.
The argument involves in a key way the large scale curvature,
and thus the quantum gravitational Wilson loop.}
\beq
G \; \approx \; G_c  \; = 0.6230 \, a^2  \;\; .
\label{eq:gc_phys}
\eeq
From the known laboratory value of Newton's constant $G$, 
$ l_P \equiv \sqrt{ \hbar G / c^3 } $$= 1.616199(97) \times 10^{-33} cm$
one then obtains for the fundamental lattice spacing 
$a = 1.2669 \, \sqrt{G_c} \, \equiv \, l_P $, or
\beq
a \; = \; 2.0476 \, \times \, 10^{-33} \, cm \;\; ,
\label{eq:a_phys}
\eeq
and from it a value for the  cutoff $\Lambda \simeq 1/a $.
This last result then allows one to restore the correct dimensions in all dimensionful
quantities.
\footnote{
Note that in general the edge lengths are fluctuating and their average is
close to, but not equal to, one.
Nevertheless (for $\lambda_0=1$) one finds for the {\it average} lattice spacing 
in units of $a$ $ < \! l^2 \! > \; \equiv \; l_0^2 \; = [ \, 2.398(9) \, a \, ]^2 $,
so that $a$ and $l_0$ are quite comparable.}

Note that from Eqs.~(\ref{eq:corr_pow}), (\ref{eq:corr_exp}) and (\ref{eq:chi_corr}) one has
\beq
\chi_{\cal R} (G) \, \sim \, { \int d^4 x \int d^4 y < \sqrt{g} R (x) \, \sqrt{g} R (y) >_c
\over < \int d^4 x \sqrt{g} > } 
\, \mathrel{\mathop\sim_{ G \rightarrow G_c}} \; A_{\chi} \, ( G - G_c )^{d \, \nu - 2}
\, \sim \,  \xi^{ 2 / \nu - d }  \; .
\label{eq:chi_corr2}
\eeq
The last scaling result follows from the fact that the curvature
fluctuation is also the second derivative of the free energy with respect
to $k$ [see Eq.~(\ref{eq:chir_z})], and that for the free energy the standard
scaling assumption \cite{scaling} in the vicinity of the ultraviolet fixed point reads 
$ F_{sing} (G) = - (1/V) \log Z_{L \, sing} \sim \xi^{-d} $ [see Eq.~(\ref{eq:f_sing})]
with $\xi (G) $ given in Eq.~(\ref{eq:xi_g}).
This then allows the fundamental exponent $\nu$ to be computed much more easily, and more accurately, than from the distance-dependence of the curvature correlation function of Eq.~(\ref{eq:corr_exp}).
One useful consequence of the basic scaling result of Eq.~(\ref{eq:chi_corr2}) is 
that the power $n$ in Eq.~(\ref{eq:corr_pow}) is related to the
correlation length exponent $\nu$ in four dimensions by $n=d-1/\nu = 4-1/\nu$
[for the definition of $n$ see Eq.~(\ref{eq:n_pow})].
Numerical evaluations of the path integral are consistent with
$\nu=1/3$ \cite{ham15}, which then leads to the simple result
$n=1$ for the invariant curvature correlation in Eq.~(\ref{eq:corr_pow}).
For the local average curvature of Eqs.~(\ref{eq:avr_cont}) and (\ref{eq:avr_latt}),
now expressed in terms of the correlation length $\xi$,
one then obtains the rather simple result
\beq
{\cal R} ( G ) \;\; \mathrel{\mathop\sim_{ G \rightarrow G_c}} \;\;
\xi^{ 1 / \nu - d } \;\; \sim \;  1 / \xi  \;\; ,
\label{eq:r_xi}
\eeq
whereas, from Eq.~(\ref{eq:chi_corr}), the corresponding result for the curvature
fluctuation is also quite simple
\beq
\chi_{\cal R} (G) \;\; \mathrel{\mathop\sim_{ G \, \rightarrow \, G_c }} \;\;
\xi^{ 2 / \nu - d }  \; \sim \;  \xi^2  \;\; 
\label{eq:chi_xi}
\eeq
in four dimensions, $d=4$. 
The above results are rather helpful in establishing a direct connection between
the correlation length $\xi$ on the one hand, 
and the average local curvature ${\cal R}$ and its fluctuation $\chi_{\cal R}$ on
the other hand..

Figure 1 and Table I present a detailed comparison between the lattice value
for the universal  exponent $\nu$, and other approaches.
The latter include the calculation of $\nu$ in the framework of the $2+\epsilon$ expansion
for gravity in the continuum \cite{epsilon,wei79,eps} carried out to two loop order \cite{aid97},
which gives $\nu^{-1} = d-2 + {3 \over 5} (d-2)^2 + O( (d-2)^3 )$.
Note that the scaling exponent $\nu$ is expected to
be {\it universal}, and therefore characteristic of quantum gravity
(the unique theory of a massless spin two particle in four dimensions \cite{fey}),
and therefore independent of specific features of the regularization
procedure (lattice, dimensional regularization, momentum cutoff etc.).
The same does not apply to the critical point and to the critical amplitudes,
which are generally regularization dependent.
\footnote{
In statistical field theory $y=\nu^{-1}$ is usually referred to as the leading thermal 
(as opposed to magnetic) exponent. 
Under a real space renormalization group transformation with scale $b$ one has
for the corresponding relevant operator $O' = b^y \, O $.
The results presented here point to the existence of a single
relevant operator in the vicinity of the ultraviolet fixed point, so that the 
corresponding operator $O$ is associated, as expected, with the local scalar curvature.
}

Another popular approach to the calculation of the universal exponent $\nu$ 
is based on a truncated renormalization group approach in the 
continuum in four dimensions.
This gave values initially around $\nu^{-1} \simeq 2.8 $ \cite{reu98,litim} with 
some sizeable uncertainties;
it is beyond the scope of this work to go into details regarding the features
of each one of these calculations, so only a few representative cases will be mentioned here.
Recent improved functional renormalization group calculations tend to 
generally fall roughly in the region $\nu^{-1} \simeq 2.0$ to $3.5 $.
Studies using a bi-metric parametrization gave $\nu^{-1} \simeq 4.7 $ \cite{reu10}, 
and later $\nu^{-1} \simeq 3.6 $ in \cite{cod13,reu14}.
In \cite{fal14,fal15} it was argued that only fluctuations should be included 
that have an on-shell meaning, in which case one finds $\nu^{-1} \simeq 3.0 $, much
closer to the lattice results.
In \cite{per15,per16} systematic studies were done of the dependence of the exponent
$\nu$ on the metric parametrization and its influence on the functional measure contribution,
giving generally for the leading exponents $\nu^{-1} \simeq 4,2 $ to lowest order. A similar 
value $\nu^{-1} \simeq 3.0 $ was found using a geometric flow in the linear approximation in \cite{dem14}. Another systematic large parameter space investigation of gauge fixing and measure choices was done in \cite{gie15}, with estimates eventually falling within the above mentioned 
range $\nu^{-1} \simeq 2.0 \dots 3.5 $.


\begin{figure}
\begin{center}
\includegraphics[width=0.7\textwidth]{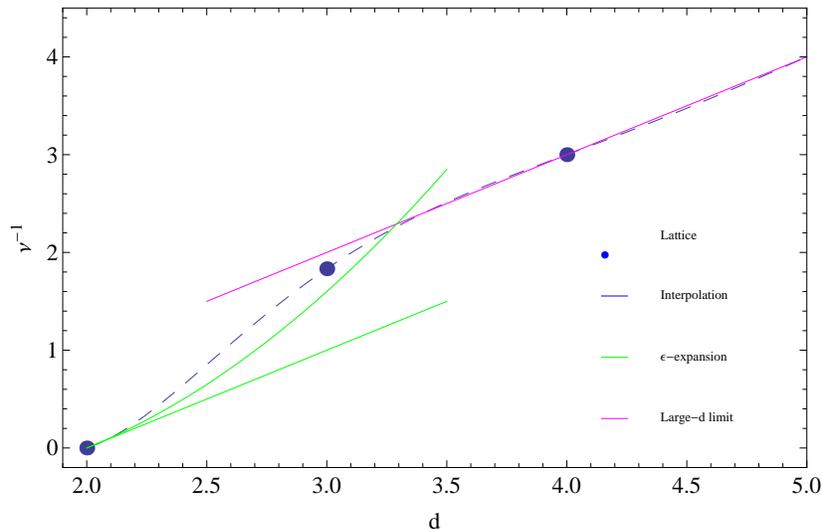}
\end{center}
\caption{
Universal scaling exponent $\nu$ determining the running of $G$ 
[see Eqs.~(\ref{eq:grun_k}), (\ref{eq:grun_box}) and (\ref{eq:grun_r})]
as a function of spacetime dimension $d$.
Shown are the results in $2+1$ dimensions obtained from the
exact solution of the lattice Wheeler-DeWitt equation \cite{htw12,ht16},
the numerical result in four dimensions \cite{ham15}, 
the $2+\epsilon$ expansion result to one \cite{eps} and two loops \cite{aid97},
and the large $d$ result $\nu^{-1} \simeq d-1 $ \cite{larged}.
For actual numerical values see Table I.
}
\end{figure}

The graph in Figure 1 also includes the known exact result for quantum gravity 
in {\it three} spacetime dimensions, obtained from the exact solution of the 
Wheeler-DeWitt equation in
$2+1$ dimensions, which gives $\nu^{-1} = 3/2 $ exactly \cite{hw11,htw12,ht16}.
The latter universal exponent should be compared to the old numerical Euclidean lattice result 
in three dimensions  $\nu^{-1} =1.72(10) $ \cite{gra3d}, to the 
$2+\epsilon$ result of $\nu^{-1}= 1+3/5+ \dots= 8/5=1.60$ \cite{aid97},
and finally to the Einstein-Hilbert truncation results mentioned previously, which
in three dimensions cluster around $\nu^{-1} \approx 2.3 $ \cite{per10} and
$\nu^{-1} \approx 1.6 \dots 2.0 $  \cite{dem14}, again in general agreement 
with the trend found for the lattice results in the same number of dimensions.

\begin{table}

\begin{center}
\begin{tabular}{|l|l|}
\hline\hline
\\
Method used to compute the exponent $\nu$ in d=4 & Universal Exponent $\nu$
\\
\\ \hline \hline
Euclidean Lattice Quantum Gravity \cite{ham15} &  $\nu^{-1} = 2.997(9)  $
\\ \hline
Perturbative $2+\epsilon$ expansion to one loop \cite{eps} & $\nu^{-1}= 2$ 
\\ \hline
Perturbative $2+\epsilon$ expansion to two loops \cite{aid97} & $\nu^{-1} = 22/5 = 4.40$
\\ \hline
Einstein-Hilbert RG truncation \cite{litim} &  $\nu^{-1} \approx 2.80$
\\ \hline
Recent improved Einstein-Hilbert RG truncation \cite{fal14} &
$\nu^{-1} \approx 3.0 $ 
\\ \hline
Geometric argument \cite{larged} $\rho_{vac \; pol} (r) \sim r^{d-1}$ & $\nu^{-1} = d-1 = 3$
\\ \hline
Lowest order strong coupling (large G) expansion \cite{loops} &
$\nu^{-1} = 2 $
\\ \hline
Nonlocal field equations with $G(\Box)$ for the static metric
\cite{eff} &  $\nu^{-1} = d-1$ for $ d \geq 4$
\\ \hline \hline
\end{tabular}
\end{center}
\label{grav1}

\center{\small {\it
TABLE I.  Comparison of estimates for the universal gravitational scaling
exponent $\nu$, based on a variety of different analytical and
numerical methods.
These include the numerical results of \cite{ham15},
The $2+\epsilon$ expansion for pure gravity carried
out to one and two loops \cite{eps,aid97}, an estimate for the leading
exponent in a truncated renormalization group 
expansion \cite{litim,fal14}, a simple argument based on
geometric features of the quantum vacuum
polarization cloud for gravity \cite{larged}, and finally the value 
obtained from consistency of the exact solution to the nonlocal field equation with a $G(\Box)$
for the case of the static isotropic metric \cite{eff,static}.
\medskip}}


\end{table}

Other lattice and continuum methods can be used to provide an estimate for
the exponent $\nu$ in various spacetime dimensions.
Results worth mentioning here include a simple argument
based on the geometric features of the graviton vacuum polarization cloud, which gives 
$\nu = 1 / (d-1)$ for large $d$ \cite{larged} (shown as a straight line in the graph of Figure 1),
and the lowest order estimate for $\nu$ from the first nontrivial order in the strong
coupling expansion of the gravitational Wilson loop \cite{loops}, which gives $\nu =1/2 + \cdots $.
And, finally, the result of \cite{eff,static}, where it was found that a consistent exact solution to 
the nonlocal effective field equations of Eq.~(\ref{eq:grun_field}) (discussed here later on) 
for the static isotropic metric in $d \ge 4 $  can only be found provided  $\nu = 1 /(d-1)$ exactly, in
agreement with the geometric argument mentioned earlier.



\vskip 40pt

\section{Renormalization Group Running of Newton's G}

\label{sec:run}

\vskip 20pt

The results discussed so far are helpful in establishing a direct connection between the
fundamental gravitational correlation  length $\xi$ and various diffeomorphism 
invariant averages such as the average local curvature and its fluctuation.
In this framework one can view the result of Eq.~(\ref{eq:xi_g}) as equivalent to stating that the 
Callan-Symanzik renormalization group beta function has a non-trivial zero at $G_c$.
Generally the cutoff independence of the nonperturbative mass scale 
$m=1/\xi$ in Eq.~(\ref{eq:xi_g})  implies
\beq
\Lambda \, { d \over d \, \Lambda } \, m ( \Lambda , G (\Lambda )) 
\; = \; 0 \;\; .
\label{eq:callan_lambda}
\eeq
Moreover, if one defines the dimensionless function $F(G)$ via
\beq
\xi^{-1} \; \equiv \; m \; = \; \Lambda \; F ( G( \Lambda) ) \;\; ,
\label{eq:f_function} 
\eeq
then, from the usual definition of the Callan-Symanzik beta function
$ \beta (G) \, = \, \partial G (\Lambda) / \partial \log \Lambda $,
one obtains
\beq
\beta (G) \; = \; - \, { F (G) \over F' (G) } \;\; .
\label{eq:callan_gen}
\eeq
It follows that the renormalization group $\beta$-function, and thus
the running of $G (\mu) $ with scale, can be defined some
distance away from the nontrivial  fixed point;
more generally, the running of $G (\mu ) $ is obtained by solving
the differential equation
\beq
\mu \; { d \, G (\mu) \over d \, \mu } \, = \, \beta ( G (\mu) ) \; 
\label{eq:beta_gen} 
\eeq
with $\beta (G) $ obtained from Eq.~(\ref{eq:callan_gen}).
Integrating Eq.~(\ref{eq:beta_gen}) close to the nontrivial fixed point
one obtains for $G > G_c $
\beq
m_0 \, = \, 
\Lambda \, \exp \left ( { - \int^G \, {d G' \over \beta (G') } } \right )
\;\;  \mathrel{\mathop\sim_{G \rightarrow G_c }} \;\;
\Lambda \, | \, G - G_c \, |^{ - 1 / \beta ' (G_c) } \;\;\;\; ,
\label{eq:m_beta}
\eeq
with $m_0$ an integration constant of the RG equations.
It has dimensions of a mass or inverse length, so it is naturally identified with the invariant 
correlation length $\xi$ : $m_0 \propto 1/\xi$.
In particular, comparing results in Eqs.~(\ref{eq:m_beta}) and (\ref{eq:xi_g}) one obtains
\beq
\beta ' (G_c) \, = \, - 1/ \nu \;\; ,
\label{eq:nu_eps}
\eeq
which implies that the universal exponent $\nu$ is directly related to the 
derivative of the Callan-Symanzik $\beta$ function in the 
vicinity of the  fixed point at $G_c$;
computing $\nu$ determines the universal running of $G$ in the vicinity of $G_c$.
In addition, the renormalization group equations generally imply 
that the effective coupling $G(\mu)$ will
grow (anti-screening) or decrease (screening) with distance scale $ r \sim 1/ \mu$, depending on whether $G > G_c$ or $G < G_c$, respectively.
One crucial physical insight obtained from the lattice is that only the phase $G>G_c$ is physically
acceptable \cite{ham15}; the phase $G<G_c$ corresponds at large
distances to an entirely unphysical, collapsed branched polymer with no
sensible continuum limit.

From the previous discussion one infers that 
the physical mass scale $m = \xi^{-1}$ also determines the magnitude of the corrections
to scaling, and plays therefore a role similar to the
scaling violation parameter $\Lambda_{\overline{MS}} $ in QCD.
As in gauge theories, this nonperturbative mass scale emerges dynamically in spite of
the fact that the fundamental gauge boson remains strictly {\it massless} to all orders in
perturbation theory, and consequently its mass does not violate any local
gauge invariance.
Furthermore one expects, as in gauge theories, that in gravity the magnitude of $\xi$ cannot be 
determined perturbatively, and to pin down a specific value requires a fully
nonperturbative approach, as given here by the lattice formulation.
In turn, the genuinely nonperturbative physical mass parameter $m = \xi^{-1} $ of
Eq.~(\ref{eq:xi_g}) is itself a renormalization group invariant and thus {\it scale independent}. 
In the immediate vicinity of the fixed point it obeys the general renormalization group equation,
which follows from Eq.~(\ref{eq:callan_lambda}),
\beq
\mu \; { d \over d \, \mu } \; m ( \mu, G(\mu)) \; = \;
\mu \; { d \over d \, \mu } \;
\left ( \; A_m  \; \mu \; | \, G (\mu) - G_c \, |^{ \nu } \; \right )
\, = \, 0  \;\;\;\; 
\label{eq:callan} 
\eeq
with  $\mu$ an arbitrary momentum scale.
Here again, by virtue of Eq.~(\ref{eq:xi_g}), the second expression 
on the right-hand-side is only appropriate in very close proximity of the 
 fixed point at $G_c$.
Solving explicitly Eq.~(\ref{eq:callan}) for $G(q^2)$, with
$ q $ an arbitrary wave vector scale, one finally obtains for the running 
of Newton's $G$ with the action of Eqs.~(\ref{eq:ac_cont}) or (\ref{eq:ac_latt})
\beq
G(q^2) \; = \; G_c \left [ \; 1 \, + \, c_0 \, 
\left ( { m^2 \over q^2 } \right )^{1 / 2 \, \nu} \, 
+ \, O ( \left ( { m^2 \over q^2 } \right )^{1 / \nu} ) \; \right ]
\;\; .
\label{eq:grun_k} 
\eeq
Here again $m=1/ \xi$,  and the coefficient $c_0$  for the amplitude of the 
quantum correction is $ c_0 \; = \; 8 \pi \, G_c \, A_\xi^{1/\nu} $, 
with $ A_\xi = 0.80(3) $ from a numerical study of the decay of 
curvature correlation functions, and also as before $\nu=1/3$ \cite{cor94,ham15}.
Consquently the dimensionless amplitude for the leading quantum correction in the lattice running of Eq.~(\ref{eq:grun_k}) is $c_0 \approx 8.02 $.
This then completely determines the running of $G$ in the vicinity of the fixed point,
namely for scales $ r \ll \xi $.
Note that in the lattice theory of gravity only the smooth
phase with $G>G_c$ exists (in the sense that an instability develops
and spacetime collapses onto itself for $G<G_c$), which then
implies that the gravitational coupling can only 
{\it increase} with distance [+ sign for the quantum correction in Eq.~(\ref{eq:grun_k})].
In other words, a gravitational screening phase does not exist in
the lattice theory of quantum gravity.
The above situation appears to be true both for the Euclidean theory
in four dimensions, and in the Lorentzian version in $3+1$ dimensions \cite{htw12}.
A better, covariant formulation for the running of Newton's $G$ is given later, in
Eq.~(\ref{eq:grun_box}).
Note also that the domain of validity for the expressions in Eq.~(\ref{eq:grun_k}) is 
$ q \gg m \equiv 1 / \xi $ or $ r \ll \xi $; the strong infrared divergence 
at $q \simeq 0$ is largely an artifact of the current expansion, and should be 
regulated either by cutting off the momentum integrations at $ q \simeq m = 1 / \xi $,
or by the replacement on the r.h.s. $q^2 \rightarrow q^2 + m^2 $.

It is clear that the magnitude of the quantum correction in 
Eq.~(\ref{eq:grun_k}) depends crucially on the magnitude of the 
nonperturbative physical scale $\xi$. 
It will be argued later that this quantity is related, as in Yang-Mills theories,
to the gravitational condensate, physically represented by the observed cosmological constant.
Therefore at this stage it will turn out that the only physically sensible interpretation is that
the observed $\lambda_{\rm obs}$ is tentatively related to the scale $\xi$,
\beq
\third \, \lambda_{\rm obs} \; \simeq \; { 1 \over \xi^2 } \;\; .
\label{eq:lambda_xi}
\eeq
From the above perspective "short distances" are not really that short, since $\xi$ in comparison
to $G$ or the Planck length is a very large quantity, of cosmological magnitude.
\footnote{
The fundamental nonperturbative scale $\xi$ plays a crucial role in the following,
and having a precise quantitative value for it is of paramount importance
when trying to make contact with current astrophysical and cosmological observations.
Here, for concreteness, a specific value in $Mpc$ will be assumed, in line
with the most recent satellite data, see for example \cite{planck15}. 
It is nevertheless quite possible that significant updates to this value will
take place in the next few years, as increasingly sophisticated data, and 
data analysis methods, become available. 
One would nevertheless expect that various predictions, arising from
the vacuum condensate framework described here, should lead to {\it one} single
consistent value for the scale $\xi$.
In this context it is worth remembering that before 1999 astrophysical observations
were deemed to be entirely consistent with $\lambda=0$.}
It follows that the reference scale for the running of $G$ in Eq.~(\ref{eq:grun_k})  is set 
by a correlation length $\xi$ which, by  Eqs.~(\ref{eq:r_xi}) and later (\ref{eq:r_xi_amp}),
is related to the observed large-scale curvature.
In particular, the specific form for the running of $G$ with scale suggests that
no detectable corrections to classical gravity should arise either a)
until the scale $r$ approaches the very large (cosmological)
scale $\xi$ or b) until one reaches extremely short distances comparable
to the Planck length $r \sim l_p $ (at which point higher derivative
terms, light matter corrections, and string contributions come into play).
In other words, the results of Eq.~(\ref{eq:grun_k}) 
[or later in the covariant form of Eq.~(\ref{eq:grun_box})] would imply that
classical gravity is largely recovered on atomic, laboratory, solar, and
even galactic scales, or as long as the relevant distances satisfy $r \ll \xi$.



\vskip 40pt

\section{Gravitational Wilson Loop and Curvature Condensate}

\label{sec:smear}

\vskip 20pt

In gauge theories the Wilson loop is known to play a central role:
on the one hand it is a manifestly gauge invariant quantity, on the other hand it provides key physical information on the nature of the static potential between two quarks.
In gravity it is possible to construct a close analog of the gauge Wilson loop, by taking the path-ordered product of rotation matrices (describing the parallel transport of a vector, and thus specified in terms of the affine connection) along a closed loop.
Nevertheless this path ordered product is not related to the gravitational potential; the latter is obtained from a different set of observables which involve the correlations of particle world lines  {modacorr,modaloop}.
Instead the gravitational Wilson loop provides information, as already in the infinitesimal loop case, on the behavior of {\it curvature} on very large scales.

The required integration over rotation matrices (or, equivalently, the integration over the affine connection) is most easily done in a first order formulation, where the affine connection and the metric are considered as independent degrees of freedom. 
Such a formulation exists on the lattice \cite{cas89} and is therefore most suitable for computing 
the gravitational Wilson loop \cite{loops}.
As in the gauge theory case, the integration over rotation matrices is performed using an invariant Haar measure over the group, which then almost immediately leads to a (minimal) area law for the quantum gravitational Wilson loop,
\beq
W ( C ) \; \sim \; e^{ - A / \xi^2 } \;\; .
\label{eq:loop_quant}
\eeq
Note that in the above expression use has been made of the fact that the basic reference scale appearing in the area law is the correlation length $\xi$ , a well-known scaling result in gauge theories
and justified there by renormalization group arguments.
Also, $C$ denotes the closed path that defines the loop; 
a more precise definition of the gravitational loop \cite{loops} will be given further below.
Suffice it to say here that the use of the Haar measure over rotations assumes and implies large 
local fluctuations in the metric, and thus in the affine connection, which is certainly justified for 
large $G$, where gravitational fluctuations in different spacetime regions decouple.

On the other hand, a macroscopic semiclassical observer is led to relate the parallel transport 
of a coordinate  vector around a very large closed loop, via Stoke's theorem,
to the value of the locally measured curvature.
This then leads immediately to the semiclassical result \cite{loops}
\beq
W ( C ) \; \sim \; e^{- A \cdot R } \;\; ,
\label{eq:loop_class}
\eeq
where $R$ is a measure of the slowly varying local macroscopic curvature;
again a more precise definition will be given further below.
Comparing the quantum result of Eq.~(\ref{eq:loop_quant}) to the semiclassical
result of Eq.~(\ref{eq:loop_class}) (which is feasible since both contain the minimal area $A$ of the loop
in question) then provides a more or less direct relationship between the local large scale,
semiclassical curvature $R$ and the correlation length $\xi$, namely $R \sim 1 / \xi^2 $, 
a result already alluded to earlier in Eq.~(\ref{eq:lambda_xi}).

This last set of considerations in turn provides a further key ingredient in quantum gravity, namely
the correspondence between the macroscopic semiclassical curvature and the invariant correlation length $\xi$.
One immediate consequence is that the scale for quantum effects in Eq.~(\ref{eq:grun_k})
is related to the observed cosmological constant, which in quantum gravity
acts effectively as an infrared regulator.
Thus potentially serious infrared divergences associated with the masslessness of the graviton are regulated by this new nonperturbative scale $\xi$, a mechanism which is
similar to the way infrared divergences regulate themselves dynamically in QCD and 
non-Abelian lattice gauge theories.
Consequently the scale $\xi$ plays a role which seems analogous to the scaling violation parameter
$\Lambda_{\overline{MS}}$ in QCD; one important difference is that the running of $G$, 
due to the existence of a nontrivial  fixed point, is not logarithmic. 
Instead the correct scale dependence of $G$ is given by Eq.~(\ref{eq:grun_k}) and thus follows 
a power law, with an exponent $\nu$ related to the derivative of the beta function at 
the fixed point in G. 

A second crucial consequence is that the scale for quantum effects is not given by Newton's constant; it is given instead by the size of $\xi$, which because of its relationship to the cosmological constant is a very large, cosmological scale of the order of $10^{28} cm $.
It would seem therefore that such quantum effects will only become detectable when one explores 
the nature of gravity on cosmological scales comparable to $\xi$.
The running of $G$ is exceedingly tiny on solar system and galactic scales, but nevertheless increases dramatically as one approaches distance scales which are comparable to the observed
cosmological constant $\lambda$.
What then remains to be done is therefore to incorporate the above running of $G$ into a set of generally covariant equations which can then be applied to the calculation of quantum corrections to known classical gravity results at very large distances.
This will be discussed later.
\footnote{
Note that in gauge theories the correlation length $\xi$ can be determined directly numerically by investigating the decay of Euclidean correlation functions of suitable local operators as a function of the separation distance. Generally these correlation functions are dominated by the lightest particle with a given spin. In the case of gravity such a detailed and complete analysis has not been performed yet,
although it is in principle feasible, just as it is in lattice QCD. 
One complication that arises in the case of gravity is the fact that correlation functions between invariant local operators have to be computed at a fixed geodesic distance \cite{cor94}.
The latter of course fluctuates, depending on the choice of background metric configuration 
used in evaluating the Feynman path integral.
}

It is important at this stage to understand where the Wilson loop relationship in Eqs.~(\ref{eq:loop_quant}) and (\ref{eq:loop_class}) is coming from.
A precise definition of the gravitational Wilson loop was given in \cite{npb400,modacorr,loops}.
First note that infinitesimal transport loops appear already, for example, in the definition of the 
correlation function for the scalar curvature, Eq.~(\ref{eq:corr_cont}).
Here what will be considered instead is the parallel transport of a vector around a loop $C$
which is {\it not} infinitesimal.
In the following this loop will be assumed to be close to planar,
a well-defined geometric construction described in detail in \cite{loops}.
First define the total rotation matrix ${\bf U}(C)$ along the path $C$ via a 
path-ordered (${\cal P}$) exponential of the integral of the
affine connection $ \Gamma^{\lambda}_{\mu \nu}$,
\beq
U^\mu_{\;\; \nu} (C) \; = \; \Bigl [ \; {\cal P} \, \exp
\left \{ \oint_{C}
\Gamma^{\cdot}_{\lambda \, \cdot} d x^\lambda
\right \}
\, \Bigr ]^\mu_{\;\; \nu}  \;\; .
\label{eq:rot_cont}
\eeq
The lattice action itself already contains contributions from infinitesimal
loops, but more generally one might want to consider near-planar, 
but noninfinitesimal, lattice closed loops $C$.
To make the above expression well defined it needs to be put on a lattice. 
There one defines a finite product of elementary rotations defined along 
a given lattice path
\beq
U^{\mu}_{\;\; \nu} (C) \; = \;
\Bigl [ \prod_{s \, \subset C}  U_{s,s+1} \Bigr ]^{\mu}_{\;\; \nu}  \; .
\label{eq:latt_wloop_a}
\eeq
The introduction of such rotation matrices in the Regge-Wheeler lattice
was discussed in detail in \cite{reg61,whe64,loops}, and a first order lattice
formulation for gravity based on it was given in \cite{cas89}; 
the following discussion will be based this well understood formalism.
A coordinate scalar can then be defined by contracting the above rotation matrix ${\bf U}(C)$ 
with a unit length area bivector $\omega_{\alpha\beta} (C )$, representative 
of the overall geometric features of the loop.
Now if the parallel transport loop in question is centered at the point
$x$, then one can define the operator $W_C (x)$ by
\beq
W_C (x) \; = \; \omega_{\mu\nu}(C,x) \; U^{\mu\nu} (C,x) 
\label{eq:loop_x}
\eeq
with the near-planar loop centered at $x$ and of linear size $r_C$.
Of course for an {\it infinitesimal} loop, involving an infinitesimal lattice
path $C_0$ of linear size $\sim a$, the overall rotation matrix is given by 
\beq
U^{\mu}_{\;\; \nu} (C_0) \; = \; 
\; \simeq \;  \Bigl [ \, e^{ \, \half \, R \cdot A } \Bigr ]^{\mu}_{\;\; \nu} 
\; \simeq \;  \Bigl [ \, e^{ \delta \cdot \omega } \Bigr ]^{\mu}_{\;\; \nu} 
\eeq
where now $\omega_{\mu\nu} (C_0) $ is the area bivector associated with the
infinitesimal loop of area $\sim a^2$, and $\delta $ the corresponding deficit angle;
here $R$ is lattice Riemann tensor at the hinge (triangle) in question,
and $A_{\mu\nu}$ the corresponding area bivector.
Then an invariant correlation function between two such operators is given by
\beq
G_{C} (d) \; = \; < W_C (x) \; W_C (y) \; \delta ( | x - y | - d ) >_c
\;\; ,
\label{eq:sme_loop}
\eeq
with the two loops separated by some fixed geodesic distance $d$.
Of course for {\it infinitesimal} loops one recovers the expressions given
earlier in Eqs.~(\ref{eq:corr_cont}) and (\ref{eq:corr_latt}).


\begin{figure}
\begin{center}
\includegraphics[width=0.7\textwidth]{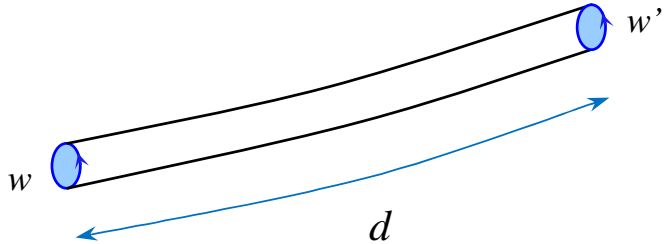}
\end{center}
\caption{
Correlation function of two infinitesimal parallel transport loops, 
separated by a geodesic distance $d$.
This correlation corresponds to the one defined in Eqs.~(\ref{eq:corr_cont})
and (\ref{eq:corr_latt}).
In the strong coupling limit one needs, in order to get a non-zero correlation, 
to fully tile the minimal tube connecting the two infinitesimal initial and final loops.
}
\end{figure}

In general one needs to specify the relative orientation of the two loops.
So, for example, one can take the first loop in a plane perpendicular to the
direction associated with the geodesic connecting the two points, and
the same for the second loop;
the parallel transport of a vector along this geodesic will then be
sufficient to establish the relative orientation of the two loops.
Nevertheless if one is interested in the analog (for large loops) of the scalar
curvature, then it will be adequate to perform a weighted sum over all
possible loop orientations at both ends.
This is in fact precisely what is done for infinitesimal loops of size $ r_C \sim
a $, if one looks carefully at the way the Regge lattice action was originally defined.

It is possible to give a more quantitative description for the behavior of the loop-loop
correlation function given in Eq.~(\ref{eq:sme_loop}), at least in the strong coupling limit.
The following estimate is based on the previous results and definitions, and
is further illuminated by the important analogy and correspondence of lattice gravity
to lattice non-Abelian gauge theories outlined in detail in \cite{larged,loops}.
First it will be convenient to assume that the two (near planar) loops are of comparable shape
and size, with overall linear sizes $ r_C \sim L/(2 \pi) $ and perimeter $ P \simeq L $.
In addition, the two loops are separated by a distance $ d \gg L $, 
and for both loops it will be assumed that this separation is much
larger than the lattice spacing, $ d \gg a$ and $L \gg a$.
Then to get a nonvanishing correlation in the strong coupling, large $G$
limit it will be necessary to completely tile a tube connecting the two
loops, due to the geometric minimal area law arising from the use of the uniform (Haar)
measure for the local rotation matrices at strong coupling, again as discussed 
in detail in \cite{loops}.
Quite generally in this limit one expects an {\it area law} for the correlation between 
gravitational Wilson loops (see also the discussion for the Wilson loop itself
given later below), which here takes the form
\beq
G_{C} (d) \; \simeq \; 
\exp \left \{ - \, { L \cdot d \over \xi^2 } \right \} \; 
= \; \exp \left \{ - \, { A (L,d) \over \xi^2 } \right \} \; ,
\label{eq:sme_loop_asy}
\eeq
with $ A (L,d) $ the minimal area of the tube connecting the two loops.
Consistency of the above expression with the corresponding result for small
(infinitesimal) loops given in previously in Eq.~(\ref{eq:corr_exp})
requires that for small loops (small $L$) the value of $L$ saturates to $\xi$, $ L \simeq \xi $,
so that the correct exponential decay is recovered for small loops.

This result is not unexpected, since $\xi$ can only come into play only for distances
much larger than the fundamental lattice spacing $a$.
Consequently the asymptotic decay of correlations for
large loops is somewhat different in form as compared to the decay of
correlations for infinitesimal loops, with an additional factor of $\xi$ appearing
for large loops; nevertheless in both cases one has the expected minimal area law.
In other words, the results of Eqs.~(\ref{eq:corr_pow}) and
(\ref{eq:corr_exp}) only apply to infinitesimal loops, which probe the
parallel transport on infinitesimal (cutoff) scales; these
results then need to be suitably amended when much larger loops, 
of semiclassical significance, are considered.


\begin{figure}
\begin{center}
\includegraphics[width=0.7\textwidth]{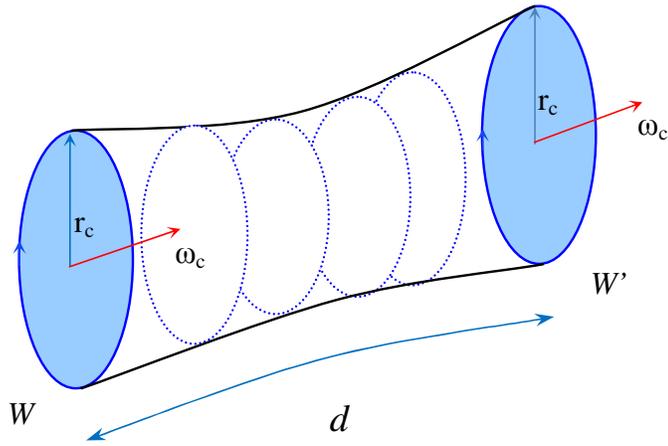}
\end{center}
\caption{
Correlation function for two large parallel transport loops of size
$r_c$ and orientation $\omega_c$, separated by a geodesic distance $d$.
This correlation corresponds to the one defined in Eq.~(\ref{eq:sme_loop}).
In the strong coupling limit one needs, in order to get a non-zero correlation, 
to fully tile the tube connecting the two large initial and final loops.
}
\end{figure}

The above result applies to strong coupling, $G \gg G_c$.
As one approaches the critical point at $G_c$ more than one exponential will
contribute, in analogy to Eq.~(\ref{eq:corr_pow}) for the single plaquette correlation.
If the single loop contribution is proportional, as in the area law of
Eq.~(\ref{eq:sme_loop_asy}), to $\exp (- m^2 L d )$ with $m=1/ \xi$, 
then for a spectral function 
$ \rho (\mu ) \, = \, 2 \, B \, \mu^{\beta -1} / \Gamma (\beta / 2)$ 
one obtains, in the limit of a small infrared cutoff $m \equiv 1/\xi$,
\beq
G_C (d)  \; = \; \int_m^\infty \, d \mu \; \rho (\mu) \; e^{- \mu^2 L d} 
\; \simeq \;
{B  \over ( L \, d )^{ \beta / 2 } } \, - \, 
{ B \, m^\beta \over \Gamma ( 1 + \beta / 2 ) } \, + \, \cdots \;\; .
\label{eq:spectral_w}
\eeq
Consistency of this expression with the infinitesimal loop result of
Eq.~(\ref{eq:corr_pow}) then fixes $ \beta = \alpha / 2$ and $ B = a^\alpha \, A$.
For large loops one obtains
\beq
G_C (d) \; = \;  A \; \left ( {a  \over L } \right )^{ \alpha } \cdot { 1 \over d^\alpha }
\cdot \left [ \, 1 \, - \, { ( d \, L \, m^2 )^\alpha \over \Gamma ( 1 + \alpha ) }
\, + \, \cdots \right ]
\label{eq:spectral_w1}
\eeq
Furthermore from $\alpha=2 \, (4-1/\nu)=2$ , as in Eq.~(\ref{eq:corr_pow}),  one has 
$\beta= 2 \alpha = 4$,
which gives  for large (non-infinitesimal) loops of linear size $L$ the following result,
valid in the vicinity of the fixed point $ G \simeq G_c $,
\beq
G_C (d) \; = \; A \; \left ( {a  \over L } \right )^{ 2} \cdot { 1 \over d^2 } \cdot
\left [ \, 1 \, - \, \half \, d^2 \, L^2 \, m^4 \, + \, \cdots \right ]
\label{eq:spectral_w2}
\eeq
This last function describes the correlation of large, {\it macroscopic} parallel
transport loops of linear size $ L \gg a $, separated by an invariant distance $d$.
Note the additional suppression, by a factor of $ (a/L)^2 $ when compared
to the {\it infinitesimal} loop correlation function of Eqs.~(\ref{eq:corr_pow})
and (\ref{eq:spectral_m});
for a macroscopic (semi-classical) parallel transport loop
one has $ L \sim \xi \gg a $.
Note that due to the dimensions of the $R-R$ curvature correlation function
[see Eq.~(\ref{eq:corr_cont})] the constant
$A$ has dimensions of one over length squared, $A \sim A_0 / a^2 $ with
$A_0$ dimensionless.
Furthermore, it is important to note that when the correlation of
larger (i.e. non-infinitesimal) loops are considered the power law
decay is unchanged, only the amplitude gets modified
[compare Eq.~(\ref{eq:spectral_w2}) on the one hand, and Eqs.~(\ref{eq:corr_pow}) and (\ref{eq:spectral_m}) given earlier for infinitesimal loops;
in both cases the dependence on the separation is $\sim 1/d^2$.
Again, to clarify, Eq.~(\ref{eq:sme_loop_asy}) describes the exponential
"large distance" ($ d \gg \xi $) behavior of the loop correlation function, 
whereas Eq.~(\ref{eq:spectral_w2}) describes the power law
"short distance" ($ d \ll \xi $) behavior of the same loop correlation function.
So the above result is analogous to what was found in Eq.~(\ref{eq:corr_exp}) 
describing there, on the one hand, the exponential "large distance" ($ d \gg \xi $) behavior
of the microscopic curvature (infinitesimal loop) correlation function, and on the other
hand, from Eq.~(\ref{eq:corr_pow}), the power law
"short distance" ($ d \ll \xi $) behavior of the same correlation function.

One crucial ingredient needed in pinning down the magnitude of
the quantum correction for $G ( q^2 )$ in Eqs.~(\ref{eq:grun_k})
or (\ref{eq:grun_box}), as well as the result for the loop correlation function
of Eq.~(\ref{eq:spectral_w2}),  is the actual value of the genuinely nonperturbative
reference scale $\xi$.
It was argued in \cite{loops} that, in analogy to ordinary
gauge theories, the gravitational Wilson loop itself provides precisely such an insight.
The main points of the argument are rather simple, and can thus
be reproduced in just a few lines.
In analogy to the gauge theory case, these arguments
rely generally on the concept of universality, the existence
of a universal correlation length at strong coupling,
and the use of the Haar invariant measure to integrate
over large fluctuations of the metric, or of the fundamental local parallel 
transport matrices.
Following \cite{modacorr,modaloop}, in \cite{loops} the vacuum expectation 
value corresponding to the gravitational Wilson loop is naturally defined as
\beq
< W(C)> \; = \; < \tr \left [ \, \omega (C) \; U_1 \; U_2 \; ... \;  U_n \right ] > \;\; .
\label{eq:wloop_latt}
\eeq
Here the $U$'s are elementary rotation matrices, whose form is 
determined by the affine connection, and which therefore
describe the parallel transport of vectors around a loop $C$; see
also Eq.~(\ref{eq:latt_wloop_a}).
Again here $\omega_{\mu\nu} (C) $ is a constant unit bivector, characteristic of the overall
geometric orientation of the loop, giving the notion of a normal to the loop.
In the continuum the combined rotation matrix ${\bf U}(C)$ is given by the
path-ordered (${\cal P}$) exponential of the integral of the
affine connection $ \Gamma^{\lambda}_{\mu \nu}$, as in
Eq.~(\ref{eq:rot_cont}),
so that the previous expression represents a suitable
discretized and regularized lattice form.
It can then be shown \cite{loops} that quite generally
in lattice gravity, and for sufficiently strong coupling, one 
obtains universally an area law for large near planar loops as advertised in Eq.~(\ref{eq:loop_quant})
\beq
<W(C)> \; \simeq \; \exp \, ( - \, A_C / {\xi }^2 ) \;\; ,
\label{eq:wloop_latt1}
\eeq
where $A_C$ is the geometric minimal area of the loop as spanned by
a given perimeter.
\footnote{
In Wilson's lattice formulation \cite{wil74} this is a standard textbook result 
for non-Abelian gauge theories, 
see for example \cite{frampton,peskin}, specifically Eq. (22.3)
 in the second reference.
There $\xi$ represents the gauge field correlation length, or the inverse of
the lowest glueball mass; 
here following \cite{loops}
the gravitational result is written in the same invariant scaling form involving the
fundamental nonperturbative correlation length $\xi$.}
This last result relies on a modified first order
formalism for the Regge lattice theory \cite{cas89}, in which
the lattice metric degrees of freedom are separated out into local Lorentz
rotations and tetrads.
Moreover, the result of Eq.~(\ref{eq:wloop_latt}) is in fact rather universal, 
since it can be shown to hold in all known lattice
formulations of quantum gravity at least in the strong coupling (large $G$) regime.
In \cite{loops} an explicit expression for the correlation length $\xi$ appearing in 
Eq.~(\ref{eq:wloop_latt1}) was given in the strong coupling limit.
There one finds 
$ \xi = 4 / \sqrt{ k_c \, \vert \log ( k / k_c ) \vert \, + \, O (k^2)}$.
For $k$ close to $k_c$ this then gives immediately
$ \xi \, \simeq \, 4 \, \vert k_c - k \vert^{ - 1/2 } $ and thus, to this
order, $\nu = \half $ in Eq.~(\ref{eq:xi_g}).
Nevertheless, the discussion of the previous sections and
the numerical solution of the full lattice theory suggests that 
the correct expression for $\xi$ to be used in
Eq.~(\ref{eq:wloop_latt1}) should be the one in Eq.~(\ref{eq:xi_g}),
with $\nu =1/3 $ [Eq.~(\ref{eq:Gc})], $k_c$ given in 
Eq.~(\ref{eq:Gc}) and amplitude $A_\xi = 0.80(3) $.

One then needs to make contact between these results
and a semiclassical description, which requires that one 
connects the nonperturbative result of Eq.~(\ref{eq:wloop_latt1}) to
a suitable semiclassical physical observable.
By the use of Stokes's theorem, semiclassically the parallel
transport of a vector round a very large loop depends on the exponential of a 
suitably coarse-grained Riemann tensor over the loop.
In this semiclassical picture one has for the combined
rotation matrix ${\bf U}$
\beq
U^\mu_{\;\; \nu} (C) \; \sim \;
\Bigl [ \;
\exp \, \left ( \half \,
\int_{S(C)}\, R^{\, \cdot}_{\;\; \cdot \, \lambda\sigma} \, d A^{\lambda\sigma} \;
\right )
\, \Bigr ]^\mu_{\;\; \nu}  \;\; ,
\label{eq:rot-cont1}
\eeq
where $ A^{\lambda\sigma} $ is an area bivector,
$ A^{\lambda\sigma}_{C} = \half \oint_C dx^\lambda \, x^\sigma $.
The above semiclassical procedure then gives for the loop in question
\beq
W(C) \, \simeq \, \tr \left \{ \, \omega (C) \, \exp \left ( \, \half \,
\int_{S(C)}\, R^{\, \cdot}_{\;\; \cdot \, \lambda\sigma} \, d A^{\lambda\sigma}_{C} \; 
\right ) \right \} \; .
\label{eq:wloop_curv}
\eeq
Here $\omega_{\mu\nu} (C) $ is a constant unit bivector,
characteristic of the overall geometric orientation of the 
parallel transport loop.
For a slowly varying semiclassical curvature, the $R$ contribution can be taken out of
the integral, so that the remaining integral depends on the
overall large loop with some minimal area $A_C$, for a given perimeter $C$. 
Then, by directly comparing coefficients for the two area terms in 
Eqs.~(\ref{eq:wloop_latt1}) and (\ref{eq:wloop_curv}),
one concludes that the average large-scale curvature is of 
order $ + 1 / \xi^2 $, at least in the strong coupling limit \cite{loops}.
Since the scaled cosmological constant can be viewed as a measure of 
the intrinsic curvature of the vacuum, the above argument
then leads to an effective positive cosmological constant for this phase, corresponding to
a manifold which behaves semiclassically 
as de Sitter ($\lambda > 0$) on very large scales \cite{loops}.
For related interesting ideas see also \cite{she12}.

The above arguments then lead to the following key connection between the macroscopic 
(semiclassical) average curvature and the nonperturbative correlation length $\xi$
of Eqs.~(\ref{eq:corr_exp}),  (\ref{eq:grun_k}), (\ref{eq:grun_r})  and (\ref{eq:grun_box}), namely
\beq
\langle \, R \, \rangle_{\rm large \; scales} 
\;\; \sim \; + \, 6 / \xi^2 \;\; ,
\label{eq:xi_r}
\eeq 
at least in the strong coupling (large $G$) limit.
It is important to note that the result of Eq.~(\ref{eq:xi_r})
applies to parallel transport loops whose linear size $r_C$ 
is much larger than the lattice spacing, $ r_C \gg a $;
nevertheless in this limit the answer for the macroscopic
curvature in Eq.~(\ref{eq:xi_r}) becomes independent of
the loop size or its minimal area \cite{loops}.
Furthermore, these arguments lead, via the classical field equations,
to the identification of $1 / \xi^2$ with the observed (scaled)
cosmological constant $\lambda_{\rm obs}$, 
\footnote{Up to a constant of
proportionality, expected to be of order unity.}
\beq
\third \;  \lambda_{\rm obs} \; \simeq \;  + \, { 1 \over \xi^2 }  \; .
\label{eq:xi_lambda}
\eeq 
In this picture the latter is regarded as the quantum {\it gravitational condensate},
a measure of the vacuum energy, and thus of the intrinsic curvature of
the vacuum \cite{loops}.
It is nonzero as a result of nonperturbative graviton condensation.

The above considerations can finally contribute to providing a quantitative handle 
on the physical {\it magnitude} of the nonperturbative scale $\xi$.
From the observed value of the cosmological constant 
(see for ex. the 2015 Planck satellite data \cite{planck15}) one
obtains a first estimate for the absolute magnitude of the scale $\xi$,
\beq
\xi \; \simeq \; \sqrt{3 / \lambda} \; \approx \; 5320 \, Mpc \;\; .
\label{eq:xi_mpc}
\eeq
Irrespective of the specific value of $\xi$, this would indicate
that generally the recovery of classical GR results happens
for distance scales much smaller than the correlation length $\xi$.
\footnote{
The value for $\lambda$, and therefore $\xi$, relies on a multitude of
current cosmological data, which nowadays is usually analyzed in the 
framework of the standard $\Lambda CDM$ model. 
Included in the usual assumptions is the fact that Newton's $G$
does {\it not} run with scale. 
If such an assumption were to be relaxed, it would affect 
a number of cosmological parameters, including $\lambda$,
whose value could then change significantly. 
In the following the estimate of Eq.~(\ref{eq:xi_mpc}) will be 
used as a sensible starting point.}
In particular, the Newtonian potential is expected to acquire a tiny 
quantum correction from the running of $G$ [see Eq.~(\ref{eq:grun_box})]
\beq
V(r) \; = \; - \, G(r) \cdot { m_1 \, m_2 \over r } \;\; ,
\label{eq:pot}
\eeq
For example, in the case of the static isotropic metric one finds that $G(r)$ is given 
explicitly by \cite{static}
\beq
G \; \rightarrow \; G(r) \; \equiv \; 
G \, \left ( 1 \, + \, 
{ c_0 \over 3 \, \pi } \, m^3 \, r^3 \, \log \, { 1 \over  m^2 \, r^2 }  
\, + \, \dots
\right )
\label{eq:grun_r}
\eeq
with $m \equiv 1/ \xi$, so that quantum effects become negligible on distance scales $r \ll \xi$.
\footnote{The quantum gravity correction is reminiscent of the Uehling term in
QED; nevertheless the latter is purely logarithmic, and the infrared cutoff
there is provided by the smallest mass scale appearing in QED loop corrections, 
the renormalized electron mass. In quantum gravity the role of the
infrared cutoff is played by the graviton mass, which in perturbation theory
(as in Yang-Mills theories) stays strictly zero to all orders, due to local coordinate invariance.}

One might think perhaps that the running of $G$ envisioned here might lead to 
observable consequences on much shorter, galactic length scales.
That this is not the case can be seen, for example, from the following argument.
For a typical galaxy one has an overall size $\sim 30 \, kpc$, giving for the quantum correction
the estimate, from Eq.~(\ref{eq:grun_r}) for the static potential, 
$( 30 \, kpc / 5320 \times 10^3 \, kpc )^3 \sim 1.79 \times 10^{-16} $ which is tiny due 
to the large size of $\xi$ [see Eq.~(\ref{eq:xi_mpc})].
It seems therefore unlikely that such a correction will be detectable at these scales,
or that it could account, in part, for anomalies in the galactic rotation curves.
The above argument nevertheless shows a certain sensitivity of the results to the 
value of the scale $\xi$;
thus an increase in $\xi$ by a factor of two tends to reduce the effects of $G(\Box)$ by 
$2^3 = 8$, as can be seen from Eq.~(\ref{eq:grun_box}) with $\nu=1/3$
and the fact that the amplitude of the quantum correction 
is always proportional to the combination $c_0 / \xi^3$.
Figure 4 shows the expected qualitative behavior for the
running $G(q)$ over scales slightly smaller or comparable to $\xi$.
The main uncertainty arises from estimating the physical 
magnitude of $\xi$ itself [see Eq.~(\ref{eq:xi_mpc})].
Specifically, from Eq.~(\ref{eq:grun_k}) the lattice prediction 
at this point is for roughly a 5\% effect on
scales of $ 0.148 \times 5320 \, Mpc \approx 790 \, Mpc $, and a 
10 \% effect on scales of $ 0.187 \times 5320 \, Mpc \approx 990 \, Mpc $.

Figure 5 shows how the lattice running of $G(q)$, given earlier in Eq.~(\ref{eq:grun_k}),
compares to the continuum analytical self-consistent
Hartee-Fock solution to Dyson's equations
for quantum gravity, obtained recently in \cite{hf18} :
\beq
G_{HF} (q^2) \; = \; G_c \left [ \; 1 \, - \, { 3 \, m^2 \over 2 \, q^2 } \, 
\log \left (\, { 3 \, m^2 \over 2 \, q^2 } \right ) \; \right ]
\;\; .
\label{eq:grun_hf} 
\eeq
Here again $m$ is related to the gravitational correlation length via 
$m \equiv 1/ \xi$, see Eqs.~(\ref{eq:xi_lambda}) and (\ref{eq:xi_mpc}).
One notes therefore that the Hartree-Fock approximation to the self-consistent equation
for the graviton vacuum polarization tensor also predicts an infrared rise of $G(q)$
(antiscreening), and furthermore unambiguously determines
the amplitude of the quantum correction ($c_0$, here equal to $3/2$).
In this approximation the mean field result for the exponent is
$ \nu = 1 /(d-2) $, so that the power is equal to two for Eq.~(\ref{eq:grun_hf})
in four dimensions.
So there are two main differences that stand out compared to the lattice result of
Eq.~(\ref{eq:grun_k}), namely that the power is two here and not three, and the fact that 
here there is an additional, slowly varying $\log (q) $ component.
\footnote{
One of the earliest applications of the Hartree-Fock approximation
to solving Dyson's equations for propagators and vertex functions
was in the context of the BCS theory for a superconductor.
Later it was applied to a (perturbatively non-renormalizable) 
relativistic theory of a self-coupled Fermion, where it provided the 
first convincing
evidence for a dynamical breaking of chiral symmetry and the
emergence of Nambu-Goldstone bosons \cite{njl61}.
}



\begin{figure}
\begin{center}
\includegraphics[width=0.7\textwidth]{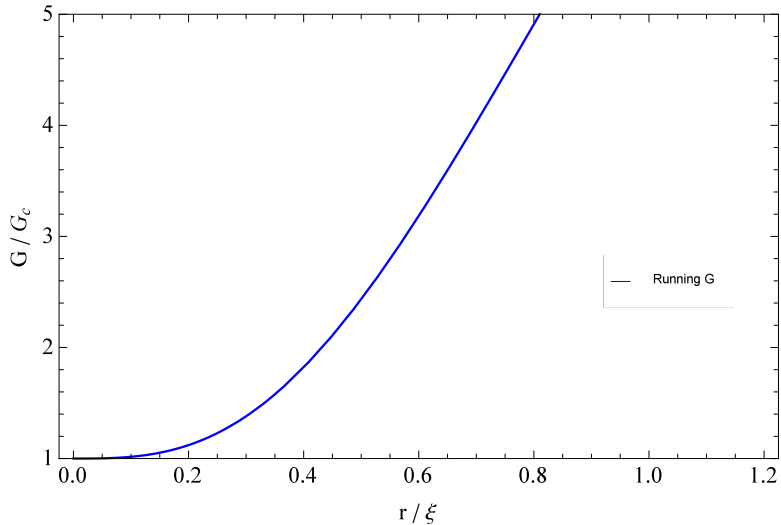}
\end{center}
\caption{
Running gravitational coupling $G(r)$ versus $r$, obtained
from $G(q)$ in Eq.~(\ref{eq:grun_k}) by setting $q \sim 1/r $
with an exponent $\nu=1/3$.
In view of Eq.~(\ref{eq:lambda_xi}), lattice quantum gravity calculations 
imply a slow rise of $G$ with distance scale, with roughly a 5\% effect on
scales of $ \approx 790 \, Mpc $, and a 10 \% effect on scales of $ \approx 990 \, Mpc $.
In this plot $G_c$, the short distance fixed point value for Newton's constant, corresponds
quite closely to the known laboratory value.
}
\end{figure}

The above results also suggest that the curvature on very small
scales behaves rather differently from the curvature on very large
scales, due to the quantum fluctuations eventually averaging out.
Indeed when comparing the result of Eq.~(\ref{eq:r_xi}) to the one in
Eq.~(\ref{eq:xi_r}) one is lead to conclude that the following change
has to take place when going from small (linear size $\sim l_p$) to large 
(linear size $ \gg l_p $) parallel transport loops 
\beq
\langle \, R \, \rangle_{\rm small \; scales} \;\; \sim \; {1 \over l_p \, \xi } 
\;\;\;\;\; \rightarrow \;\;\;\;\;
\langle \, R \, \rangle_{\rm large \; scales} \;\; \sim \; {1 \over \xi^2 } \; .
\label{eq:r_small_large}
\eeq
An intuitive way of understanding the above result is
that on small scales the strong local fluctuations in the 
metric/geometry lead to large values for the average 
rotation of a parallel-transported vector.
But then on larger scales these short distance fluctuations
tend to average out, and the {\it combined} overall rotation is 
much smaller, by a factor of $ {\cal O} ( l_p / \xi)$,
\beq
Z_R \; = \; { l_P \over \xi } \;\; .
\label{eq:z_r}
\eeq
The above quantity should then be regarded as an essential and necessary
``renormalization constant'' when comparing curvature on different
length scales, and specifically when going from very small 
(size $\sim l_P $) to large (size $ \gg l_P $) parallel transport loops.
See also the earlier discussion preceding Eq.~(\ref{eq:sme_loop_asy}), 
about the issue of comparing correlations of large loops 
versus correlations of small (infinitesimal) loops.

To conclude this section, one can raise the legitimate concern of how 
these results are changed by quantum fluctuations of various matter fields;
so far all the results presented here apply to pure gravity without any
matter fields.
Therefore here, and in the rest of the paper, what has been applied is basically the 
{\it quenched approximation}, wherein gravitational loop effects (perturbative and
nonperturbative) are fully accounted for, but matter loop corrections
are initially neglected. 
When adding matter fields coupled to gravity (scalars, fermions,
vector bosons, spin-3/2 fields etc.) one would expect, for example, the value for $\nu$
to change due to vacuum polarization loops containing these fields.
A number of arguments can be given though for why these effects
should not be too dramatic, unless the number of light matter fields
is rather large \cite{ham15}.
\footnote{
One would expect that significant changes to the result of Eqs.~(\ref{eq:grun_k}) and 
(\ref{eq:grun_box}) will arise from matter fields which are light
enough to compete with gravity, and whose Compton
wavelength is therefore comparable to the scale of the gravitational vacuum
condensate, or observed cosmological 
constant $\lambda$, namely $m^{-1} \sim 1/ \sqrt{\lambda/3}$.
At present the number of candidate fields that could fall into this
category is rather limited, with the photon and a near-massless 
gravitino belonging to this category.
The results presented here correspond to the quenched approximation
for quantum gravity, where all graviton loop effects are included,
but matter (and radiation) loops are neglected.
Matter fields are still present in the theory but are treated as quantum mechanical 
static sources.
In the $2+\epsilon$ perturbative expansion for quantum gravity
one encounters factors of $25-c$ in the renormalization groups
$\beta$ function, where $c$ is the central charge 
associated with the (massless) matter fields \cite{eps,aid97}, which would suggest
that matter loop and radiation corrections are indeed rather small.
In four dimensions similar factors involve $48-c$ \cite{nie81,hug80},
which would again lend support to the argument that such effects should be 
rather small in four dimensions.}

Note that the above results for the gravitational condensate
in many ways parallel what is found in
non-Abelian gauge theories, where for example one has
for the color condensate $< F_{\mu\nu}^2 > \, \simeq 1/ \xi^4 $
\cite{cam89,xji95,bro09,dom14}.
In QCD this last result is obtained from purely
dimensional arguments, once the existence of a fundamental
correlation length $\xi$ (which for QCD is given by the inverse 
mass of the lowest spin zero glueball) is established.
Accordingly, for gravity one would in fact expect simply on the basis 
of purely dimensional argument that the large scale curvature (corresponding to
the graviton condensate)
should be related to the fundamental correlation length by
$ \langle \, R \, \rangle \, \simeq \, 1 / \xi^2 $, as in Eq.~(\ref{eq:r_xi}).
This then points to a fundamental relationship between 
the nonperturbative scale $\xi$ (or inverse renormalized mass) and 
a nonvanishing vacuum condensate for both of these theories, nonperturbative
quantum gravity and QCD,
\beq
\langle \, R \, \rangle \;  \simeq \;  { 1 \over \xi^2 }
\;\;\;\;\;\;\;\;\;\;\;\;
\langle \, F_{\mu\nu}^2 \, \rangle  \;  \simeq \;  { 1 \over \xi^4 } \; .
\label{eq:vev}
\eeq
In gauge theories an additional physically relevant example is provided by the fermion condensate,
\beq
\langle \, \bar \psi \psi \, \rangle \, \simeq \, { 1 \over \xi^3 } \; ,
\label{eq:vev_fer}
\eeq
arising as a non-trivial consequence of the renormalization group,
confinement and chiral symmetry breaking in SU(3) gauge theories \cite{ham81}; 
for a recent review on current values see for ex. \cite{mcn13}.
Note that in all three cases the power of $\xi$ is fixed by the canonical
dimension of the corresponding field, one over length square in the
case of the curvature, an observation which can be seen to provide further
support to the identification in Eqs.~(\ref{eq:xi_r}) and (\ref{eq:xi_lambda}),
which arise from considering the gravitational Wilson loop.
The actual physical values for the QCD condensates are well known;
current lattice and phenomenological estimates cluster around
$ \langle \, { \alpha_S \over \pi } \, F_{\mu\nu}^2 \, \rangle  \;
\simeq \;  (440 \, MeV)^4 $ and 
$ \langle \, \bar \psi \psi \, \rangle \, \simeq \, (290 \, MeV)^3 $ \cite{bro09,dom14}.
On the other hand, modifications to the static potential in gauge theories
are best expressed in terms of the running coupling constant 
$\alpha_S (\mu) $, whose scale dependence is determined by the 
celebrated beta function of $QCD$. 
There the relevant scale is the nonperturbative $ \Lambda_{\overline{MS}} \approx 210 \, MeV$ 
whose size is comparable to $\xi$, $ \Lambda_{\overline{MS}} \simeq \xi^{-1}$.
More specifically, in gauge theories the inverse of the 
correlation length $\xi$  corresponds to the lowest mass excitation,
the scalar glueball with mass $m_0 = 1 / \xi$.
If the lightest scalar $0^{++}$ glueball has a mass of approximately
$m = 1750 MeV$ (which then fixes $\xi=1/m$ at about $0.1 Fm$ ), then $\Lambda_{\overline{MS}}$ 
in QCD is about eight times smaller.
But of course one important difference between nonperturbative
gravity and QCD is the fact that in the former the  cutoff still appears explicitly,
hidden in the physical value of Newton's constant $G$ (which is dimensionful).
So there exists then a second dynamically generated scale $\xi$, whose magnitude is
not directly related to the value of $G$; instead it reflects
how close the bare $G$ is to the  fixed point value $G_c$.
Finally it should be added that the idea of a graviton condensate as described here,
and thus arising from non-perturbative gravity effects, is not entirely
new.
Interesting papers looking into such effects - but based on entirely different assumptions and
therefore from a rather different perspective -
can be found for example in \cite{mar13} and \cite{kuh14}.



\begin{figure}
\begin{center}
\includegraphics[width=0.7\textwidth]{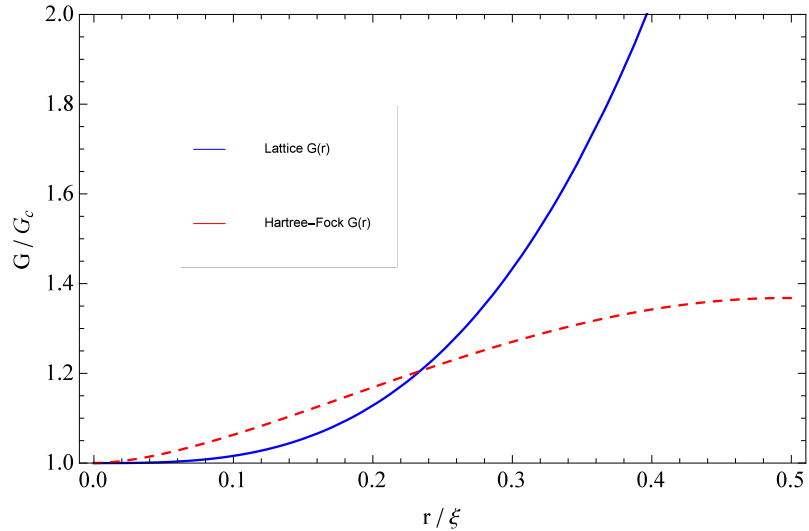}
\end{center}
\caption{
Running gravitational coupling $G(r)$ versus $r$, obtained
from $G(q)$ in Eq.~(\ref{eq:grun_k}) by setting $q \sim 1/r $
with an exponent $\nu=1/3$.
Note that the approximate Hartree-Fock analytical result of Eq.~(\ref{eq:grun_hf}) (red line) 
initially rises more rapidly for small $r$.
The nonperturbative scale $\xi$ is related to the gravitational vacuum condensate,
as in Eq.~(\ref{eq:lambda_xi}).
}
\end{figure}



\vskip 40pt

\section{Effective Field Equations}

\vskip 20pt

The result of Eq.~(\ref{eq:grun_k}) expresses the renormalization group
running of Newton's $G$ as a function of momentum scale.
As it stands, the expression in Eq.~(\ref{eq:grun_k}) does 
not satisfy general covariance, and needs to be promoted to a more useful and acceptable form.
It follows that in order to apply consistently the above result to an arbitrary background geometry, a 
fully covariant formulation is required.
One way of describing the running of Newton's $G$ is by a set of effective nonlocal field equations 
with a $G(\Box)$ \cite{eff,lambda}.
A second option is to formulate a fully covariant effective gravitational action with a running
$G(\Box)$, also discussed in detail in \cite{eff,lambda}.
In the following both options will be discussed.
This general approach, originally pioneered by Vilkovisky \cite{vil84}, has had great success in
incorporating in a gauge independent way the renormalization group running of the 
coupling in gauge theories such as QED and QCD.
\footnote{
A precursor to the effective action approach involving the running of $G$ can be found 
to some extent in the Brans-Dicke scalar-tensor modification of Einstein's gravity \cite{bra61}, 
where 
Newton's constant is promoted to a field-dependent quantity, as a result of incorporating
in a partial way the requirements of Mach's principle regarding the origin of inertia.
Nevertheless the two approaches remain quite different, in the sense that the discussion here
is strictly limited to what is expected on the basis of a quantum field theoretic treatment
of Einstein's gravity with a cosmological constant, and suitable additional known matter fields.
A Brans-Dicke type of extension of the current quantum theory is possible, 
say by the addition of
a dynamical massless scalar field with suitable covariant couplings, but will not be considered here.}

An effective field theory approach can be derived by writing
down an effective action, involving either a $G(\mu)$ or a $G (\Box)$ \cite{eff,lambda}.
Based on the previous discussion, a suitable effective action, describing the residual effects of 
quantum gravity on very large distance scales, is of the form
\beq
I_{\rm eff} \, [ g_{\mu\nu} ] \; = \; - \, { 1 \over 16 \pi \, G (\mu) }
\int d^4 x \, \sqrt g \; \Bigl ( R \, - \, { 6 \over \xi^2 } \Bigr )
\, + \, I_{\rm matter} \, [ g_{\mu\nu} , \dots ] \;\; ,
\label{eq:grun_ac}
\eeq
with $G(\mu)$ a very slowly varying (on macroscopic scales) Newton's
constant, in accordance with Eqs.~(\ref{eq:grun_k}) or
(\ref{eq:grun_box}), and amplitude $c_0 \approx 2 \times 8.02 (55) \approx 16.0 $ \cite{ham15}.
Note that the effective action of Eq.~(\ref{eq:grun_ac}) is obtained from the
one in Eq.~(\ref{eq:ac_cont}) by a suitable field rescaling, in accordance with the discussion
preceding Eq.~(\ref{eq:metric_scale}).
Here again the nonperturbative scale $\xi$ appears therefore both in
the running of $G$ and in the cosmological constant term with $\third \lambda = 1/\xi^2$. 
Nevertheless it was found that if a covariant $G(\Box)$ is used in the above effective 
action the resulting effective field equations are rather complicated
and hard to solve in practice, due to the fractional exponents appearing in $G(\Box)$
\cite{eff,lambda}.

For reasons that will become clearer later on, in most of the upcoming discussion a 
different route will be followed, based on the much simpler (and thus more manageable) 
effective field equations based nevertheless again on $G(\Box)$.
In either case it seems natural to perform the standard quantum mechanical replacement
$ q^2 \rightarrow - \Box $, where $\Box ( g_{\mu\nu} )$ is the
covariant D'Alembertian for a given background metric $g_{\mu\nu} (x) $ \cite{eff},
\beq
\Box (g) \; = \; g^{\mu\nu} \, \nabla_\mu \nabla_\nu \; .
\label{eq:box}
\eeq
This then leads to a consistent covariantly formulated running of $G$, with
\beq
G(\Box) \; = \; G_c \left [ \; 1 \, + \, c_0 \, 
\left ( { 1 \over - \xi^2 \, \Box  } \right )^{1 \over 2 \nu} \,  + \, \dots \; \right ] \;\; 
\label{eq:grun_box} 
\eeq
Note that the precise form of the covariant $\Box$, and thus of $G(\Box)$, depends on the 
tensor nature of the object it acts on \cite{eff}.
Numerical studies of lattice quantum gravity give for the exponent $\nu=1/3$ and for the quantum
amplitude $c_0 \approx 16.0 $ \cite{ham15}, which (fortunately, or unfortunately)
leaves very little ambiguity regarding the result of Eq.~(\ref{eq:grun_box}).
As noted earlier, one way of viewing physically the result of Eq.~(\ref{eq:grun_box})
is that quantum gravitational fluctuations generate anti-screening, with an initially
very slow running of $G$, as shown earlier in Figure 4.
The anti-screening arises because of the radiative dressing of the source by a virtual gravitation graviton cloud, in analogy to the screening of a bare charge in QED by the virtual electron-positron cloud. 
In a sense, therefore, the above corrections describe the gravitational analog of the running coupling constant in QED.

Generally, fractional powers of inverse d'Alembertians require careful handling.
This can be done either by computing the effect of integer powers $ \Box^n $ and then
analytically continue the result to fractional negative values $n \rightarrow - 1 / 2 \nu $,
or by using a regulated parametric integral representation
\beq
\left ( { 1 \over - \, \Box \, (g) + \mu^2 } \right )^{1/ 2 \nu} 
\, = \, 
{ 1 \over \Gamma ( { 1 \over 2 \nu } ) } \,
\int_0^\infty d \alpha \; \alpha^{  1 / 2 \nu - 1 } \;
e^{  - \alpha \, [ - \Box (g) + \mu^2 ] } \; ,
\label{eq:gbox_exp}
\eeq
where $\mu \rightarrow 0 $ is a suitable infrared regulator, here again with exponent $\nu=1/3$.
Note that for $\nu=1/3$ the quantum correction in Eq.~(\ref{eq:grun_box})
proportional to $c_0$ always includes a $1/\xi^3 $, which
therefore naturally sets the overall scale for the leading quantum correction, irrespective of
the background geometry considered.
Then a suitable set of manifestly covariant effective field equations with a running
$G(\Box)$ takes the form \cite{eff}
\beq
R_{\mu\nu} \, - \, \half \, g_{\mu\nu} \, R \, + \, \lambda \, g_{\mu\nu}
\; = \; 8 \pi \, G  ( \Box )  \; T_{\mu\nu}  \;\; 
\label{eq:grun_field}
\eeq
with the additional nonlocal contribution coming from the quantum correction
in the $G(\Box)$ of Eq.~(\ref{eq:grun_box}).
It is important to note that the nonperturbative scale $\xi$ enters 
the effective field equations in {\it two} places, first in the cosmological
constant term $\third \lambda = 1/\xi^2$ of Eq.~(\ref{eq:lambda_xi}) dues to the
non-vanishing vacuum condensate $<R> \, \neq 0$ discussed earlier,
and also in the running of $G$ of Eq.~(\ref{eq:grun_box}) where it sets the reference scale.
A clear implication here is that those two scales, which in principle
could be entirely unrelated, appear to be one and the same in the present 
renormalization group context.
The nonlocal, but manifestly covariant, effective field equations of Eq.~(\ref{eq:grun_field}) 
can then, at least in principle, be solved for a number of physically relevant metrics.
For the specific case of a static isotropic metric it is possible to obtain 
an exact expression for $G(r)$ in the limit $r \gg 2 M G$ \cite{eff}, a result given
previously in Eq.~(\ref{eq:grun_r}).
Not unexpectedly, generally all three expressions in Eqs.~(\ref{eq:grun_k}), 
(\ref{eq:grun_box}) and (\ref{eq:grun_r})
are consistent with a gradual slow increase in $G$ with distance
$r$, and thus with a modified Newtonian potential in the same limit.
\footnote{
It is useful to observe here, quite generally and independent of the lattice results, that one
finds it difficult to implement a weakly running cosmological constant, if general 
covariance is to be maintained at the level of the effective field
equations. If the running of $\lambda$ is implemented via a $\lambda (\Box)$, then
because of $ \nabla_\lambda \, g_{\mu\nu} =0$ one also has
$ \Box^n \, g_{\mu\nu} =0$, which makes it nearly impossible to maintain
general covariance and have a nontrivial running $\lambda (\Box)$, 
as pointed out in \cite{lambda}.}

Naturally the next step is a systematic examination of the nature of solutions to the
full effective field equations of Eq.~(\ref{eq:grun_field}), with $G ( \Box )$ involving
the covariant d'Alembertian of Eq.~(\ref{eq:box}), 
acting there on the second rank tensor $T_{\mu\nu}$.
A scale-dependent Newton's constant is then expected to lead to small modifications
of the standard cosmological solutions to the Einstein field equations,
proportional to the amplitude $c_0$.
Already the action of $\Box (g) \, = \, g^{\mu\nu} \, \nabla_\mu \nabla_\nu $ is rather complicated 
on second rank tensors; one has
\bea
\nabla_{\nu} T_{\alpha\beta} \, = \, \partial_\nu T_{\alpha\beta} 
- \Gamma_{\alpha\nu}^{\lambda} T_{\lambda\beta} 
- \Gamma_{\beta\nu}^{\lambda} T_{\alpha\lambda} \, \equiv \, I_{\nu\alpha\beta}
\nonumber
\eea
and
\beq 
\nabla_{\mu} \left ( \nabla_{\nu} T_{\alpha\beta} \right )
= \, \partial_\mu I_{\nu\alpha\beta} 
- \Gamma_{\nu\mu}^{\lambda} I_{\lambda\alpha\beta} 
- \Gamma_{\alpha\mu}^{\lambda} I_{\nu\lambda\beta} 
- \Gamma_{\beta\mu}^{\lambda} I_{\nu\alpha\lambda}  \; .
\label{eq:box_on_tensors}
\eeq

Of course, one of the simplest applications is to the Friedmann-Lema\^itre-Robertson-Walker 
(FLRW) framework applied to the standard homogeneous isotropic metric
\beq
d \tau^2 \; = \;  dt^2 - a^2(t) \left \{ { dr^2 \over 1 - k\,r^2 } 
+ r^2 \, \left ( d\theta^2 + \sin^2 \theta \, d\varphi^2 \right )  \right \} \;\;\;\; k =0, \pm1 \; ;
\label{eq:frw}
\eeq
in the following only the case $k=0$ (spatially flat universe) will be discussed.
In this framework a popular choice for $T_{\mu\nu}$ is the perfect fluid form, 
\beq
T_{\mu \nu} = \left [ \, p(t) + \rho(t) \, \right ] u_\mu \, u_\nu + g_{\mu \nu} \, p(t)
\label{eq:tmunu_perf}
\eeq
for which one needs to compute the action of $\Box^n$ on $T_{\mu\nu}$,  and 
then analytically continues the answer to negative fractional values of $n = -1/2 \nu $.
The results of \cite{eff,static,lop07}  then show, among other things,  that a 
nonvanishing pressure contribution is generated in 
the effective field equations, even if one initially assumes a pressureless fluid, $p(t)=0$.
Specifically, for a universe filled with nonrelativistic 
matter ($p$=0) one obtains the following set of effective Friedmann equations,
\beq
{ k \over a^2 (t) } \, + \,
{ \dot{a}^2 (t) \over a^2 (t) }  
\; = \;  { 8 \pi \, G(t) \over 3 } \, \rho (t) \, + \, { \lambda \over 3 }
\label{eq:fried_tt}
\eeq
for the $tt$ field equation, and
\beq
{ k \over a^2 (t) } \, + \, { \dot{a}^2 (t) \over a^2 (t) }
\, + \, { 2 \, \ddot{a}(t) \over a(t) } 
\; = \;  - \, { 8 \pi \, \delta G (t) \over 3 } \, \rho (t) 
\, + \, \lambda
\label{eq:fried_rr}
\eeq
for the $rr$ field equation.
In the above expressions, the running of $G$ appropriate for 
the Robertson-Walker metric is
\beq
G (t) \, \equiv \, G_0 \left ( 1 + { \delta G(t) \over G_0 } \, \right ) 
\, = \, G_0 \left [ 1 + c_t \, 
\left ( { t \over \xi } \right )^{1 / \nu} \, + \, \dots \right ] 
\label{eq:grun_t}
\eeq
with $c_t \simeq 0.450 \; c_0 $ [for an amplitude $c_0$ appearing in Eq.~(\ref{eq:grun_box})] 
for the tensor box operator \cite{eff}.
From the above form of $\delta G(t)$ one sees that again the amplitude of the quantum correction
is proportional  to the combination $c_0 / \xi^3$ for $\nu=1/3$.
Furthermore, the running of $G$ induces an effective pressure term in the second 
($rr$) equation, due to the presence of an induced relativistic fluid, whose origin lies in
the quantum gravitational vacuum-polarization contribution.
Another noteworthy feature of the new effective field equations is the additional power-law
acceleration contribution, on top of the standard exponential one due to the $\lambda$ term.

On way of viewing the results is that the effective field equations with a running $ G $, here Eqs.~(\ref{eq:fried_tt}) and (\ref{eq:fried_rr}), can be recast in an equivalent form by 
defining a vacuum-polarization pressure $p_{vac}$ and density $\rho_{vac}$, 
such that in the FLRW background one has
\beq
\rho_{vac} (t) = {\delta G(t) \over G_0} \, \rho (t)  \;\;\;\;\;\;\;\;\;\;\;\; 
p_{vac} (t) = { 1 \over 3} \, {\delta G(t) \over G_0} \, \rho (t) \; .
\label{eq:rhovac_t}
\eeq
From this viewpoint, the inclusion of a vacuum-polarization contribution in the FLRW 
framework amounts to a replacement 
$ \rho(t) \rightarrow \rho(t) + \rho_{vac} (t)  $,  $ p(t) \rightarrow p(t) +  p_{vac} (t) $
in the original field equations.
Then, just as one introduces a parameter $w$, describing the matter equation of state, 
\beq
p (t) = w \,  \rho(t)
\label{eq:w_def}
\eeq
with $ w=0 $ for nonrelativistic matter, one can do the same here for the remaining  quantum contribution by setting
\beq
p_{vac} (t) = w_{vac} \; \rho_{vac} (t) \; .
\label{eq:wvac_def}
\eeq
The original calculations \cite{eff}, and more recently
\cite{ht10} (which included metric perturbations) give $ w_{vac}= \third $.
Note that it was shown in \cite{eff} that this result is obtained {\it generally} for the 
given class of $G(\Box)$ considered,  and is  not tied to a specific choice for the 
universal exponent $\nu$, such as $\nu=\third$.

More generally, the procedure of defining a $\rho_{vac}$ and a $p_{vac}$ contribution,
arising from quantum gravitational vacuum-polarization effects, is not necessarily
restricted to the FLRW background metric case.
One can always decompose the full source term in the effective nonlocal
field equations of Eqs.~(\ref{eq:grun_box}) and Eq.~(\ref{eq:grun_field}), making use of
\beq
G(\Box) = G_0 \, \left ( 1 \, +  {\delta G(\Box) \over G_0} \right ) 
\;\;\;\;\;\;  {\rm with} \;\;\;\;\;
{\delta G(\Box) \over G_0} \equiv c_0 \left ( { 1 \over \xi^2 \Box } \right )^{1 / 2 \nu}  \; ,
\label{eq:grun_box_1}
\eeq
 as two contributions,
\beq
{ 1 \over G_0 } \, G(\Box) \; T_{\mu\nu}  \, = \, 
\left ( 1 + {\delta G(\Box) \over G_0}  \right ) \, T_{\mu\nu}  \, \equiv \,
T_{\mu\nu}   +  T_{\mu\nu}^{vac} \; .
\label{eq:tmunu_vac}
\eeq
The latter then involves the nonlocal part
\beq
T_{\mu\nu}^{vac} \, \equiv \,  {\delta G(\Box) \over G_0} \, T_{\mu\nu} \; .
\label{eq:tmunu_vac1}
\eeq
Consistency of the full nonlocal field equations now requires that the {\it sum}
be covariantly conserved,
\beq
\nabla^\mu \left ( T_{\mu\nu}   +  T_{\mu\nu}^{vac}  \right ) = 0 \; .
\eeq
In general one cannot expect that the contribution $ T_{\mu\nu}^{vac} $
will always be expressible in the perfect fluid form of Eq.~(\ref{eq:tmunu_perf}), even if the
original $ T_{\mu\nu} $ for matter (or radiation) has such a form.
The former will in general contain, for example, nonvanishing shear stress contributions, 
even if they were originally absent in the matter part \cite{ht10}.
Indeed in a number of cases of physical interest one deals quite generally 
with a background metric that is slightly perturbed,
$ g_{\mu \nu} = \bar{g}_{\mu \nu} + h_{\mu \nu} $.
Consequently the covariant d'Alembertian operator 
$ \Box \; = \; g^{\mu\nu} \, \nabla_\mu \nabla_\nu  $
acting on second rank tensors [such as the $T_{\mu\nu}$ in Eq.~(\ref{eq:grun_field})]
needs to be Taylor expanded in the small perturbation $h$,
\beq
\Box (g) \, = \, \Box^{(0)}  + \Box^{(1)} (h) + O (h^2) \; .
\eeq
For $G(\Box)$ itself one obtains the expansion
\beq
G(\Box) \, = \, G_0 \,  \left [ 
1 + \, { c_0 \over \xi^{1 / \nu} } \,  
\left ( \Box^{(0)} + \Box^{(1)} (h) + O(h^2) \right )^{- 1/ 2 \nu}  + \dots
\right ]  \; ,
\label{eq:gbox_h}
\eeq
to which one can apply the binomial formula $ (A+B)^{-1} = A^{-1} - A^{-1} B A^{-1}  + \dots $.
This then allows one to work out in some detail a number of predictions that arise
from the original manifestly covariant effective field equations, Eq.~(\ref{eq:grun_field}).
It is also customary to later expand the relevant fields (metric perturbations, matter perturbations etc.)
in Fourier modes, with the small $k$ modes as the leading contribution, and higher modes
treated later again as perturbations.
This nevertheless significantly complicates further the analysis of the results \cite{ht10,ht11}, 
given the intrinsic dependence on scale $k$ of the quantum correction in $G$, 
see for example Eq.~(\ref{eq:grun_k}) for $G(q)$ or its equivalent covariant form of 
$G(\Box)$ in Eq.~(\ref{eq:grun_box}).



\vskip 40pt


\section{Large Scale Curvature and Matter Density Correlations}

\label{sec:matter}

\vskip 20pt

Quantum gravity, and the existence of a nontrivial quantum condensate for the curvature, lead to a number of specific physical predictions, which are in principle observationally testable.
Many of these quantum correction effects can in principle be calculated, given the effective long-distance quantum corrected gravity theory formulated in Eqs.~(\ref{eq:grun_field}) and (\ref{eq:grun_box}).
The most salient effects include a running of Newton's constant $G$ with scale on very large (cosmological) scales;
the modification of classical results for (relativistic) matter density perturbations and the associated growth exponents; 
a non-vanishing so-called slip function in the conformal Newtonian gauge;
quantum effects that lead to nontrivial curvature, and therefore matter density, correlations at large distances,  with the latter parameterized by a set of exponents characterizing the decay of correlation functions and their amplitudes, all of which are in principle calculable.
As in the case of QCD and Yang-Mills theories, one expects essentially no adjustable parameters. Summarizing what has been stated before, one has that the running of $G$ 
in Eq.~(\ref{eq:grun_field}) is completely
determined by the universal exponent $\nu$, the nonperturbative quantum 
amplitude $c_0$ and the correlation length $\xi$, which in turn is related to either the 
vacuum expectation value of the curvature, or to what is equivalent to it, the observed
non-vanishing large-scale cosmological constant $ \lambda_{obs} $.

Much of what has been discussed so far was relates to the fact 
that in a quantum theory of gravity the gravitational constant $G$ runs
with scale, in accordance with Eq.~(\ref{eq:grun_box}).
But there are additional consequences, which arise from the fact
that in general gravitational correlations do not follow
free field (Gaussian) predictions and which will therefore be the subject of this
section.
One would expect such correlations to have some observational relevance, and
one such example is the curvature correlation function of Eqs.~(\ref{eq:corr_cont})
and (\ref{eq:corr_pow}), with power $2n=2 (d-1/\nu)=2$.
But it will be useful to first examine some local averages.
For the average local curvature ${\cal R}(k)$ one has from
Eq.(\ref{eq:r_xi}), using Eq.(\ref{eq:xi_g}) and $\nu =1/3$,
\beq
{ < \int d^4 x \, \sqrt{ g } \, R(x) >
\over < \int d^4 x \, \sqrt{ g } > } \; \sim \; \xi^{1/\nu-d} \; \sim
\; { A'_{\cal R} \over a \; \xi } \;\; ,
\label{eq:r_xi_amp} 
\eeq
as given earlier in Eq.~(\ref{eq:r_xi}).
Lattice calculations allow one to extract various amplitude coefficients, such as the one in the
above expression.
The dimensionless amplitude $ A'_{\cal R} $ in Eq.~(\ref{eq:r_xi_amp}) 
is expected to be $O(1)$ in lattice units, and numerically 
one finds \cite{ham15} $ A'_{\cal R} \, = \, 3.40(13)$.
On the other hand for the curvature fluctuation $ \chi_{\cal R} (k) $ one has 
from Eqs.~(\ref{eq:chi_cont}) and (\ref{eq:chi_xi})
\beq
{ < ( \int d^4 x \, \sqrt{g} \, R )^2 > - < \int d^4 x \, \sqrt{g} \, R >^2
\over < \int d^4 x \, \sqrt{g} > } \; \sim \; 
\xi^{ 2 / \nu - d }  \; \sim \;  A'_{\chi}  \, \xi^2 / a^2 \;\; ,
\label{eq:chi_xi2}
\eeq
also given earlier in Eq.~(\ref{eq:chi_xi}).
Note that in both Eqs.~(\ref{eq:r_xi_amp}) and (\ref{eq:chi_xi2}) the correct dimensions have been restored, by inserting suitable powers of the lattice spacing $a$ (curvature has dimensions of
one over length squared); 
also a specific value for this lattice spacing was given earlier in Eq.~(\ref{eq:a_phys}).
For the dimensionless amplitude in Eq.~(\ref{eq:chi_xi2}) one finds numerically 
$ A'_{\chi} \, = \, 2.22(9) $ \cite{ham15}, again in general agreement with the prejudice 
that nonvanishing dimensionless critical amplitudes should be $O(1)$.

These results in turn provide some useful information related to the 
local curvature correlation function at a fixed geodesic distance 
[see Eqs.~(\ref{eq:corr_cont}) and (\ref{eq:corr_latt})].
By scaling one obtains immediately (from $\nu=1/3$) for the power appearing in 
Eq.~(\ref{eq:corr_pow})
\footnote{
One can contrast this result with what one finds in weak field perturbation theory.
There one finds \cite{modacorr}
$ < \! \sqrt{g} R (x) \sqrt{g} R (y) \! >_c
\; \sim \; < \! \partial^2 h (x) \partial^2 h (y) \! > 
\; \sim \; 1/ \vert x-y \vert^6 $ and thus $2n=6$, so the result here is quite different.
If one defines in the usual way an anomalous dimension $\eta$ for the 
graviton propagator in momentum space, $< \! h \, h \! > \, \sim 1/k^{2-\eta}$, 
one finds from the lattice calculation
$\eta = d-2-2/\nu$ or $\eta =-4$ in four dimensions for $\nu=1/3$,
which deviates rather significantly from the Gaussian or perturbative value.
Such a large deviation is already observed in the $2+\epsilon$ expansion
[see Eq.~(\ref{eq:nu_eps})] and is not peculiar to lattice quantum gravity;
in the context of gravity such an interesting possibility was already discussed 
some time ago in \cite{fubini}.}
\beq
2 n \, = \, 2 \, (d - 1 / \nu) = 2 \;\; .
\label{eq:n_pow}
\eeq
For the local curvature-curvature correlation 
function of Eq.~(\ref{eq:corr_cont}) at ``short distances'' $r \ll \xi$ (and again for $\nu=1/3$)
one then obtains the rather simple result
\beq
< \sqrt{g} \; R(x) \; \sqrt{g} \; R(y) \; \delta ( | x - y | -d ) >_c
\;\; \mathrel{\mathop\sim_{d \; \ll \; \xi }} \;\;
{ 1 \over d^{\, 2 ( 4 - 1/\nu ) } } \; \sim \; { A_0 \over a^2 \, d^2 }
\;\; .
\label{eq:corr_pow1}
\eeq
As before, in the last term the correct dimensions have been restored
by inserting suitable powers of the lattice spacing $a$.
The dimensionless amplitude $A_0$ of Eq.~(\ref{eq:corr_pow1})
 is related to the amplitude in Eq.~(\ref{eq:chi_xi2}) because of 
Eq.~(\ref{eq:chi_corr}), and one finds from the numerical solution \cite{ham15}
$ A_0 \, \equiv \,  A'_{\chi} / 2 \pi^2 \,  = \, [ 0.335(20)  ]^2 $,
so that the dimensionless curvature correlation function
normalization constant is $ N_R \, \equiv \, \sqrt{A_0} = 0.335(20) $.
Again as expected, this amplitude is close to $O(1)$ in units
of the  cutoff (fundamental lattice spacing) $a$.
Note that the two-point function result of Eq.~(\ref{eq:corr_pow2}),
and related to it the scaling dimension $n=1$ of Eq.~(\ref{eq:n_pow}),
also determines the form of the reduced three-point curvature correlation function
\beq
< \sqrt{g} \; R(x_1) \; \sqrt{g} \; R(x_2) \; \sqrt{g} \; R(x_3) >_{c \, R}
\;\; \mathrel{\mathop\sim_{d_{ij} \; \ll \; \xi }} \;\;
{ C_{123}  \over d_{12} \, d_{23} \, d_{31}  } \;\; .
\label{eq:corr_triple}
\eeq
with $C_{123}$ a constant, and relative geodesic distances 
$d_{ij} = \vert x_i - x_j \vert $ etc.
The relevance and measurement of nontrivial three- and four-point matter density 
correlation functions in cosmology was discussed in detail some time ago in \cite{pee93}.

It is instructive at this stage to compare the above result for the local curvature correlation 
given in Eq.~(\ref{eq:corr_pow1}) to the expression for the local average 
curvature of Eq.~(\ref{eq:r_xi_amp}): both expressions still contain explicitly
the size of the microscopic, {\it infinitesimal} parallel transport loop $ \sim a \sim l_P$,
which originates in the fact that both these quantities make explicit reference 
to infinitesimal (ultraviolet cutoff sized) parallel transport loops.
The explicit appearance of the ultraviolet cutoff in these averages can be explained
by the appearance of residual short distance divergences associated with such
small infinitesimal loops.
At the same time, a comparison of the result of Eqs.~(\ref{eq:r_xi_amp})
and  (\ref{eq:r_xi_amp}) for the local curvature
with the corresponding result of Eq.~(\ref{eq:xi_r}) for the large scale,
{\it macroscopic} curvature suggests a substantial changeover when going
from small (size $\sim a$) to large (size $ \gg a$) parallel transport loops.
As discussed earlier, the results can be summarized by the statement
that the curvature  on large (macroscopic) scales is much smaller (by a factor $1/\xi$)
than the curvature on small (Planck length) scales, due to a dramatic averaging
out of the fluctuations.

As described earlier in Eqs.~(\ref{eq:r_small_large}) and (\ref{eq:z_r}), one then 
expects that the transition from infinitesimal to macroscopic loops (linear size $\gg a$)
can be affected in the correlation function of 
Eq.~(\ref{eq:corr_pow1})  by the replacement of $ a^2 \rightarrow \xi^2 $.
This then would give for large (macroscopic size $\gg a $) parallel transport loops
a modified form of the correlation function of Eq.~(\ref{eq:corr_pow1})
\beq
< \sqrt{g} \; R(x) \; \sqrt{g} \; R(y) \; \delta ( | x - y | -d ) >_c
\;\; \mathrel{\mathop\sim_{d \; \ll \; \xi }} \;\; 
{A_1 \over \xi^2 \, d^2 } \;\;\; ,
\label{eq:corr_pow2}
\eeq
with the general expectation that the overall amplitudes nevertheless be comparable,
$ A_1 \approx A_0 $.
Note that the universal power $n=2$ is unchanged compared to Eq.~(\ref{eq:corr_pow1}),
only the amplitude has been modified, in accordance with the earlier 
gravitational Wilson loop result of Eqs.~(\ref{eq:spectral_w1}) and (\ref{eq:spectral_w2}).
Unfortunately so far these large loop correlations have not been computed explicitly, 
but nevertheless the above ideas should become testable by explicit numerical 
simulations in the near future.

The next step is to determine whether the knowledge of the curvature correlation, as
given in Eqs.~(\ref{eq:corr_pow1}) or (\ref{eq:corr_pow2}), can be translated into
information regarding other two-point correlations subject to astrophysical measurement.
First consider what can be stated purely at the classical level.
There one can use the field equations to directly relate the local curvature to the
local matter mass density. 
From Einstein's field equations with $\lambda=0$ 
\beq
R_{\mu\nu} - \half \, g_{\mu\nu} \, R \; = \; 8 \pi \, G \, T_{\mu\nu} \;\;
\label{eq:einstein}
\eeq
for a perfect fluid one then obtains for the Ricci scalar, in the limit of
negligible pressure,
\beq
R (x) \; \simeq \; 8 \pi \, G \; \rho (x) \;\; .
\label{eq:trace}
\eeq
This last result then relates the local fluctuations in the
scalar curvature $ \delta R(x)$ to local fluctuations in the matter 
density $\delta \rho (x)$, and could therefore
provide a potentially useful connection to the quantum result 
of Eqs.~(\ref{eq:corr_pow}),  (\ref{eq:corr_pow1}) and (\ref{eq:corr_pow2}).
Note that the same kind of reasoning would apply alternatively to a 
$T_{\mu\nu}$ describing radiation, which would then be relevant 
for a radiation-dominated early universe.
Of course, in the Newtonian limit the above result simplifies to
Poisson's equation
\beq
\Delta \, h_{00} ( {\bf x}, t ) \; = \; 8 \pi \, G \, \rho ( {\bf x}, t ) \;\; ,
\label{eq:poisson}
\eeq
where $h_{00}= 2 \phi $ and $\rho$ are the macroscopic gravitational field and
the macroscopic mass density, respectively.

Now, in the current cosmology literature \cite{pee93,wei08}
it is customary to describe matter density 
fluctuations in terms of the matter density contrast correlation function
\footnote{In cosmology the (dimensionless) galaxy
matter density two-point function is usually referred to as $\xi(r)$,
but here it seems desirable to avoid confusion with the gravitational 
correlation length $\xi$.}
\beq 
G_\rho (r) \; = \; < \! \delta \rho (r) \; \delta \rho (0) \! > \;\; .
\label{eq:galaxy_corr}
\eeq
The latter is related to its Fourier transform $P(q)$ by
\beq
G_\rho (r) \; = \; { 1 \over 2 \pi^2 } \int_\mu^\Lambda dq \, q^2 \, P(q) \, \;\; .
{\sin q r \over q r } \; ,
\eeq
It has to contain, in general, both an infrared regulator ($\mu$) and
an  ultraviolet cutoff ($\Lambda$), to make sure the integral stays convergent.
If one assumes, as is sometimes the case, that the power spectrum $P(q)$ is 
described by a simple power law
\beq
P(q) \; = \; { a_0 \over q^s } \;\; ,
\label{eq:power_spec}
\eeq
(where $n_s = -s $ is commonly referred to as the spectral index),
then one finds in the scaling regime $ 1/\mu \gg r \gg 1/ \Lambda $ for
the real-space density contrast correlation function
\beq
G_\rho (r) \; = \; c_s \, a_0 \cdot { 1 \over r^{3-s}  } \;\; ,
\label{eq:galaxy_corr1}
\eeq
provided the index $s$ satisfies $ 0 < s < 3 $, and 
here $c_s \equiv \Gamma (2-s) \, \sin ( \pi s / 2 ) \, / \, 2 \pi^2 $
(terms containing the ultraviolet cutoff $\Lambda$ appear as well, but they are proportional to 
$ \sin ( \Lambda r) $ and $ \cos ( \Lambda r) $, oscillate rapidly and average out to zero).
This then leads to the identification of exponents [see Eq.~(\ref{eq:n_pow}) and Eq.~(\ref{eq:corr_pow1})] $ d - 1  - s \; = \; 2 \, n \equiv  2 \, ( d - 1 / \nu ) $ or 
\beq
s \; = \; { \textstyle{2 \over \nu }\displaystyle } - d - 1 \; = \; 1  \;\;\;\; (d=4) \; .
\label{eq:s_nu}
\eeq
For the specific value $s=1$ one has $P(q)=a_0 / q $, and therefore in position space
\beq
G_\rho (r) \; = \; \; { a_0 \over 2 \pi^2 } \cdot { 1 \over r^2 }
\;\;\;\;\;\; (s=1) \;\; ,
\label{eq:grho_a0}
\eeq
which is in fact, as will be discussed later, also consistent with the result given earlier in Eq.~(\ref{eq:corr_pow1}) for the invariant, connected curvature correlation function.

Generally, all of the above expressions get modified for very small wave vector $ q \sim m $,
as should be clear by now from the discussion in the previous sections, where the 
appearance of a dynamically generated infrared cutoff (as in QCD, and more generally in 
non-Abelian gauge theories) stood out as an essential ingredient.
Such an infrared cutoff is either introduced explicitly in the wave vector integrations, 
or, alternatively, the power spectrum is infrared regulated at small $q$ by replacing 
$q^2 \rightarrow q^2 + \mu^2 $.
\footnote{
In perturbative QCD the replacement $q^2 \rightarrow q^2 + \mu^2 $ partially accounts
for the existence of (nonperturbative) infrared renormalon effects.
In practice including these effects turns out to be, not surpirisingly, 
phenomenologically quite successful, see for ex. \cite{renormalons,ric79}
and references therein.}
One then writes more appropriately, instead of Eq.~(\ref{eq:power_spec}),
\beq
P(q) \; = \; { a_0 \over ( q^2 + \mu^2 )^{s/2} } \;\; .
\label{eq:power_spec_reg}
\eeq
Again, by Fourier transforms one then obtains for the correlation in real space
\beq
G_\rho (r) \; = \; { a_0 \over 2^{ s+1 \over 2}  \, \pi^{3 \over 2} \, \Gamma( { s \over 2 } ) } \,
\left ( { \mu \over r } \right )^{ 3 - s \over 2 } \; K_{s - 3 \over 2} ( \mu \, r )  
\label{eq:grho_a0_reg}
\eeq
which reproduces Eq.~(\ref{eq:galaxy_corr1}) for short distances $ r \ll \mu^{-1} $.
For very large spatial separations $r \gg \mu^{-1} $ the asymptotic behavior
of $G_\rho (r)$ is now  given instead by
\beq
G_\rho (r) \; \mathrel{\mathop\sim_{ r \ll \xi  }}  \;
a_0 \, c'_s \; \mu^{1- { s \over 2 } } \, { 1 \over r^ {2 - { s \over 2 } } } \; e^{- \mu r } 
\label{eq:grho_a0_reg_asy}
\eeq
with amplitude $ c'_s = 1/ 2^{1+ { s \over 2 } } \Gamma( { s \over 2 } ) $.
In view of the previous discussion, it is natural to identify here the infrared  cutoff 
$\mu \equiv m = 1 / \xi$, and such a choice will be implicit from now on 
throughout the following discussion.
Note the correspondence of the result of Eq.~(\ref{eq:galaxy_corr1}) with the short distance
curvature correlation result of Eq.~(\ref{eq:corr_pow}),
as well as the same type of correspondence of Eq.~(\ref{eq:grho_a0_reg_asy}) with the 
large distance curvature correlation result of Eq.~(\ref{eq:corr_exp}).

In practice, observational data for such matter density correlations
is commonly presented in the following compact form \cite{pee93}
\beq
G_\rho (r) \; = \; \left ( { r_0 \over r } \right )^\gamma \; ,
\label{eq:grho_r0}
\eeq
with an empirically determined exponent $\gamma$,
and a scale $r_0$ fitted to astrophysical (usually galactic cluster) observations.
For $\gamma$ close to two, one has by comparing Eq.~(\ref{eq:grho_a0}) to Eq.~(\ref{eq:grho_r0})
$ a_0 \, = \, 2 \pi^2 \, r_0^2 $.
It seems therefore rather tempting at this stage to try to connect the observational
result of Eq.~(\ref{eq:grho_r0}) to the quantum
correlation function in Eq.~(\ref{eq:corr_pow1}).
One then expects for the matter density fluctuation correlation a power law decay as well, of the form
\footnote{
Note that in weak field perturbation theory one finds, by virtue of the equations of motion,
$ < \! \rho (x) \, \rho (y) \! >_c  \; \sim \; $
$ < \! \partial^2 h (x) \, \partial^2 h (y) \! >  \; \sim \; $
$1/ \vert x-y \vert^6 $, 
so again the result here is rather different.
}
\beq
< \delta \rho ({\bf x},t) \; \delta \rho ({\bf y},t') > \;\;\; 
\mathrel{\mathop\sim_{ \vert {\bf x}-{\bf y} \vert \; \ll \; \xi }} \;\;
{1 \over a^2 (t)} \cdot {1 \over a^2 (t')} \cdot
{1 \over \vert {\bf x}-{\bf y} \vert^2 } \;\; 
\label{eq:corr_pow3}
\eeq
where $a(t)$ here stands for the cosmological scale factor.
\footnote{
The scale-factor dependent prefactor involving $a(t)$ and $a(t')$ has such a 
simple form only in a matter-dominated universe; for $\lambda \neq 0$ the $a$-dependent
prefactor is a bit more complicated, nevertheless it still reduces to unity
for equal times $t=t'=t0$ \cite{wei08}.}
The last correlation function can be made dimensionless by 
suitably dividing it by the square of some average matter 
density $\rho_0 \approx 0.3089 \rho_c $ with $\rho_c = 3 H_0^2 / 8 \pi G $
and $H_0^{-1} \approx 4430 \, Mpc$ for $h =0.677$ (using again, for concreteness,
the Planck 2015 data \cite{planck15}).
By comparing powers and coefficients in Eqs.~(\ref{eq:corr_pow1}) 
and (\ref{eq:grho_r0}) one finds
\beq
\gamma \; = \; 2\, ( d - 1/ \nu ) \; =\; 2 \;\;\;\; (d=4) \; .
\eeq
For the reference scale $r_0$ in this last equation one derives the quantitative estimate
\beq
r_0 \; = \; { 1 \over 8 \pi \, G \, \rho_0 }  \cdot { \sqrt{A_0} \over  a } \; ,
\label{eq:r0}
\eeq
with $ \sqrt{A_0} \simeq 0.335 (20) $ the dimensionless amplitude for
the curvature correlation  function of Eq.~(\ref{eq:corr_pow1}), 
and $a$ the lattice spacing given in Eq.~(\ref{eq:a_phys}).

The preceding arguments still contains nevertheless a fundamental flaw,
related to the use, at this stage in unmodified form, of the curvature
correlation function result of Eq.~(\ref{eq:corr_pow1}).
As discussed previously, that form applies to
the correlation of {\it infinitesimal} (Planck length, or cutoff size) loops, 
which would not seem to be appropriate for the macroscopic 
(or semiclassical) parallel transport loops, such as
the ones that enter the field equations (\ref{eq:einstein}) 
and (\ref{eq:trace}), and which thus relate locally the macroscopic
$\delta R (x) $ to the $\delta \rho (x)$.
It would therefore seem desirable to be able to correct for the
fact that the parallel transport loops sampled in
Eq.~(\ref{eq:trace}) are much larger than the infinitesimal
ones sampled in the correlation function in
Eq.~(\ref{eq:corr_pow1}).
As in Eqs.~(\ref{eq:r_small_large}) and
(\ref{eq:z_r}), the transition to macroscopic loops (linear size $\gg a$)
can be affected in Eq.~(\ref{eq:corr_pow1}) 
by the replacement of $ a^2 \rightarrow \xi^2 $.
This then gives for large (macroscopic size $\gg a $) parallel transport loops 
[see Eq.~(\ref{eq:corr_pow2})]
\beq
< \sqrt{g} \; R(x) \; \sqrt{g} \; R(y) \; \delta ( | x - y | -d ) >_c
\;\; \mathrel{\mathop\sim_{d \; \ll \; \xi }} \;\; 
{A_1 \over \xi^2 \, d^2 } \;\;\; ,
\label{eq:corr_pow2a}
\eeq
with the expectation of a comparable amplitude $ A_1 \approx A_0 $.
This last result then leads to the following improved estimate for 
the macroscopic matter density correlation of Eq.~(\ref{eq:galaxy_corr}),
\beq
G_\rho (r) \; = \; \left ( { 1 \over 8 \pi \, G } \right )^2 \; 
{ 1 \over  \rho_0^2 } \cdot
{ A_1 \over \xi^2 \; r^2 } \; ,
\eeq
so that comparing to Eq.~(\ref{eq:grho_r0}) one finds again 
for the exponent $ \gamma = 2 $, and for the length scale $r_0$
the improved value
\beq
r_0 \; = \; { 1 \over 8 \pi \, G \, \rho_0 } \cdot 
{ \sqrt{A_1}  \over \xi } \; \approx \; 0.25 \; \xi \;
\label{eq:r0_xi}
\eeq
which now seems more in line with observational data.
For the Fourier amplitude $a_0$ in Eq.~(\ref{eq:power_spec}) one obtains the estimate
\beq
a_0 \; = \; 2 \pi^2 \, r_0^2 \; \approx \; 1.23 \; \xi^2  \; .
\label{eq:a0_xi}
\eeq
Observed galaxy density correlations give indeed for the exponent 
in Eq.~(\ref{eq:grho_r0}) a value close to two, namely
$\gamma \approx 1.8 \pm 0.3 $ for distances in the $0.1\, Mpc$
to $50\, Mpc$ range \cite{pee93,bahcall}, and  for the
length scale $r_0 \approx 10 \, Mpc $, which is about two orders of
magnitude smaller than the result of Eq.~(\ref{eq:r0_xi})
[using $\xi =\sqrt{3 / \lambda} \approx 5320 \, Mpc$, see 
Eq.~(\ref{eq:xi_mpc})].
More recent estimates for the exponent $\gamma$, going up to distance
scales of $100 \, Mpc$, range between $1.79$ and $1.84$
\cite{baugh,longair,tegmark,durkalec,wang,coil}.
The conclusions are similar if one looks at the galaxy power spectrum data,
which also suggests $ a_0 \approx ( 30 \, Mpc )^2 $ assuming an exponent
$s=1$ exactly, again consistent 
in view of Eq.~(\ref{eq:a0_xi}) with $\xi \approx 30 \, Mpc $, a value that 
seems rather low (again by two orders of
magnitude) in view of the original identification of $\xi$ as the vacuum condensate
scale, Eqs.~(\ref{eq:xi_r}) and (\ref{eq:xi_mpc}).
Nevertheless at this point the (perhaps admittedly rather naive)
identification given in Eqs.~(\ref{eq:r0_xi}) and (\ref{eq:a0_xi}),
while intriguing, is possibly entirely accidental.
It largely bypasses any considerations regarding the actual physical origin 
(beyond the simple arguments given here) of the 
galaxy correlation function in Eq.~(\ref{eq:r0}), including the model-dependent
form and evolution of primordial density perturbations, the detailed nature of
linear and non-linear relativistic matter density perturbation theory 
for a given comoving background etc.

Let us note here that the previous discussion focuses on the relationship 
between the curvature
correlation and the matter density correlation, in accordance with
Eqs.~(\ref{eq:corr_pow}), (\ref{eq:corr_pow1}) and the field equation result
Eq.~(\ref{eq:trace}).
Nevertheless for sufficiently small $q$ it is no longer legitimate to
assume that Newton's constant $G$ is constant, as was done in 
Eq.~(\ref{eq:trace}), and which later affects the results
of Eqs.~(\ref{eq:r0}) and (\ref{eq:a0_xi}) to the extent that they involve $G$.
Instead one should make use of the effective field equations of
Eq.~(\ref{eq:grun_field}), which involve a running $G(\Box)$, or
more simply make use of $G(q)$ as given in Eq.~(\ref{eq:grun_k}).
When the above replacement is performed, one finds
\beq
P(q) \; = \; 
{ A_1 \over 32 \, G(q)^2 \, \rho_0^2 \, \xi^2 } \cdot
{ 1 \over ( q^2 + m^2 )^{s/2} }
\; \equiv \; 
{ A_1 \over 32 \, G_0^2 \, \rho_0^2 \, \xi^2 } \cdot
{ 1 \over ( q^2 + m^2 )^{s/2} } \cdot
\left [ 
1 + c_0  \left (  { m^2 \over q^2 + m^2 } \right )^{3/2}  
\right ]^{-2} 
\label{eq:power_spec_reg_g}
\eeq
with as before $m^{-1}  \equiv \xi $ and (in view of the preceding discussion) still $s=1$.
Then the most important modification, as can be seen quite clearly in Figure 5, is the rather dramatic
decrease in magnitude (due to the $ 1/G(q)^2 $ factor) of $P(q)$ for small ${\bf q}$,
with a clear turnover at $ q = \sqrt{ 5^{2/3} \, c_0^{2/3} - 1 } \, / \xi \approx 4.195 / \xi $.

In view of the more complex behavior of $P(q)$, it is clearly no longer possible to associate
a single spectral index with $P(q)$.
Following Eq.~(\ref{eq:power_spec}) one can nevertheless define an {\it effective} 
spectral index $ s( q ) $ via
\beq
s(q) \; = \; - \, { \partial \, \log P(q) \over \partial \, \log q } \; .
\label{eq:power_index}
\eeq
This quantity tends to $s=1$ for $q \gg 1 / \xi $ but then dips below zero for $ q \simeq 1 / \xi $.
A plot of $s(q)$ is shown in Figure 6.
One would expect that such a drastic turnover, caused by the quantum running of $G (q) $ on 
very large scales, should become visible in future cosmological observations.
From a practical perspective, it might make more sense to treat the numerical
amplitude $A_1$ in Eq.~(\ref{eq:corr_pow2}) as a free parameter,
given the uncertainties, discussed earlier, associated in its determination from a 
first principle lattice calculation [see the discussion following Eq.~(\ref{eq:corr_pow2a})].
In other words, it would seem that so far, based on existing results, the $q$-dependence of 
$P(q)$ is more credible at this stage than its overall normalization.
A much more detailed and recent discussion of the expected behavior of the 
matter power spectrum $P(k)$ and its extrapolation to the CMB regime,
based on the vacuum condensate picture of quantum gravity presented here, can
be found in the recent paper \cite{ham18}.

In conclusion, the main results of this section can be summed up as follows. 
The vacuum condensate picture of quantum gravity leads to three main predictions 
for matter density correlations, of which the first one is
that the power appearing in Eq.~(\ref{eq:grho_r0}) should be exactly $\gamma = 2$
for ``short distances'' $ r \ll \xi $ (or $q \gg 1/\xi$),
and that the reference length scale $r_0$ appearing in the same equation
should be related to $\xi$, as in Eq.~(\ref{eq:r0_xi}). 
The second prediction is that the power spectrum exponent in Eq.~(\ref{eq:power_spec})
should be exactly $s=1$ again for $q \gg 1/\xi$, and that the 
amplitude $a_0$ in the same equation should be related to $\xi$ as in Eq.~(\ref{eq:a0_xi}).
The third prediction is that the power spectrum $P(q)$ as a function of $q$ should exhibit 
a marked break for $ q \sim \xi^{-1}$, as given later in Eqs.~(\ref{eq:power_spec_reg}) and
(\ref{eq:power_spec_reg_g}), and shown in Figures 5 and 6.
\footnote{
A further complication arises in the cosmological context from the fact that at late times 
$H_0^{-1} \equiv {\dot a} / a \simeq t_0 \simeq 0.79 \, \xi $ so that $H_0^{-1}$
and $\xi$ are quite comparable in magnitude (which follows from the FRW evolution
of the scale factor for a universe dominated by a $\lambda$ term), whereas at early times
a new, much shorter scale appears $H^{-1} \equiv {\dot a} / a \ll \xi $.
Such effects are expected to play a role as well, and have not been taken into account here yet.}
More generally, the results outlined here and in the previous sections suggest that existing,
and future, astrophysical and cosmological data should be re-analyzed  in terms
of a wider $ G(q) $ scale dependent form, similar to the one in Eq.~(\ref{eq:grun_k}), and of the
general (but nevertheless rather simple) two-parameter type $ G(q) = G_0 \, [ 1+ (q_0/q)^p \, ] $, with 
$q_0^{-1} \gg 1 \, Mpc $ a wave vector reference scale, and $p$ a positive power.



\begin{figure}
\begin{center}
\includegraphics[width=0.7\textwidth]{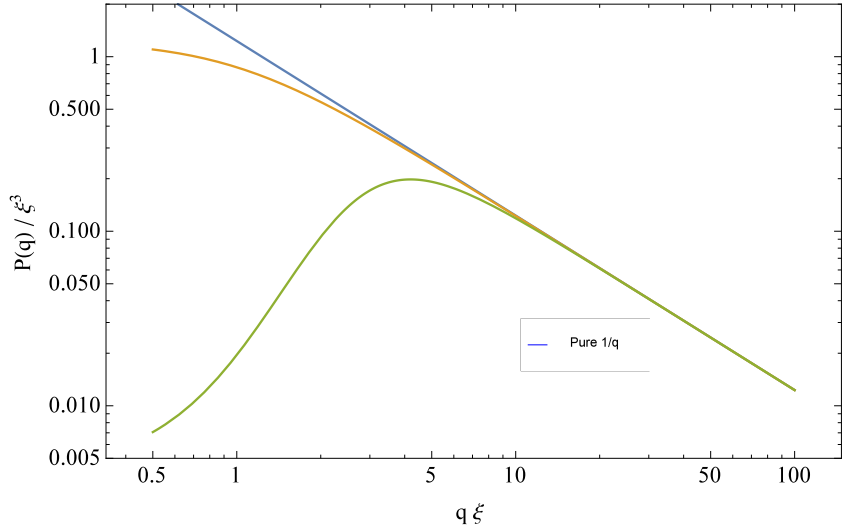}
\end{center}
\caption{
Qualitative behavior of the (gravitational quantum fluctuation-induced) matter density 
power spectrum $P(q)$ with a running Newton's $G(q)$, as given explicitly in Eq.~(\ref{eq:power_spec_reg_g}).
Here it is compared to the results of Eqs.~(\ref{eq:power_spec}) and (\ref{eq:power_spec_reg})
for a  constant $G$.
Also, the spectral exponent is $s=1$ and the amplitude is $a_0$, as discussed in the text 
[see Eqs.~(\ref{eq:s_nu}) and (\ref{eq:a0_xi})]; in the plot the $q$ wave vector is measured 
for convenience in units of $\xi $.
Note the rather marked turnover for small $q \approx 4.20 / \xi  $ due to the running of 
$G$, as discussed in the text.
The nonperturbative scale $\xi$ is related to the gravitational vacuum condensate,
as in Eqs.~(\ref{eq:xi_lambda}) and (\ref{eq:xi_r}).
}
\end{figure}



\begin{figure}
\begin{center}
\includegraphics[width=0.7\textwidth]{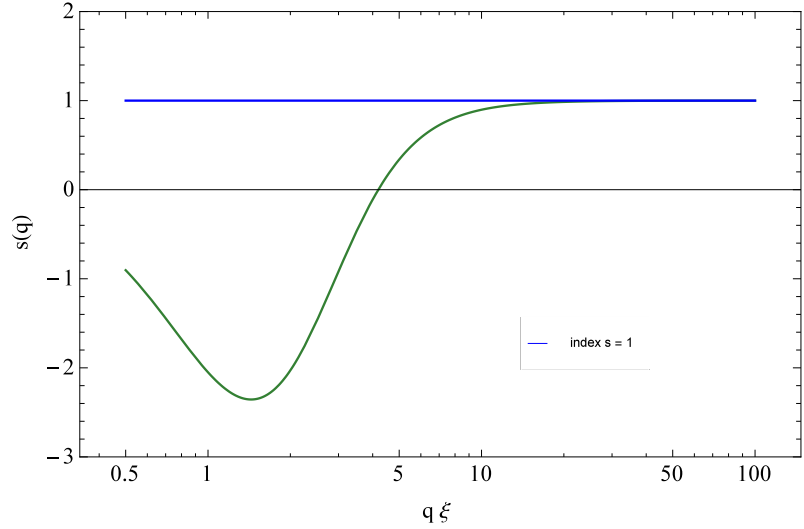}
\end{center}
\caption{
Effective spectral index $s(q)$ as defined in Eqs.~(\ref{eq:power_spec}) 
and (\ref{eq:power_index}).
The horizontal line at the top is the $s=1$ value, corresponding to a constant (scale-independent)
Newton's $G$.
The spectral index approaches the value $s=1$ for large $q \gg 1/\xi $,
but dips below zero for $ q \simeq 1 / \xi $.
}
\end{figure}


\vskip 60pt

\section{Gravitational Slip Function with $G(\Box)$}

\label{sec:conf}

\vskip 20pt

A running of Newton's $G$ gives rise to significant long-distance effects, which fundamentally 
modify the classical field equations of general relativity at very large distance scales. 
The following section provides an update on the results originally presented in \cite{ht11},
especially in view of the recent high accuracy lattice results presented in 
\cite{ham15}, and in
particular related to the new improved estimate for the quantum amplitude $c_0$ of Eq.~(\ref{eq:grun_box}).
It is common practice to describe relativistic effects in cosmology within the framework
of the {\it comoving} gauge, where the metric is written as 
\beq
g_{\mu \, \nu} = \bar{g}_{\mu \, \nu} + h_{\mu \, \nu} \; ,
\eeq
with background metric $ \bar{g}_{\mu \, \nu} = {\rm diag} \left(-1, a^2, a^2, a^2 \right) $,
fluctuations such that $ h_{0i} = h_{i0} = 0 $, with the remaining $h_{ij}$'s decomposed
into a trace and stress part
\beq
h_{ij} ({\bf k},t) \; = \;  a^2
\left[ \, { 1 \over 3 }\,  h \, \delta_{ij} 
+ \left( {1 \over 3} \, \delta_{ij} - {k_i \, k_j \over k^2} \right) \, s  \right] 
\label{eq:stress_def}
\eeq
so that $ Tr(h_{ij}) = a^2 \, h $.
In this gauge the metric is then parametrized by the scale factor $ a(t) $ and the two additional
functions $s$ and $h$.

There are nevertheless instances where effects which deviate from standard Newtonian physics
are more easily described within the context of the {\it conformal Newtonian} gauge, 
where the metric is parametrized by two scalar potentials $ \psi $ and $\phi $
\cite{mab95,ber01}.
It is of some interest to explore possible cosmological consequences of a running Newton's 
constant $ G ( \Box ) $ in this gauge, as discussed recently in \cite{ht10}.
In this gauge one sets for the metric $ g_{\mu \, \nu} = \bar{g}_{\mu \, \nu} + h_{\mu \, \nu} $ with
$ \bar{g}_{\mu \, \nu} =a^2 \, diag \left(-1, 1, 1, 1 \right) $, $ h_{0i} = h_{i0} = 0 $, 
and furthermore
\beq
h_{00} = a^2 \, \left(- \, 2 \, \psi \right)   \;\;\;\;\;\;\;\; 
h_{ij} = a^2 \, \left(- \, 2 \phi \right) \, \delta_{ij} \; ,
\eeq
with the components of $h_{\mu\nu} (x) $ again considered here as a small perturbation.
The line element is then given by
\beq
d\tau^2 = - g_{\mu \, \nu} \, dx^\mu \, dx^\nu \; = \;
a^2 \; \Big\lbrace \left( 1 + 2 \, \psi \right) \, dt^2 
- \left(1 - 2 \, \phi \right) \, dx_i \, dx^i \Big\rbrace \; .
\eeq
Then gravitational {\it slip function} $\eta (x) $ is defined as 
\beq
\eta \; \equiv \; { \psi - \phi \over \phi } \; .
\label{eq:eta_def}
\eeq
In classical General Relativity one has $\phi (x) = \psi (x) $ giving then $\eta (x) =0$, 
which makes the quantity $\eta$ a rather useful parametrization for deviations 
from the classical theory (whatever their origin might be).

Generally, quantum corrections arising from a running of $G(\Box)$ give rise to additional terms 
in the field equations, which no longer ensure that $\phi=\psi$.
In practice, at some stage of the calculation one needs to compute higher order contributions 
from the $h_{ij}$'s  which requires one to expand, for example, 
$G(\Box)$ in the metric perturbations $h_{ij}$,
Consequently the covariant d'Alembertian $ \Box = g^{\mu\nu} \nabla_\mu \nabla_\nu$
has to be Taylor expanded in the small field perturbation $h_{\mu\nu}$,
\beq
\Box (g) \, = \, \Box^{(0)}  + \Box^{(1)} (h) + O (h^2) \; ,
\eeq
and similarly for $G(\Box)$ as in Eq.~(\ref{eq:gbox_h}), 
which requires the use of the binomial expansion for the operator
$ (A+B)^{-1} = A^{-1} - A^{-1} B A^{-1}  + \dots $.
One has therefore
\beq
G(\Box) = G_0 \, \left \{
1 + \, { c_0 \over \xi^{1 / \nu} } \, 
\left [
\left ( { 1 \over \Box^{(0)} } \right )^{1 / 2 \nu} 
- {1 \over 2 \, \nu} \, { 1 \over \Box^{(0)} } \cdot \Box^{(1)} (h) \cdot
\left ( { 1 \over \Box^{(0)} } \right )^{1 / 2 \nu} \, + \dots
\right ]
\right \}  \; ,
\label{eq:gbox_hs_expanded}
\eeq
where the superscripts $(0)$ and $(1)$ refer to zeroth and first order in this 
weak field expansion, respectively.
It is also customary in these treatments to expand all relevant fields in spatial Fourier modes.
One sets for the matter density, pressure and velocity, as well as for the metric,
\bea
\delta \rho ({\bf x},t) & = & \delta \rho_{\bf q} (t) \, e^{i \, {\bf q} \, \cdot \, {\bf x}}
\;\;\;\;\;\;\;\;
\delta p ({\bf x},t) = \delta p_{\bf q} (t) \, e^{i \, {\bf q} \,\cdot \, {\bf x}}
\nonumber \\
\delta {\bf v} ({\bf x},t) & = & {\delta {\bf v}}_{\bf q} (t)  \, e^{i \, {\bf q} \, \cdot \, {\bf x}}
\;\;\;\;\;\;\;\;
h_{ij} ({\bf x},t) = h_{ {\bf q} \, ij} (t)\, e^{i \, {\bf q} \, \cdot \, {\bf x}} 
\label{eq:fourier}
\eea
with ${\bf q}$ the comoving wave number, and similarly for any of the other fields as well.
An additional approximation is then to limit at first the treatment to the leading ${\bf q} = 0 $ mode, 
and leave the more challenging $O ( {\bf q}) $ corrections for a  later calculation.
Nevertheless here the nonlocal nature of the quantum corrections makes the calculation 
of the slip function $\eta$, as well as other quantities, technically rather
difficult due to the intrinsic non-locality of $G(\Box)$ in Eq.~(\ref{eq:grun_box}), 
which will require at some stage a number of physically motivated approximations,
so that a partial, useful answer can be obtained.
Furthermore, the general result for the slip function $\eta$ is most easily presented in a form where the 
metric perturbation $h_{ij}$  is decomposed into the comoving trace ($h$) and a stress ($s$) part;
then in the original comoving gauge the spatial metric perturbation is written as in Eq.~(\ref{eq:stress_def}).
It is useful here to record the (rather straightforward) relationship between perturbations in the
standard comoving and conformal Newtonian gauge, namely 
\beq
\psi = - \, { 1 \over 2 q^2 } \, a^2 \, \left( \ddot{s} + 2 \, {\dot{a} \over a} \, \dot{s} \right)
\;\;\;\;\;\;
\phi = - \, {1 \over 6} \, \left( h + s \right) + \, {a^2 \over 2 \, q^2} \, {\dot{a} \over a} \, \dot{s} \; .
\label{eq:phi_transf_cN_pbls}
\eeq
Further details regarding the various choices of gauge (comoving, synchronous and conformal
Newtonian) and their mutual relationship, as they apply to this specific problem,
can be found in \cite{ht11}.

The zeroth order (in the metric perturbation $h_{\mu\nu}$) the Friedman equations in the presence of a running $G$ are given in the comoving gauge by
\bea
3 \, {{\dot{a}}^2 (t) \over {a}^2 (t)} 
& = & 
8 \pi \, G_0 \left ( 1 + {\delta G(t) \over G_0} \right ) \, \bar{\rho}(t) + \lambda 
\nonumber \\
{{\dot{a}}^2 (t) \over {a}^2 (t)} + 2\, {\ddot{a} (t) \over a (t)} 
& = & 
- 8 \pi \, G_0 \, \left ( w + w_{vac} {\delta G(t) \over G_0} \right )\, \bar{\rho} (t) + \lambda  \; ,
\label{eq:fried_run1}
\eea
with $w=0$ for non-relativistic matter and $w_{vac}= \third$ for the graviton
vacuum polarization contribution, together with the energy conservation equation
\beq
3 \, {\dot{a} (t) \over a (t)} \, 
\left [ \left (1+w \right ) + \left (1+ w_{vac} \right )\,{\delta G(t) \over G_0} 
\right ] \bar{\rho} (t) 
+ { \dot{\delta G}(t) \over G_0} \, \bar{\rho} (t)
+ \left ( 1 + {\delta G(t) \over G_0} \right )\, \dot{\bar{\rho}} (t) = 0 \; .
\label{eq:encons_zeroth_w_1}
\eeq
These equations need to be solved first, in order to obtain an expression for the background 
scale factor $a(t)$, and for the background average matter density ${\bar{\rho}} (t)$.
Of course for $\delta G (t) = 0$ the above equations reduce to the standard Friedman equations
with cosmological constant $\lambda$.

When restricted to the ${\bf q} =0 $ mode (the zeroth order, dominant infrared mode) one obtains for the slip function an expression which is a function of $G(t)$ or $G(a)$, where $t$ is the cosmological time and $a(t)$ the corresponding scale factor.
The detailed dependence of $G$ on $t$ or $a(t)$ (and other parameters) is in turn quite
sensitive to the specific choice of cosmological parameters, such as the rate of expansion, the relative contribution of dark matter versus dark energy, etc.
In the end one finds the following ${\bf q}=0$ result for the slip function $\eta$ \cite{ht11}
\beq
\eta \; = \;  - \, 16 \pi \, G_0 \, {\delta G \over G_0} \cdot {1 \over 2 \nu} \, {8 \over 3} 
\cdot { \int \! s \, \mathrm{d}t \over \dot{s}} \; \bar{\rho} \; ,
\label{eq:eta_t}
\eeq
with $\delta G (t) $ given in Eq.~(\ref{eq:grun_t}) and $\nu=1/3$.

A specific numerical value for $\eta$ depends on various assumptions introduced in order
to concretely evaluate the above expression.
One of the simplest cases corresponds to the limit of a vanishing cosmological
constant, $\lambda \simeq 0$.
In view of Eq.~(\ref{eq:xi_lambda}),  this last limit 
corresponds to a very large $\xi$.
For a general perfect fluid with equation of state $p= w \rho $ one then has
for the scale factor $a(t) = a_0 (t/t_0 )^{2/3(1+w)} $ and  for average matter density
$\rho (t) = 1 / [6 \pi G t^2 (1+w)^2 ] $.
Then from Eq.~(\ref{eq:eta_t}) one obtains for pure non-relativistic matter, $w=0$,
\beq
\eta \; = \; { 32 \over 3 } \, c_t \, \left ( { t \over \xi } \right )^3 \, \log \left ( { t \over \xi } \right )
+ {\cal O} (t^4)  \; , \;\;\;\;\;\; (\lambda =0) \; ,
\eeq
and more generally for $w \neq 0 $
\beq
\eta \; = \; { 16 \over 3 } \, { c_t \over w \, (1-w) } \, \left ( { t \over \xi } \right )^3 \, 
+ {\cal O} (t^6)  , \;\;\;\;\;\; (\lambda =0) \; .
\eeq
In both cases the amplitude $c_t = 0.45 \, c_0 $ for the tensor box operator \cite{eff}, 
with $c_0$ entering the expression for
$G(\Box)$ of Eq.~(\ref{eq:grun_box}), and $c_0 \approx 16.04 $, giving therefore
an overall coefficient $c_t  \approx 7.2$.
Another extreme, but nevertheless equally simple, case is a pure cosmological constant 
term, which can be modeled by taking the limit $w=-1$.

To analyze the more general case of a non-vanishing cosmological constant $\lambda \neq 0 $  
combined with non-relativistic matter ($w=0$),  the following form \cite{ht10}
for the slip function expressed in terms of the scale factor $a(t)$ turns out to be more useful 
\beq
\eta (a) \; = \; {16 \over 3 \, \nu} \, {\delta G (a) \over G_0} \, \log \! \left ( {a \over a_\xi} \right )  \; .
\label{eq:eta_a}
\eeq
The integration constant $a_\xi$ is  fixed by the requirement that
the scale factor $a \rightarrow a_\xi$ for $t \rightarrow \xi $ [see Eqs.~(\ref{eq:grun_box}),
(\ref{eq:grun_t}) and (\ref{eq:grun_a}) below for the definitions of $a_\xi$ ].
In other words, by switching to the variable $a(t)$ instead of $t$, the quantity $\xi$ has
been traded for $a_\xi$. 
In practice the quantity $a_\xi$ is generally
expected to be slightly larger than the scale factor "today", i.e. for $t=t_0$.
As a result the correction in Eq.~(\ref{eq:eta_a}) is expected to be {\it negative} today.
What is then needed is the general relationship between
$t$ and $a(t)$ (for nonvanishing cosmological constant $\lambda$) so that a 
quantitative estimate for the slip function $\eta$
can be obtained from Eq.~(\ref{eq:eta_a}).
Specifically one is interested in the value of $\eta$ for a current matter fraction 
$\Omega \simeq 0.31$, as suggested by current astrophysical measurements
(according to the recent Planck 2015 data, see \cite{planck15}).
To obtain $\delta G(a)$ one then makes use of $\delta G(t) $ from Eq.~(\ref{eq:grun_t}), 
and then substitutes the correct (matter-fraction dependent) relationship between 
$t$ and $a$ \cite{ht11}, which among other things contains the constant 
\beq
a_\xi \; = \; 
\left( {1 \over \theta} \right)^{1\over 3} \mathrm{Sinh}^{2 \over 3} \! \left[ {3 \over 2}\right]  \; ,
\label{eq:axi_1}
\eeq
with the parameter $\theta$ defined as
\beq
\theta \; \equiv \; { \lambda \over 8 \pi G_0 \bar \rho_0 } \; = \; { 1 - \Omega \over \Omega }
\label{eq:theta}
\eeq
with $\bar \rho_0$ the current ($t=t_0$) matter density, and $\Omega$ the
current matter fraction.
In practice one is interested in a matter fraction of around $0.31$, 
giving $\theta \simeq 2.23 $, which is quite a bit larger than
the zero cosmological constant value of $\theta = 0$.
This then gives roughly $ t_0 / \xi = 0.794 $ and $ a_\xi \equiv a( t=\xi ) = 1.268 $
in Eq.~(\ref{eq:eta_a}).

The last step left is to make contact with observationally accessible quantities,
by expanding  in the redshift $ z $ related to the scale factor by $ a^{-1} = 1 + z $.
For exponent $ \nu = 1/3 $ and matter fraction $\Omega=0.31$ one 
obtains for the gravitational slip function "today" ($t=t_0$) 
\beq
\eta( t=t_0, {\bf q}=0 ) \; = \; - \, c_0 \, f \, \left ( { t_0 \over \xi } \right )^3  \; ,
\label{eq:eta_z}
\eeq
with $c_0 \approx 16.04 $.
Here $f$ is a numerical constant equal to $f=1.11$ for pure 
non-relativistic matter ($\lambda=0$ and $w=0$), and $f=1.71$ for the current
observed matter fraction $\Omega \approx 0.31 $ (and thus $\lambda \neq 0$ but 
still $w=0$).
Note that the correction is always negative, and since
$t_0 \approx 0.794 \, \xi $ the above ${\bf q}=0 $ [see the mode expansion in Eq.~(\ref{eq:fourier})] 
correction seems rather large in magnitude, $ \eta ({\bf q} =0 ) \approx -13.73 $.
Nevertheless one should keep in mind that the above result only corresponds to the extreme limiting case of comoving wavevector ${\bf q} \simeq 0$ .
In analogy to the quantum mechanical particle (the graviton) in a box, the lowest possible mode
corresponds to $q \simeq \pi / L $ where $L$ is the linear size of the box. 
Here the role of the $L$  is played by the correlation length $\xi$, so that the lowest
possible mode corresponds in fact to $ q \simeq 1 / \xi $, for which then the result
of Eq.~(\ref{eq:eta_z}) applies.

Generally the quantum contributions to the slip function are {\it scale dependent}, and thus will be proportionately reduced if one looks at scales which are significantly smaller compared to the largest scale in the problem, namely $\xi$.
The calculation of the slip function $\eta ({\bf q})$  over a wider range of scales is clearly a 
significantly more complicated problem, which has not been addressed yet;
as stated previously all calculations performed so far \cite{ht11} only correspond to limit of 
$ {\bf q} \rightarrow 0 $ , for which the above analytical estimate has been obtained.
Nevertheless, if the corresponding relevant astrophysical length scale is denoted by $l_0$, then
at such a scale one must have, simply by scaling from Eqs.~(\ref{eq:grun_k}) and (\ref{eq:grun_box}),
\beq
\eta ( l_0 ) \; = \;  \eta( t=t_0, {\bf q}=0 ) \times \left ( { l_0 \over t_0 } \right )^3  
\; \approx \;  - 13.73 \, \left ( { l_0 \over \xi }  \right )^3  \; ,
\label{eq:eta_l0}
\eeq
with $\eta( t=t_0, {\bf q}=0 )$ given earlier in Eq.~(\ref{eq:eta_z}).
In other words, the result for $\eta$  is reduced by the ratio of the relevant length scale compared
to $t_0$ or $\xi$, to the third power since the exponent $\nu=1/3$.
As a practical example, if one were to look at the value of the slip function $\eta ({\bf q}) $ on scales which are an order of magnitude less than the reference scale $\xi$, then this would reduce,
in accordance with Eqs.~(\ref{eq:grun_k}) and (\ref{eq:grun_box}) (where the correction is always
proportional to $\xi^{-3}$),  the answer by a factor of $10^3=1000$.
Thus for distance scales $l_0 \simeq 500 \, Mpc $ 
(using again for reference $ \xi = \sqrt{3 / \lambda} = 5320 \, Mpc $) one obtains from Eq.~(\ref{eq:eta_l0}) $\eta = - 0.011 $.
For even shorter distance scales $l_0 \simeq 50 \, Mpc $ one has $\eta = - 1.14 \times 10^{-5} $,
which is now a rather small number.

In practice, "large scales" in observational cosmology correspond to $ \gg 10 \, h^{-1} \, Mpc $,
with redshift surveys going up to scales of $\sim 300 \, h^{-1} \, Mpc $,
and recent CMB surveys probing scales up to $\sim 500 \, h^{-1} \, Mpc $
(with current estimates of the scaled Hubble constant giving $h^{-1} \approx 1.476$).
As far as astrophysical observations are concerned,
current estimates for $\eta (z=0)$ obtained from CMB measurements
give values around $0.09 \pm 0.7 $ \cite{ame07,dan09},
which could be used in the future to place a direct observational bound on
the slip function $ \eta ( {\bf q} ) $ of Eqs.~(\ref{eq:eta_z}) and (\ref{eq:eta_l0}).

\vskip 40pt

\section{Matter Density Perturbations with $G(\Box)$}

\label{sec:pert}

\vskip 20pt

The classical treatment of cosmological models in General Relativity usually starts out from a given background metric (such as the Friedmann-Lema\^itre-Robertson-Walker one), and then later uses the field equations to constrain small fluctuations about that metric. 
One application of this method is the computation of the gravitational growth of matter density perturbations  $\delta \rho (t, {\bf q}) $, usually restricted in a first approximation to the lowest 
comoving spatial momentum ${\bf q}$ modes.
In this limit the growth parameter $ \delta (t) \equiv \delta \rho (t) / {\bar \rho} $ obeys, as a function 
of scale $a (t) $, a rather simple ordinary differential equation, whose solution then
provides, given suitable initial conditions, information about the matter and dark energy 
content of the current universe.
One quantity that is often brought into play is the growth index $ f(a) $, namely the derivative of the log of $\delta (a) $ with respect to the log of the scale factor $a(t)$, 
and in addition the parameter $ \gamma = \log f / \log \Omega $, which provides information
on how the growth index $f(a)$ depends on the current matter fraction $\Omega$ \cite{pee93}. 
Cosmological observation suggests that today's matter fraction is about $\Omega \approx 0.31$
\cite{planck15}, leading to a value of $ \gamma = 0.55 $, based pretty much entirely on what 
is obtained from the systematic treatment of density perturbations within the framework of classical General Relativity.

It follows that many of the calculations just described can be repeated if one assumes now that Newton's constant runs with scale, so that the standard field equations of GR get modified by the non-local term of Eq.~(\ref{eq:grun_box}).
Under the physically motivated assumption of a comparatively slowly varying (both in space and time) 
background, it is then possible to obtain a complete and consistent set of effective 
field equations, describing small perturbations for the metric trace and matter modes \cite{ht10,ht11}.
This then gives rise, within the same set of methods and approximations used in classical GR, to a set of equations for the growth amplitude.
The latter are then studied again, initially, in the limit of small $q$ wave vectors, and this in turn
leads to modified growth exponents.
In general the results are expected to be quite sensitive to the scale ${\bf q}$, but so far only the leading term as ${\bf q}$ goes to zero has been calculated analytically, due to technical difficulties which arise
from the strong non-locality of $G ( \Box )$.
The following section provides a significant update on the results presented originally in \cite{ht11},
especially in view of the recent high accuracy lattice results presented in \cite{ham15}, and in
particular the new improved estimate for the quantum amplitude $c_0$ of Eq.~(\ref{eq:grun_box}).

Besides the modified cosmic scale factor evolution due to the $G(t)$ discussed earlier
[see for ex. Eqs.~(\ref{eq:fried_rr}) and  (\ref{eq:fried_tt})]
the running of $G(\Box)$ as given in Eq.~(\ref{eq:grun_box})
also affects the nature of matter density perturbations on large scales.
In computing these effects, it is customary to introduce a perturbed 
FRW metric of the form
\beq
{d\tau}^2 = {dt}^2 - a^2 \left ( \delta_{ij} + h_{ij} \right ) dx^i dx^j \; ,
\label{eq:pert_metric}
\eeq
with $a(t)$ the unperturbed scale factor and $ h_{ij} ({\bf x},t)$ a small
metric perturbation, and $h_{00}=h_{i0}=0$ by choice of coordinates.
After decomposing the matter fields into background and fluctuation contribution, 
$\rho = \bar{\rho}+\delta \rho$, $p = \bar{p}+\delta p $, and ${\bf v} = \bar{\bf v}+\delta {\bf v}$, 
it is customary in these treatments to expand the density, pressure and metric
perturbations in spatial Fourier modes, as in Eq.~(\ref{eq:fourier}) 
with ${\bf q}$ the comoving wave number.
Then the field equations with a  $G (\Box)$ [Eq.~(\ref{eq:grun_field})]
are given, to zeroth order in the perturbations $h_{ij}$, by the unperturbed
field equations with a $G(t)$,  which in turn fixes the three background fields 
$a(t)$, $\bar{\rho} (t)$, and $\bar{p} (t) $ in accordance with
Eqs.~(\ref{eq:fried_rr}) and  (\ref{eq:fried_tt}).

At the next step, in order to obtain an equation for the matter density contrast 
$\delta (t) = \delta \rho (t) / \bar{\rho} (t)$, it is customary 
to eliminate the metric trace field $h(t)$ from the field equations.
This is first done by taking a suitable linear combination of two field equations 
to get the single equation
\bea
\ddot{h} (t) + 2 \, {\dot{a} (t) \over a (t)} \, \dot{h} (t) 
& + & 8 \pi \, G_0 \, {1 \over 2 \nu} \, c_h \,
\left ( 1 + 3 \, w_{vac} \right ) 
{\delta G(t) \over G_0} \, \bar{\rho}(t) \, h(t) 
\nonumber \\
& = & - \, 8 \pi \, G_0 \left [
 \left ( 1 + 3 \, w \right ) + 
\left ( 1 + 3 \, w_{vac} \right ) \,  {\delta G(t) \over G_0}
\right ] \bar{\rho} (t) \, \delta (t) \; .
\label{eq:grun_field_fluc_h}
\eea
Then the first order energy conservation equations to zeroth 
and first order in $\delta G$ allow one to completely eliminate 
the $h$, $ \dot{h} $ and $ \ddot{h} $ field in terms of the matter 
density perturbation
$\delta (t)$ and its derivatives.
The resulting equation for $\delta (t) $ then reads, for the simplest case of a matter dominated universe $ w = 0 $ and $w_{vac}= \third $,
 \bea
\ddot{\delta}(t) && 
+ \left [ 
\left ( 2 \, {\dot{a}(t) \over a(t)}
- {1 \over 3} \, {\dot{\delta G}(t) \over G_0} \right ) 
- {1 \over 2 \nu} \cdot 2 \, c_h \cdot \left ( {\dot{a}(t) \over a(t)} \, {\delta G(t) \over G_0} 
+ 2\,{\dot{\delta G}(t) \over G_0} \right ) 
\right ] \dot{\delta}(t) 
\nonumber \\
&& + \left [
- \, 4 \pi \, G_0  \left (1 + {7 \over 3} \, {\delta G(t) \over G_0} 
- {1 \over 2 \nu} \cdot 2 \, c_h \cdot {\delta G(t) \over G_0} \right ) \bar{\rho}(t) 
\right. 
\nonumber \\
&& \;\;\;\;\;  
\left.  
- {1 \over 2 \nu} \cdot 2 \, c_h \cdot 
\left ( {{\dot{a}}^2(t) \over a^2(t)} \, {\delta G(t) \over G_0} 
+ 3 \, {\dot{a}(t) \over a(t)} \, {\dot{\delta G}(t) \over G_0} 
+ {\ddot{a}(t) \over a(t)} \, {\delta G(t) \over G_0} 
+ {\ddot{\delta G}(t) \over G_0} \right ) 
\right ] \delta(t) = 0 \; .
\nonumber \\
\label{eq:dfq_delta_Gbox}
\eea
This last equation then describes matter density perturbations to linear order,
taking into account the running of $ G(\Box) $, and was therefore one of the 
main results of \cite{ht11}.
Terms proportional to
\beq
c_h \; =  \; {11\over 3} \, \, {\dot{a} \over a}  \, { h \over \dot{h} } \;  \approx \; 7.927
\label{eq:ch_tens}
\eeq
describe the feedback of the metric fluctuations 
$h$ on the vacuum density $\delta \rho_{vac}$ and pressure $\delta p_{vac}$ fluctuations.
\footnote{
Again current cosmological estimates \cite{planck15} have been used here to provide a
sensible estimate for $c_h$.}
Eq.~(\ref{eq:dfq_delta_Gbox}) can be compared with the corresponding, and much simpler,
equation obtained for constant $G$ and non-relativistic matter $ w = 0 $ 
(see for example \cite{wei72} and \cite{pee93})
\beq
\ddot{\delta} (t) + 2 \, {\dot{a} \over a} \, \dot{\delta} (t) 
- 4 \pi \, G_0 \, \bar{\rho}(t) \, \delta(t) = 0 \; .
\label{eq:dfq_delta_G0}
\eeq
For the latter one obtains immediately for the growing mode
\beq
\delta_{\bf q} (t) = \delta_{\bf q} ( t_0) \, \left ( { t \over t_0 } \right )^{2/3} \; ,
\eeq
which is the standard result in the matter-dominated era \cite{wei72}.

To make progress in the more general case of Eq.~(\ref{eq:dfq_delta_Gbox})
one follows common practice and writes an equation for the density contrast 
$\delta(a)$ not as a function of $t$, but instead of the scale factor $a(t)$.
Consequently, instead of using the expression for $G(t)$ in Eq.~(\ref{eq:grun_t}),
one uses the equivalent expression for $G(a)$
\beq
G (a) = G_0 \left ( 1 + {\delta G (a) \over G_0}  \right ), 
\;\;\;\;\;\; {\rm with}
\;\; {\delta G(a) \over G_0} \equiv  c_a \, \left ( {a \over a_0 } \right )^{\gamma_\nu}  + \dots \; .
\label{eq:grun_a}
\eeq
Here the power is $ \gamma_\nu = 3/2 \nu $ for non-relativistic matter, 
since from Eq.~(\ref{eq:grun_t})  one has then $a(t)/a_0 \approx (t/t_0)^{2/3}$ for constant $G$; 
in the following $\nu= \third$ for which then $\gamma_\nu = 9/2$ for this case.
If on the other hand one uses a more general equation of state of the form
$p=w \rho$ then $ a(t)/a_0 = (t / t_0)^{ 2 / 3 (1+w) }$, and therefore $\gamma_\nu = 3(1+w)/ 2 \nu $.
Also, $c_a \approx c_t$ if $a_0$ is identified with a scale factor 
corresponding to a universe of size $\xi$; to a good approximation this corresponds
to the universe ``today'',  with the relative scale factor customarily normalized at
that time $t=t_0$ to $a(t_0)=1$.
Furthermore, in \cite{eff} it was found that in Eq.~(\ref{eq:grun_t})  $c_t = 0.450 \, c_0 $ 
for the second-rank tensor box case [which is the one appropriate for Eq.~(\ref{eq:grun_field})]
which in turn determines the size of the quantum amplitude in Eq.~(\ref{eq:grun_a}),
namely $ c_a =  0.450 \times (t_0/\xi)^2 \times c_0 = 3.62 $.

More generally, the zeroth order $tt$ field equation with constant $G=G_0$ can be written 
in terms of the current matter density fractions as
\beq
H^2 (a) \equiv \left ( {\dot{a} \over a} \right )^2 
= \left ( \dot{z} \over 1 + z \right )^2 
= H_0^2 \left [ \Omega \, \left ( 1 + z \right )^3 
+ \Omega_R \, \left ( 1 + z \right )^2 + \Omega_{\lambda} \right ]
\eeq
with $ a/a_0 = 1 / ( 1 + z ) $ where $ z $ is the red shift and $a_0=1$ the scale
factor today.
In this last case $ H_0 $ is the Hubble constant evaluated today, 
$ \Omega $ the (baryonic and dark) matter density, $ \Omega_R $ the space curvature
contribution corresponding to a curvature $ k $ term, and $\Omega_{\lambda} $
the dark energy or cosmological constant part, all again measured {\it today}.
In the absence of spatial curvature $k=0$ one has then
\beq
\Omega_{\lambda} \equiv {\lambda \over 3 \, H_0^2}
\;\;\;\;\;\;
\Omega \equiv { 8 \, \pi \, G_0 \, \bar{\rho}_0 \over 3 \, H_0^2 } 
\;\;\;\;\;\;\;\;\;\; 
\Omega + \Omega_{\lambda} = 1 \; .
\label{eq:omega_def}
\eeq
Then in terms of the scale factor $a(t)$ the equation for matter density
perturbations for constant $G=G_0$, Eq.~(\ref{eq:dfq_delta_G0}), becomes
\beq
{\partial^2 \delta(a) \over \partial a^2 } 
+ \left [ {\partial \log H(a) \over \partial a} 
+ {3 \over a } \right ] \, {\partial \, \delta(a) \over \partial a} 
- 4 \pi \, G_0 \,{1 \over a^2 H(a)^2}\, \bar{\rho} (a) \, \delta (a) = 0 \; .
\eeq
The quantity $H(a)$ is most simply obtained from the FLRW field equations
\beq
H (a) = \sqrt{ {8 \pi \over 3} \,  G_0 \, \bar{\rho} (a) + {\lambda \over 3} } \; ,
\label{eq:Ha}
\eeq
which can in principle be solved for the scale factor $a(t)$, leading to
\beq
t- t_0 = \int
{ da \over a \, \sqrt{ 
{8 \pi \over 3} \,  G_0 \, \bar{\rho}_0 \, \left ({a_{0} \over a} \right )^3  
+ {\lambda \over 3}  } } \; .
\label{eq:a_t}
\eeq
It is customary at this stage to introduce a parameter $\theta$ describing the cosmological 
constant fraction as measured today,
\beq
\theta \equiv { \lambda \over 8 \, \pi \, G_0 \, \bar{\rho}_0 }
\, =  \, { \Omega_{\lambda} \over \Omega}  \, = \,  { 1 - \Omega \over \Omega }  \; .
\label{eq:theta_def}
\eeq
In practice one is mostly interested in the observationally favored case of
a current matter fraction $\Omega \approx 0.25$ [more recent data \cite{planck15} suggest a slightly
larger value of 0.31], for which then $\theta \approx 3$.
In terms of $\theta$ the equation for the density contrast $\delta (a)$ for constant 
$G$ can then be recast in the form 
\beq
{\partial^2 \delta \over \partial a^2 }
+  { 3  \, ( 1 + 2 \, a^3 \, \theta ) \over 2 \, a \, ( 1 + a^3 \, \theta  ) }  \,
{\partial \, \delta \over \partial a} 
- { 3 \over 2 \, a^2 \, ( 1 + a^3 \, \theta ) } \; \delta  = 0 \; ,
\label{eq:dfq_delta_G0_a}
\eeq
with the growing solution to the above equation given explicitly by 
\beq
\delta_0 (a) =
c_1 \cdot a \cdot {}_2 F_1 \, \left ( {1 \over 3}, 1; {11 \over 6}; - \, a^3 \, \theta \right )  
\eeq
with  $ c_1 $ a multiplicative constants and ${}_2 F_1 (a,b;c,z) $ the Gauss hypergeometric function.
The subscript $0$ in $\delta_0 (a)$ means that the solution here is appropriate for 
the case of constant $G=G_0$.

To evaluate the correction to $\delta_0 (a) $ coming from the terms proportional to
$c_a$ from $G(a)$ in Eq.~(\ref{eq:grun_a}) one sets
\beq
\delta (a)  \propto \delta_0  (a) \, \left [ \, 1 + c_a \, {\cal F} (a) \, \right ] \; ,
\label{eq:fa_corr}
\eeq
where $ {\cal F} (a) $ is a function to be determined, and then
inserts the resulting expression in Eq.~(\ref{eq:dfq_delta_Gbox}), written as a differential equation in the scale factor $a (t)$.
One only needs to write down the differential equations for density perturbations 
$ \delta (a) $ up to first order in the fluctuations, so it is sufficient to 
obtain an expression for Hubble constant $ H (a) $ from the 
$ tt $ component of the effective field equation to zeroth order in the fluctuations,
\beq
H (a) = \sqrt{{8 \pi \over 3} \,  G_0 \left ( 1 
+ {\delta G (a) \over G_0} \right ) \, \bar{\rho} (a) + {\lambda \over 3}} \;\; .
\label{eq:H_in_efe}
\eeq
In this last expression the exponent is $ \gamma_\nu = 3/2 \nu \simeq 9/2$ for
a matter dominated background universe $w=0$, and more generally 
$ \gamma_\nu = 3 (1+w)/2 \nu $; even the use of the general Eq.~(\ref{eq:a_t})
is possible and should be explored (see discussion later).
After various substitutions and insertions have been performed, one obtains
a second order linear differential equation
for the correction ${\cal F} (a)$  to $\delta (a)$, as defined in Eq.~(\ref{eq:fa_corr}).
The resulting equation can then be solved for ${\cal F} (a)$,
giving the desired density contrast $\delta (a)$ as a function of the 
parameter $\Omega$.
The explicit form for the equation for $\delta (a)$ is of the form
\beq 
{\partial^2 \delta \over \partial a^2 } 
+  A (a) \, {\partial \delta \over \partial a} + B(a) \, \delta \, = 0 \; .
\label{eq:dfq_delta_Gbox_a}
\eeq
with the two coefficient functions $A(a)$ and $B(a)$ given by rather complicated functions \cite{ht11}.

The solution of the above differential equation for the matter density contrast in the 
presence of a running Newton's constant $G(\Box)$ then leads to an explicit
form for the function $\delta (a) = \delta_0 (a) \, [ 1 + c_a {\cal F} (a) ] $.
From it, an estimate of the size of the corrections coming from the new terms
due to the running of $G$ can be obtained.
It is clear from the previous discussion, and from the form of $G(\Box)$, that
such corrections are expected to become increasingly important towards the present
era $t \approx t_0 $ or $a \approx 1 $.

Specifically, in Ref. \cite{ht10} a value for the density perturbation growth 
index parameter $\gamma$ was computed in the presence of $G(\Box)$.
When discussing the growth of density perturbations in classical GR \cite{pee93} it is 
customary
at this point to introduce a scale-factor-dependent {\it growth index} $ f (a) $ defined as 
\beq
f (a) \equiv { \partial \log \delta (a) \over \partial \log a }  \; ,
\label{eq:fa_def}
\eeq
where $ \delta (a) $ is the matter density contrast discussed above.
In principle, the latter is obtained from the solution to the general differential equation for $ \delta (a)$,
such as the one in Eqs.~(\ref{eq:dfq_delta_G0_a}) or (\ref{eq:dfq_delta_Gbox_a}).
Nevertheless, one is mainly interested in the neighborhood
of the present era, $a (t) \simeq a_0 =1 $,
which leads to the definition of the {\it growth index parameter} $ \gamma $ via
\beq
\gamma \equiv  \left. { \log f \over \log \Omega  }  \right \vert_{a=a_0} \; .
\label{eq:gamma_def}
\eeq
The latter has been the subject of increasingly accurate cosmological observations, for
some recent references see for example \cite{smi09,vik09,rap09}.
The solution of the differential equation for $\delta (a)$ with a $G(\Box)$ then gives 
an explicit value for the $\gamma$ parameter, for any values of the current
matter fraction $\Omega$.
Nevertheless, because of present observational constraints, one is mostly interested in the range
$\Omega \approx 0.25$.
Without a running Newton's constant $G$  [$ G = G_0 $, thus $c_a=0$ in 
Eq.~(\ref{eq:grun_a})] one finds $ f (a=a_0=1) = 0.4625 $ and
$ \gamma = \gamma_0 = 0.5562 $ for the standard $\Lambda CDM$ scenario with 
$ \Omega = 0.25 $.
On the other hand, when the running of $G (\Box)$ is taken into account, 
one finds from the solution to Eq.~(\ref{eq:dfq_delta_Gbox}) for the growth 
index parameter $ \gamma $ at matter density fraction $\Omega \approx 0.25 $
some significant corrections \cite{ht10}.


What is needed next is an estimate for the magnitude of 
the coefficient $c_a$ in Eq.~(\ref{eq:grun_a}) for $G(a)$ in terms of $c_t$ in 
Eq.~(\ref{eq:grun_t}) for $G(t)$, and ultimately in terms of $c_0$ in 
the original Eq.~(\ref{eq:grun_box}).
One has $ c_a = (t_0/\xi)^3 \cdot 0.45 \cdot c_0 $, with
$t_0$ corresponding to "today" so that $ t_0/\xi \approx 0.794 $, and $c_0 = 16.04$;
the additional factor of $0.45$ arises in relating the tensor $G(\Box)$ in 
Eqs.~(\ref{eq:grun_box}) to the $G(t)$ appropriate for the FRW background 
metric in Eq.~(\ref{eq:grun_t}), as computed in \cite{eff}.
Then in Eq.~(\ref{eq:grun_a}) one has
$c_a = 0.501 \cdot 0.45 \cdot 16.04 = 3.62 $, which gives a substantial
overall amplitude.
To quantitatively estimate the actual size of the correction in the above expressions
for the growth index parameter $\gamma$, and make some preliminary comparison
to astrophysical observations, some additional information is needed.


At first, one notices that all calculations done so far refer to the case of comoving wave 
number ${\bf q} = 0 $ in Eq.~(\ref{eq:fourier}).
If those numbers were used directly, one would obtain rather large ${\cal O}(1)$ 
quantum corrections to the growth parameter $\gamma$,
\beq 
\gamma =  \gamma_0 \, - \, \gamma_c \; .
\label{eq:gamma_rel}
\eeq
where $\gamma_0$ is the classical GR value, and $\gamma_c$ the leading quantum correction
in the limit ${\bf q}=0$ (which incidentally, in all cases looked at so far, turns out to be negative).
To obtain corresponding results for ${\bf q } \neq 0 $ would then require a new, and significantly more
complex, calculation which has not been done yet.


Nevertheless it seems clear that one can apply a simple scaling argument to obtain the more general result by a significantly shorter route.
One notes that the quantum correction in Eq.~(\ref{eq:gamma_rel}) is,
by virtue of the explicit form of $G(\Box)$ in Eqs.~(\ref{eq:grun_k}) and (\ref{eq:grun_box}),
always proportional to the inverse of the nonperturbative reference length scale cubed, 
$ \propto \, 1/\xi^3 $.


In the case of a matter-dominated universe ($w=0$ and $\lambda=0$) the results are
as follows.
For this case, $ a(t) = a_0 ( t / t_0 )^{2/3} $ helps relate the
$G(a)$ in Eq.~(\ref{eq:grun_a}) to $G(t)$ in Eq.~(\ref{eq:grun_t}).
One then solves the differential equation for $\delta (a) $,
Eq.~(\ref{eq:dfq_delta_Gbox_a}), with $G(a)$ given in Eq.~(\ref{eq:grun_a}),
and exponent $ \gamma_\nu = 3/2 \nu \simeq 9/2$ relevant for
a matter dominated background universe.
One finds \cite{ht10,ht11}
\beq 
\gamma =  \gamma_0 \, - \, \gamma_1 \,  \left ( { l_0 \over \xi } \right )^3  \; ,
\label{eq:gamma_rel_l0}
\eeq
with $\gamma_0 = 0.5562 $ the classical GR value, and
$\gamma_1 = 720.3 $ the amplitude computed for the quantum correction.
Quantitatively for this case the quantum correction gives roughly a 
1 \% effect on scales of $ l_0 \approx 110 \, Mpc $, a 
5 \% effect on scales of $ l_0 \approx 180 \, Mpc $, and a
10 \% effect on scales of $ l_0 \approx 230 \, Mpc $.


The shortcomings of the results of Eq.~(\ref{eq:gamma_rel}) for $w=0$
can be partially lifted by considering the case of an equation of state with $w \neq 0 $.
In general, if $w$ is not zero, one should use instead more generally 
Eq.~(\ref{eq:a_t}) to relate the variable $t$ to $a(t)$.
The problem here is that in practice for $w \neq 0$ at least
two effective $w$'s are involved,  $w=0$ (non-relativistic matter) and $w=-1$ ($\lambda$ term).
Unfortunately, this issue later complicates considerably the problem of relating 
$\delta G(t)$ to $ \delta G(a) $, and therefore the solution to the 
resulting differential equation for $\delta (a)$.
However, as a tractable approximation, one can use in the interim 
the slightly more general result for the scale factor $a(t)$ valid for any 
$w \neq 0$, namely $ a(t) = a_0 \, ( t / t_0 )^{2/3 (1+w)}$
(the extreme case of a vacuum energy dominated cosmology,
$w=-1$, is discussed in \cite{ht10,ht11} as well).
As an example we will use here an ``effective'' value of
$w \approx - 7/9$, which would seem more appropriate for the final target 
value of a matter density fraction $\Omega \approx 0.25$.
For this choice one then obtains a significantly reduced power in 
Eq.~(\ref{eq:grun_a}), namely $\gamma_\nu = 3 (1+w)/ 2 \nu = 1 $.
Then, although Eq.~(\ref{eq:dfq_delta_Gbox}) for $\delta (t)$ remains unchanged, 
Eq.~(\ref{eq:dfq_delta_Gbox_a}) for $\delta (a)$ need to be solved with new 
parameters.
Furthermore, the resulting differential equation for $\delta (a)$, 
Eq.~(\ref{eq:dfq_delta_Gbox_a}), is still relatively easy to solve, by
the same methods discussed earlier.
For this case as stated $ \gamma_\nu =1 $ in Eq.~(\ref{eq:grun_a}),
and one obtains a somewhat smaller 
correction compared to the matter dominated case $w=0$
of Eq.~(\ref{eq:gamma_rel_l0}), namely
\beq 
\gamma =  \gamma_0 \, - \, \gamma_1 \,  \left ( { l_0 \over \xi } \right )^3  \; ,
\label{eq:gamma_rel_l0_w}
\eeq
with $\gamma_0 = 0.5562$ the classical GR value and quantum correction $\gamma_1 = 224.1 $, 
a reduction of about a factor of three when compared to the pure non-relativistic
matter ($w=0$) result of Eq.~(\ref{eq:gamma_rel_l0}).
For comparison, in the Newtonian (non-relativistic) case the correction is found
to be much smaller, by about two orders of magnitude \cite{ht10}.
There one has $c_a \approx c_t \approx 2.7 \, c_0$, so the correction to the index 
$\gamma$ becomes $ - 0.0142 \cdot 2.7 \cdot 16.04  = - 0.62 $.
Then in Eq.~(\ref{eq:gamma_rel_l0}) still $\gamma_0 = 0.5562$, and for the
quantum correction one finds again a negative value with amplitude $\gamma_1 = 0.62 $.
This last result stresses again the fact that the quantum correction is
clearly relativistic in nature: the Newtonian answer is significantly smaller.
As an example, even on scales of $l_0 \sim 10 \, Mpc$ the correction to $\gamma$ 
here is tiny, $-4.1 \times 10^{-9}$.



So far a number of general features can be observed in the results, the first one 
being the fact that generally the quantum correction to the growth index 
parameter $\gamma$ is found to be {\it negative}.
On a more quantitative level, it may be of interest at this point to compare the results of Eq.~(\ref{eq:gamma_rel_l0})
(with, for concreteness, $\gamma_0 = 0.5562$ and a negative quantum correction with
amplitude $\gamma_1 = 224.1 $) with current astrophysical observations.
Then the above quantum prediction is roughly of a 
1 \% effect on scales of $ l_0 \approx 160 \, Mpc $, a 
5 \% effect on scales of $ l_0 \approx 270 \, Mpc $, and a
10 \% effect on scales of $ l_0 \approx 340 \, Mpc $.
Observationally, the largest  galaxy clusters and superclusters studied today up to redshifts 
$ z \simeq 1 $ extend for only about, at the very most, $1/20$ the overall size of the currently
visible universe;
in such cases the correction from Eq.~(\ref{eq:gamma_rel_l0}) to the classical
GR value is expected to amount to a negative 5 \% .
Recent observational bounds on x-ray studies of large galactic clusters at distance scales 
of up to about $1.4$ to $8.5 \, Mpc$ 
(comoving radii of $\sim 8.5 \, Mpc $ and  viral radii of $\sim 1.4 \, Mpc$) \cite{smi09,vik09} 
favor values for $\gamma= 0.50 \pm 0.08 $, and more recently values for $\gamma = 0.55+0.13-0.10 $
\cite{rap09}.
\footnote{
For recent detailed reviews of the many tests of general relativity on astrophysical scales,
and a more complete set of references, see for example \cite{uza02,uza09}.}
Taking for these cases a reference scale $ l_0 = 10 \, Mpc $ in Eq.~(\ref{eq:gamma_rel_l0_w})
one obtains a correction to $\gamma$ $\simeq {\cal O} ( 10^{-6} ) $ which is
rather tiny.
It is therefore clear that the quantum effects discussed here are only relevant for very large scales,
much bigger than those usually considered, and well constrained, by laboratory, solar
or galactic dynamics tests \cite{dam06,uza02,uza09,ade03}.
For now the galactic clusters in question are not large enough 
yet to see the quantum effect of $G(\Box)$, since after all the relevant scale in 
Eq.~(\ref{eq:grun_box}) is related to $\lambda$ and is
expected to be very large, $\xi \simeq 5320 \, Mpc $ [see Eq.~(\ref{eq:xi_mpc})].

In comparing the result for the gravitational slip function in the Newtonian gauge,
as given in Eq.~(\ref{eq:eta_l0}),
\beq
\eta ( l_0 ) \; = \;  - \, 13.7 \, \left ( { l_0 \over \xi }  \right )^3  \; 
\label{eq:eta_l0_1}
\eeq
to the result of Eq.~(\ref{eq:gamma_rel_l0_w}) for the matter density growth
parameter $\gamma$ just obtained
\beq 
{ \delta \, \gamma \over \gamma_0 } \; =  \; -  403. \left ( { l_0 \over \xi } \right )^3  \; 
\label{eq:gamma_rel_l0_w1}
\eeq
one notices that the latter correction is more than an order of magnitude larger.
So it seems the bound from matter density perturbations is much more stringent than the 
one derived from the slip function.

Indeed the nonperturbative amplitude coefficient $c_0$
enters {\it all} calculations involving $G(\Box)$ with the same magnitude and sign.
One can therefore relate one set of physical results to another, 
such as the quantum correction to the slip function $\eta (z=0)$, given in
Eq.~(\ref{eq:eta_z}), to the quantum corrections to the density perturbation growth 
exponent $\gamma$,  given in Eq.~(\ref{eq:gamma_rel_l0_w}).
Then after taking the ratio the amplitude coefficient conveniently $c_0$ drops out, and
one obtains for the ratio of the quantum corrections to the matter density
perturbation growth parameter $\gamma$
to the quantum slip function $\eta$ for $t=t_0$
\beq
{ \delta \, \gamma \over \delta \, \eta } \; \simeq \;  + \, 16.3   \; .
\eeq
This last result suggest that it will be observationally more difficult
 (by an order of magnitude) to see the quantum $G(\Box)$ correction in the slip function 
$\eta ( {\bf q}) $ than in the matter density growth parameter $\gamma ( {\bf q}) $.
Nevertheless, perhaps the value of the present calculations lies in the fact
that so far a discernible trend seems to emerge from the results,
and it suggests that the quantum correction to the growth exponent 
$\gamma$ is initially rather small for small clusters, negative in sign, 
but slowly increasing in magnitude, following a cubic law with scale.



\vskip 40pt

\section{Conclusions}

\label{sec:concl}

\vskip 20pt

The vacuum condensate picture of quantum gravitation
provides in principle a series of detailed and testable predictions, 
which could either be verified or disproved
in the near future as new and increasingly accurate satellite 
observations become available.
The previous sections laid out a number of specific prediction and estimates,
many originating in a rather direct way from the gravitational scaling 
dimensions and amplitudes of invariant gravitational correlations, 
obtained originally from a variety of different nonperturbative approaches, 
including  the Regge-Wheeler lattice formulation of gravity.

One key aspect linking all these calculations together is the fact that
once the nonperturbative scale $\xi$ is set (in analogy to
the $\Lambda_{\bar MS}$ of QCD) then, in accordance with
the renormalization group, there are no further adjustable 
parameters when discussing the universal, long distance limit.
The quantum theory of gravity is therefore, like the classical theory, 
again highly constrained by general coordinate invariance.
In more than one way the current calculations are still rather incomplete; in
particular the gravitational Wilson loop and the correlation between
loops have not been studied in detail yet, and only some rather 
general properties have been inferred. 
Nevertheless the feeling is still that the underlying formulation is solid 
enough to allow future controlled and improved estimates of key 
renormalization group quantities.

The previous discussion has made it clear that the derivation of many of the 
basic results has relied heavily  on subtle - but well established (and well grounded) -
renormalization group scaling arguments, and
there is no reason yet to doubt that such arguments should
be fully applicable to gravity as well.
Particularly encouraging is the fact that by now four different approaches to 
quantum gravity give comparable results for the phase structure and 
scaling dimensions 
(see the comparison Table I, as well as Figure 1), which suggests a unique 
underlying renormalization group universality class associated with the
massless spin two field in four dimensions.
Of course one important common element in all these approaches 
is the existence of a non-trivial fixed point in $G$ of the renormalization 
group equations.
A key physical aspect that emerges from the theory is a growth of Newton's $G$ 
with scale, quantified by a new, genuinely nonperturbative correlation length $\xi$, 
with the latter intimately related to the gravitational vacuum condensate (and thus
to the physical, observed cosmological constant).
The second key physical aspect is the existence of non-trivial scaling and anomalous dimensions
for gravitational n-point functions, which leads, among others, to specific predictions for 
matter density correlations and related quantities.
In conclusion, the main aspects of this new physical picture for gravity can be summarized as follows:

\begin{enumerate}

\item[ $\circ$ ]
The vacuum condensate picture of quantum gravity contains from the start
a very limited set of parameters, and is as a result strongly constrained.
It involves a new, genuinely nonperturbative scale [the gravitational
vacuum condensate, see Eqs.~(\ref{eq:xi_r}) and (\ref{eq:xi_lambda})], which 
relates the running of Newton's $G$ to the current observed value 
of the cosmological constant, and to the long distance behavior 
of physical diffeomorphism invariant curvature correlations.

\item[ $\circ$ ]
While in principle both signs could be possible, in the strong coupling limit 
the effective, long distance cosmological constant is 
positive [see Eq.~(\ref{eq:xi_r}), the arguments preceding it, and more detailed
discussion in \cite{loops}].
The basic argument relies of the behavior of the gravitational Wilson loop :
in the same strong coupling regime it seems impossible from the lattice theory
to obtain a negative value for the effective cosmological constant,
irrespective of the choice of boundary conditions (which, incidentally, 
in the lattice context play no role in the argument).

\item[ $\circ$ ]
The theory predicts a slow increase in strength of the
gravitational coupling when very large, cosmological scales are
approached [see Eqs.~(\ref{eq:grun_k}) and (\ref{eq:grun_box})].
In this context the observed scaled cosmological constant $\lambda$
is seen to act as a dynamically induced infrared cutoff, similar to
what happens in non-Abelian gauge theories.
In principle, both the universal power and amplitude for this infrared growth
are calculable from first principles in the underlying lattice theory.

\item[ $\circ$ ]
The lattice theory appears to exclude the possibility of 
a physically acceptable phase with gravitational screening.
The perturbative, weak coupling (small $G$) phase is found to be inherently unstable
in the lattice formulation, a consequence of the conformal instability. 
Thus a genuinely semiclassical regime for quantum gravity, whereby quantum effects
can be included as small perturbations, does not seem to exist.
On the other hand for large enough quantum fluctuations (large $G$)
the conformal instability is overcome, and a new stable, anti-screening phase emerges.
The stability of quantum gravity can thus be viewed as an entropy effect, 
intimately connected to non-trivial properties of the gravitational functional measure.

\item[ $\circ$ ]
Calculations presented here give a number of specific predictions for the behavior of
invariant curvature correlations as a function of geodesic distance, 
and specifically the powers and
amplitudes involved [see Eqs.~(\ref{eq:corr_pow1}) and (\ref{eq:corr_pow2})].
Perhaps the most important result is the fact that the curvature
correlation function decays like inverse distance squared ($n=1$ and thus $s=1$ and $\gamma=2$).
This in turn can be used to relate in a standard way, via the quantum equations of motion, 
curvature correlations to matter density correlations and thus to their observed power spectrum
[see Eqs.~(\ref{eq:grho_a0}) and (\ref{eq:corr_pow3})].

\item[ $\circ$ ]
Given the exceptionally large value of the scale $\xi$ (originating from the fact that the 
observed $\lambda$ is very small compared to the scale associated with $G$) 
no observable deviations from classical General Relativity are expected on
laboratory, solar systems and even galactic scales [see Eqs.~(\ref{eq:grun_r}) and (\ref{eq:pot})].
  
\end{enumerate}

In addition, the picture outlined here points to what appears to be a deep analogy 
between the nonperturbative vacuum state of quantum gravity and 
known properties of strongly coupled non-Abelian gauge theories, 
and specifically QCD.
Indeed in QCD one also finds a nonperturbative mass parameter 
$m = 1/ \xi$ (sometimes referred to as the mass gap) which 
is known to arise dynamically without nevertheless violating any local gauge symmetries,
and is understood to be a renormalization group invariant.
That such a mass scale can be generated dynamically is a highly 
nontrivial outcome of the strong coupling dynamics of QCD,
and its associated renormalization group equations.
Over time this analogy has been of great help in illustrating properties
of quantum gravity, many of which are ultimately still based on fundamental
principles of the  renormalization group, connected with universal
scaling properties as they apply to the vicinity of a nontrivial ultraviolet fixed point.


\vspace{20pt}

{\bf Acknowledgements}

The author is grateful for useful discussions and correspondence
with professors Paul Frampton, Holger Gies, James Hartle, Giorgio Parisi,
Roberto Percacci, Gabriele Veneziano and Ruth Williams.
Large scale numerical calculations were performed on supercomputers located at
the Max Planck Institut f\" ur Gravitationsphysik (Albert-Einstein-Institut)
in Potsdam, Germany.

\newpage


\vfill


\end{document}